\documentclass[useAMS,usenatbib,fleqn,a4paper]{mn2e}
\usepackage{epsfig}
\usepackage{deluxetable} 
\usepackage{url}
\usepackage{rotating}
\usepackage{amsmath}
\usepackage{graphicx}
\usepackage{setspace}
\usepackage{psfig}
\usepackage{natbib}
\usepackage{float}
\usepackage{placeins}
\usepackage{booktabs}

\title[Gamma-ray-emitting NLSy1: the {\em Swift} view]{Gamma-ray-emitting narrow-line Seyfert 1 galaxies: the {\em Swift} view}
\author[D'Ammando]{F. D'Ammando$^{1}$\thanks{E-mail: dammando@ira.inaf.it} \\
$^{1}$INAF -- Istituto di Radioastronomia, Via Gobetti 101, I-40129 Bologna, Italy\\
}

\begin{document}

\date{Accepted. Received; in original form}

\maketitle

\label{firstpage}

\begin{abstract}

We report the analysis of all {\em Swift} observations available up to 2019 April of $\gamma$-ray-emitting narrow-line Seyfert 1 galaxies (NLSy1). The distribution of X-ray luminosities (and fluxes) indicates that the jet radiation significantly contributes to their X-ray emission, with Doppler boosting making values higher than other radio-loud NLSy1. The 0.3--10 keV photon indices are on average harder with respect to radio-quiet and radio-loud NLSy1, confirming a dominant jet contribution in X-rays. However, the lower variability amplitude with respect to blazars and the softening of the spectrum in some periods suggests that also the corona radiation contributes to the X-ray emission. In optical and ultraviolet (UV) significant flux changes have been observed on daily, weekly, and monthly time-scale, providing a clear indication of the significant contribution of the jet radiation in this part of spectrum. A strong correlation between X-ray, UV, and optical emission and simultaneous flux variations have been observed in 1H 0323$+$342, SBS 0846$+$513, PMN J0948$+$0022 as expected in case the jet radiation is the dominant mechanism. Correlated multiband variability favours the jet-dominated scenario also in FBQS J1644$+$2619 and PKS 2004--447. The summed X-ray Telescope spectra of 1H 0323$+$342, SBS 0846$+$513, PMN J0948$+$0022, and FBQS J1644$+$2619 are well fitted by a broken power law with a break around 2 keV. The spectrum above 2 keV is dominated by the non-thermal emission from a beamed relativistic jet, as suggested by the hard photon index. A Seyfert-like feature like the soft X-ray excess has been observed below 2 keV, making these $\gamma$-ray-emitting NLSy1 different from typical blazars.  
  
\end{abstract} 

\begin{keywords}
radiation mechanisms: non-thermal -- galaxies: active -- galaxies: jets -- galaxies: Seyfert -- ultraviolet: galaxies -- X-rays: galaxies
\end{keywords}

\section{Introduction}
  
Narrow-line Seyfert 1 galaxies (NLSy1) are a subclass of active galactic nuclei (AGN) identified by their optical properties \citep[e.g.,][]{pogge00}, in particular relatively narrow permitted optical emission lines \citep[i.e., full width at half maximum, FWHM H$\beta$ $<$ 2000 km $^{-1}$;][]{goodrich89}. Their permitted lines are only slightly broader than forbidden lines, with a weak [O$_{\rm\,III}$]$\lambda$5007 emission line (i.e., [O$_{\rm\,III}$]/H$\beta$ $<$ 3), a criterion more typical for Seyfert 1 galaxies with respect to Seyfert 2 galaxies \citep{osterbrock85}, and strong Fe$_{\rm\,II}$ emission lines \citep{goodrich89}. Such properties place NLSy1 at the lower end of the line width distribution for the Seyfert 1 galaxies, distinguishing them from the broad-line Seyfert 1 galaxies (BLSy1). Only a small fraction of NLSy1 \citep[$<$ 7$\%$;][]{komossa06, zhou06, rakshit17} are classified radio loud (i.e. $R$ $>$ 10, being radio loudness $R$ defined as ratio of rest-frame 1.4 GHz and 4400 \AA\, flux densities), while $\sim$15$\%$ of quasars are~radio loud \citep[e.g.,][]{kellerman16}. The number of NLSy1 with radio loudness higher than 100 is even more small ($\sim$2-3\%).

In X-rays, NLSy1 galaxies exhibit an extreme behaviour in terms of flux and spectral variability with respect to BLSy1 galaxies \citep[e.g.,][]{turner99}. ROSAT observations of NLSy1 in the 0.1--2.4 keV energy range have shown steeper slopes with respect to BLSy1 \citep{boller96}. This phenomenon has been confirmed by Advanced Satellite for Cosmology and Astrophysics (ASCA) 
observations over an extended energy range \citep[i.e., 0.3--10 keV, e.g.,][]{brandt97, vaughan99}. This may be an indication that the primary source of X-ray emission, the corona, is different in NLSy1 with respect to BLSy1 in terms of geometry and temperature. Moreover, a remarkable variability has been observed in X-rays for NLSy1 \citep[e.g., IRAS 13224--3809, PHL 1092;][]{boller97, brandt99}, showing the most rapid and largest amplitude variations seen in radio-quiet objects. 

\noindent In addition, NLSy1 show remarkably strong soft X-ray excess \citep[e.g.,][]{boller96, leighly99, crummy06, zhou06}. The origin of the soft X-ray excess is a  matter of debate both in radio-quiet and radio-loud AGN \citep[e.g.,][]{gierlinski04, piconcelli05}. Different scenarios have been proposed to explain this feature: thermal emission from the disc \citep[e.g.,][]{turner89}, relativistically blurred reflection from the accretion disc \citep[e.g.,][]{ballantyne01}, relativistically smeared absorption from disc wind \citep[e.g.,][]{gierlinski04}, or a warm Comptonization component \citep[e.g.,][]{magdziarz98}. Usually, it is not possible to clearly distinguish between the different models on a statistical basis. Another possibility is that the soft excess originates from the jet itself, with the high-energy tail of the synchrotron emission extends to the soft X-ray range \citep[e.g.,][]{sambruna00}. Alternatively, bulk Comptonization of disc or broad-line region (BLR) photons by cold electrons moving along the jet has been proposed as a possible mechanism to produce the soft excess \citep[]{celotti07}.

The extreme properties of NLSy1 may be related to an extreme value of fundamental physical parameters related to the accretion process. One possibility is that NLSy1 have a supermassive black hole (SMBH) with relatively low masses (i.e., 10$^{6}$--10$^{7}$ M$_\odot$) with respect to BLSy1 with similar luminosities. However, the BH masses estimated by means of the virial method and usually reported in literature have been challenged by several authors for different reasons. In case of highly accreting AGN, the~effects of radiation pressure on the BLR clouds should be higher with the consequence of having underestimated masses~\citep{marconi08}. In~case of a disc-like geometry of the BLR, as proposed for NLSy1, projection effects could explain the smaller masses estimated with the virial method in NLSy1 \citep{decarli08}. All that is in agreement with the mean BH mass of $\sim$10$^{8}$ M$_\odot$ obtained by \citet{viswanath19} for both radio-quiet and radio-loud NLSy1 by fitting their optical spectra \citep[see also][]{calderone13}. The BH mass estimates obtained by fitting accretion disc models to the spectra of NLSy1 are similar to the values obtained for BLSy1.
 
\noindent Past studies suggested that NLSy1 may not be a homogeneous class. Not only optically and X-ray selected NLSy1 seems to have different properties \citep[e.g.,][]{grupe04} but jetted and non-jetted NLSy1 \footnote{Based on the presence or not of a strong relativistic jet, see the discussion in \citet{padovani17}.} are identified in the last years \citep[e.g.,][]{foschini15,dammando16}. Observations with the Large Area Telescope (LAT) on board the {\em Fermi Gamma-ray Space Telescope} satellite have revealed NLSy1 as a new class of $\gamma$-ray-emitting AGN with several properties similar to the blazars. In particular, compared to the population of blazars, the jetted NLSy1 are similar to the flat spectrum radio quasars (FSRQ), typically at lower $\gamma$-ray luminosities \citep[e.g.,][]{dammando16,dammando19}. It is a very small class, consisting of only nine sources classified as {\it bona-fide} NLSy1 in the Fourth Source {\em Fermi}-LAT catalogue \citep[4FGL,][]{abdollahi20}. Other $\gamma$-ray-emitting AGN are associated with a candidate NLSy1 in literature \citep[see e.g. section 2 in][]{dammando19}, these sources are not considered in this work. 
 
NLSy1 are usually hosted in late-type spiral galaxies \citep[e.g.,][]{krongold01, deo06, ohta07, orban11, mathur12}, although~some of them have been found in early-type S0 galaxies \citep[e.g., Mrk 705 and Mrk 1239,][]{markarian89}. The~presence of a relativistic jet in this class of object seems to be in contrast to the paradigm that formation of relativistic jets could happen only in very massive galaxies \citep[e.g.,][]{boettcher02,marscher10} and poses intriguing questions about the nature, disc/jet connection, high-energy emission mechanisms, and formation of powerful relativistic jets in the different class of AGN. The indication of beamed emission in some NLSy1 suggests that orientation effects play a part in these sources. The X-ray band is an interesting regime, where both accretion disc, corona and jet potentially make strong contributions in these sources: thermal or reprocessed soft X-rays from the disc; inverse Compton (IC) scattering of disc photons by a hot corona electrons in hard X-rays; IC scattering of soft photons by relativistic non-thermal electrons in the jet in both soft and hard X-rays. For this reason, a systematic study of {\em Neil Gehrels Swift Observatory} observations of the nine $\gamma$-ray-emitting NLSy1 included in the 4FGL catalogue is proposed here. 

Together with the X-ray regime, simultaneous optical and ultraviolet (UV) observations obtained with the UVOT telescope on board {\em Swift} will be analysed to investigate variability properties and the connection with the X-ray activity. The comparison with the $\gamma$-ray data collected by {\em Fermi}-LAT and the study of the spectral energy distribution (SED) of the sources will be presented in a forthcoming paper. 

The paper is organized as it follows. Section~\ref{Swift_obs} describes the {\em Swift} observations and data analysis. Flux, luminosity, and spectral variability are discussed in Section~\ref{variability}, while the summed X-ray spectral analysis is presented in Section~\ref{summed}. In Section~\ref{summary} the results are summarized.

Unless stated otherwise, uncertainties correspond to 90 per cent confidence limits on one parameter of interest ($\Delta\chi^{2}$ = 2.7). The photon indices are parameterized as $N(E) \propto E^{-\Gamma}$ with $\Gamma = \alpha +1$ ($\alpha$ is the spectral index). Throughout this paper, we assume the following cosmology: $H_{0} = 71\; {\rm km \; s^{-1} \; Mpc^{-1}}$, $\Omega_{\rm M} = 0.27$, and $\Omega_{\rm \Lambda} = 0.73$ in a flat Universe \citep{planck16}.    

\section{{\em Swift} observations}\label{Swift_obs}

In order to enhance our knowledge of $\gamma$-ray-emitting NLSy1, it is necessary to have a long-term monitoring at different frequencies of these sources. In this context, the {\em Neil Gehrels Swift Observatory} satellite \citep{gehrels04} provides unique capabilities with simultaneous observations from optical to hard X-ray bands and the possibility to have several observations over a long time-scale. The {\em Swift} observations were performed with all three instruments on board: the X-ray Telescope \citep[XRT,][0.2--10.0 keV]{burrows05}, the Ultraviolet/Optical Telescope \citep[UVOT,][170--600 nm]{roming05} and the Burst Alert Telescope \citep[BAT,][15--150 keV]{barthelmy05}. 

Table~\ref{Swift_Obs} reports the name, redshift, and $\gamma$-ray luminosity in the 0.1--300 GeV energy range (based on the values reported in the 4FGL catalogue) of the sources in the sample, together with the number of {\em Swift} observations carried out up to 2019 April, the date of the first and last observations analysed, and the Galactic hydrogen absorbing column in the direction of the source.  Among the nine $\gamma$-ray-emitting NLSy1 included in the 4FGL catalogue, only TXS 2116--077 has not been observed by {\em Swift} in the period studied here.\footnote{New high-quality spectroscopic data of this source disfavour its classification as an NLSy1 \citep{jarvela20}.}

The hard X-ray flux of these sources turned out to be below the sensitivity of the BAT instrument for such short exposures and therefore the data of the single observations from this instrument will not be used. Moreover, in the hard X-ray band only one $\gamma$-ray-emitting NLSy1 is reported in the {\em Swift}-BAT 105-month catalogue~\citep{oh18}: 1H 0323$+$342, with~a photon index of 1.62 $\pm$ 0.30 and an X-ray luminosity of 2.45 $\times$ 10$^{44}$ erg s$^{-1}$ in the 14--195 keV band\footnote{In addition, PMN J0948$+$0022 is reported in the preliminary 100-month {\em Swift}-BAT Palermo catalogue \url{http://bat.ifc.inaf.it/100m\_bat\_catalog/100m\_bat\_catalog\_v0.0.htm}.}. The photon index of 1H 0323$+$342, compatible within the uncertainties with the value obtained with {\em NuSTAR} in the 3--79 keV energy range \citep[1.80 $\pm$ 0.01,][]{landt17}, is harder than the values usually observed in other NLSy1 \citep[e.g.,][]{dadina07,malizia08}.   

The analysis procedure applied to the XRT and UVOT data is described in Sections~\ref{SwiftXRT} and \ref{SwiftUVOT}, respectively.  

\begin{table*}
\centering
\caption{The sample of $\gamma$-ray-emitting radio-loud NLSy1 studied.}
\label{Swift_Obs}
\begin{tabular}{cccccccc}
\hline
\textbf{Source name} & \textbf{Redshift} & \boldmath{\textbf{L}$_{\gamma}$} & \textbf{Number {\em Swift} observations} & \textbf{First observations} & \textbf{Last observations} & \textbf{$N_{\rm\,HI}$} \\ 
                     & $z$               & (erg s$^{-1}$)                   &                                 &                    &                   & (cm$^{-2}$)       \\
\hline
1H 0323$+$342       &  0.061  & 2.1 $\times$ 10$^{44}$ & 136 & 2006-07-05 & 2018-12-13 &  1.27 $\times$ 10$^{21}$ \\
SBS 0846$+$513      &  0.584  & 3.2 $\times$ 10$^{46}$ &  32 & 2011-08-30 & 2019-04-24 &  2.91 $\times$ 10$^{20}$ \\
PMN J0948$+$0022    &  0.585  & 7.5 $\times$ 10$^{46}$ &  45 & 2008-12-05 & 2016-06-24 &  5.08 $\times$ 10$^{20}$ \\
IERS 1305$+$515     &  0.787  & 6.9 $\times$ 10$^{45}$ &   3 & 2012-06-04 & 2012-08-21 &  1.12 $\times$ 10$^{20}$ \\
B3 1441$+$476       &  0.705  & 4.7 $\times$ 10$^{45}$ &  11 & 2012-03-22 & 2012-06-28 &  1.66 $\times$ 10$^{20}$ \\
PKS 1502$+$036      &  0.408  & 1.0 $\times$ 10$^{46}$ &  16 & 2009-07-25 & 2017-02-12 &  3.93 $\times$ 10$^{20}$ \\
FBQS J1644$+$2619   &  0.145  & 2.7 $\times$ 10$^{44}$ &  11 & 2011-12-26 & 2018-01-18 &  5.14 $\times$ 10$^{20}$ \\
PKS 2004--447       &  0.240  & 1.7 $\times$ 10$^{45}$ &  32 & 2011-05-15 & 2016-10-24 &  3.17 $\times$ 10$^{20}$ \\
\hline
\end{tabular}
\end{table*}

\subsection{{\em Swift}-XRT data analysis}
\label{SwiftXRT}

All XRT observations were performed in photon counting (PC) mode, except for 1H 0323$+$342 for which observations in PC and windowed timing (WT) modes were carried out, depending on the brightness of the source. The XRT spectra were generated with the {\em Swift}-XRT data products generator tool at the UK Swift Science Data Centre\footnote{http://www.swift.ac.uk/user\_objects} \citep[for details see][]{evans09}. Spectra having count rates higher than 0.5 counts s$^{-1}$ may be affected by pile-up. To correct for this effect the central region of the image has been excluded, and the source image has been extracted with an annular extraction region with an inner radius which depends on the level of pile-up \citep[see e.g.,][]{moretti05}.

The X-ray spectra in the 0.3--10 keV energy range is fitted by an absorbed power-law model using the photoelectric absorption model \texttt{tbabs} \citep{wilms00} with an H$_{\rm\,I}$ column density consistent with the Galactic value in the direction of the source as reported in \citet[][see Table~\ref{Swift_Obs}]{kalberla05}. A large number of spectra show low number of counts (i.e. $<$ 200), therefore not allowing us to use the $\chi^{2}$ statistics. To maintain the homogeneity in the analysis, the spectra are grouped using the task \texttt{grppha} to have at least one count per bin and the fit has been performed with the Cash statistics \citep{cash79}. We used the spectral redistribution matrices in the Calibration database maintained by \textsc{HEASARC}. The X-ray spectral analysis was performed using the \textsc{XSPEC 12.9.1} software package \citep{arnaud96}. The results of the fit are reported in Tables A1--A6 in Appendix~A. Single observations with a number of counts $<$ 15, for which it is not possible to constrain the spectral parameters, are not included in the tables and are not considered for variability studies (see Section~\ref{variability}). In case of IERS 1305$+$515 and B3 1441$+$476 the number of counts in the single observations is too low to perform a spectral fit and only the summed spectrum will be fitted for these two sources (see Section~\ref{summed}). 

For each source, all XRT observations were also combined together in order to produce an average spectrum with a higher signal-to-noise ratio with respect to the single observations. The quality of the summed spectra enabled the use of $\chi^{2}$ statistics. The obtained spectra are grouped using the task \texttt{grppha} to have at least 20 count per bin. The summed spectra were then analysed using three different models: a simple power-law, a broken power-law, and a black body plus power-law model. In all cases a photoelectric absorption fixed to the  value from the H$_{\rm\,I}$ maps of \citet{kalberla05} has been added. The single power-law and broken power-law models are compared by applying an $F$-test. A model has been preferred over the other if the probability of null hypothesis obtained in the $F$-test is $<$ 0.01, corresponding to 99\% confidence interval.   

\subsection{{\em Swift}-UVOT data analysis}
\label{SwiftUVOT}

During the {\em Swift} pointings, the UVOT instrument observed the sources in its optical ($v$, $b$, and $u$) and UV ($w1$, $m2$, and $w2$) photometric bands \citep{poole08,breeveld10}. UVOT data in all filters\footnote{In case of multiple images in the same filter for the same observation, we analyse the first image, usually the one with the longest exposure.} were analysed with the \texttt{uvotsource} task included in the {\textsc HEASoft} package (v6.26.1) and the 20170922 CALDB-UVOTA release. Source counts were extracted from a circular region of 5 arcsec radius centred on the source, while background counts were derived from a circular region with 20 arcsec radius in a nearby source-free region.
The UVOT magnitudes are corrected for Galactic extinction using the $E(B-V)$ value from \citet{schlafly11} and the extinction laws from \citet{cardelli89} and converted to flux densities using the conversion factors from \citet{breeveld10}. Observed magnitudes are reported in Tables B1--B6 in Appendix~B.   

\section{Variability properties}\label{variability}

Thanks to the high number of observations spanning a period between 5 and 12 yr, it is possible to investigate the variability properties of all $\gamma$-ray-emitting NLSy1 of the sample, excluding  IERS 1305$+$515 and B3 1441$+$476, in terms of flux and spectral changes. 

\noindent In Sections~\ref{Var_flux} and \ref{Var_spec}, the results of the analysis of flux, luminosity and spectral variability, respectively, are discussed. 

\begin{figure*}
\begin{center}
\rotatebox{0}{\resizebox{!}{57mm}{\includegraphics{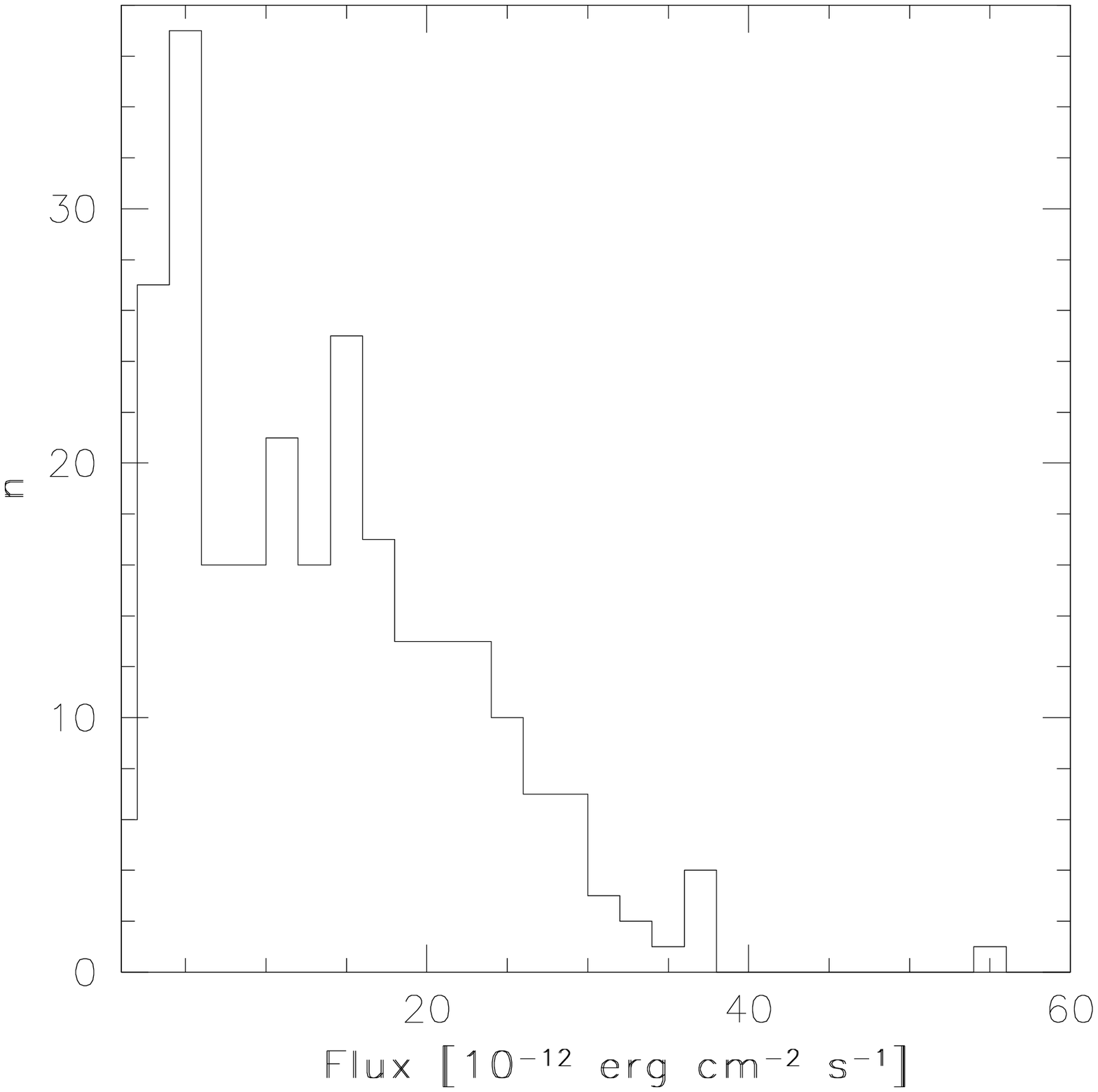}}}
\hspace{0.05cm}
\rotatebox{0}{\resizebox{!}{57mm}{\includegraphics{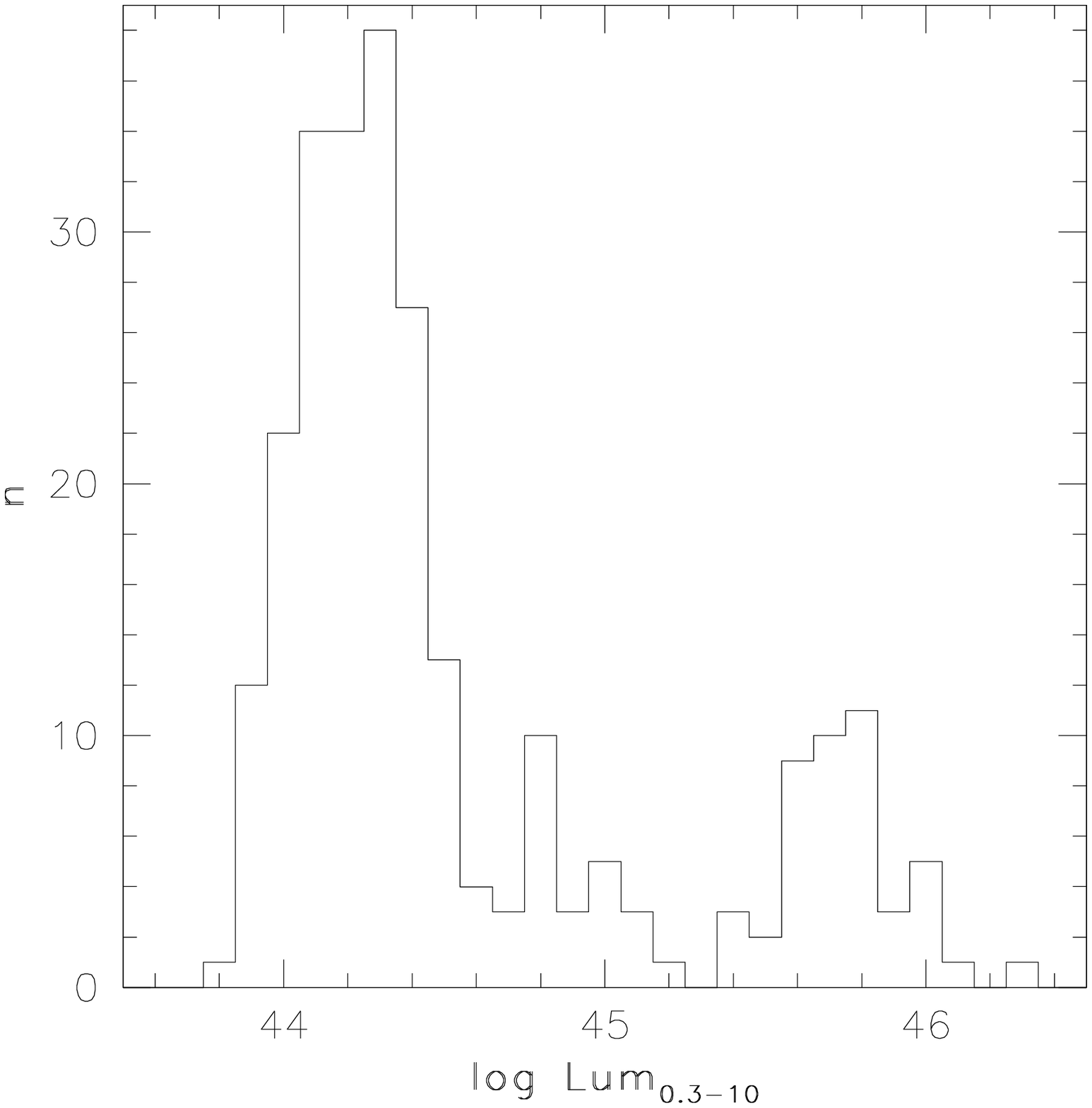}}}
\hspace{0.05cm}
\rotatebox{0}{\resizebox{!}{57mm}{\includegraphics{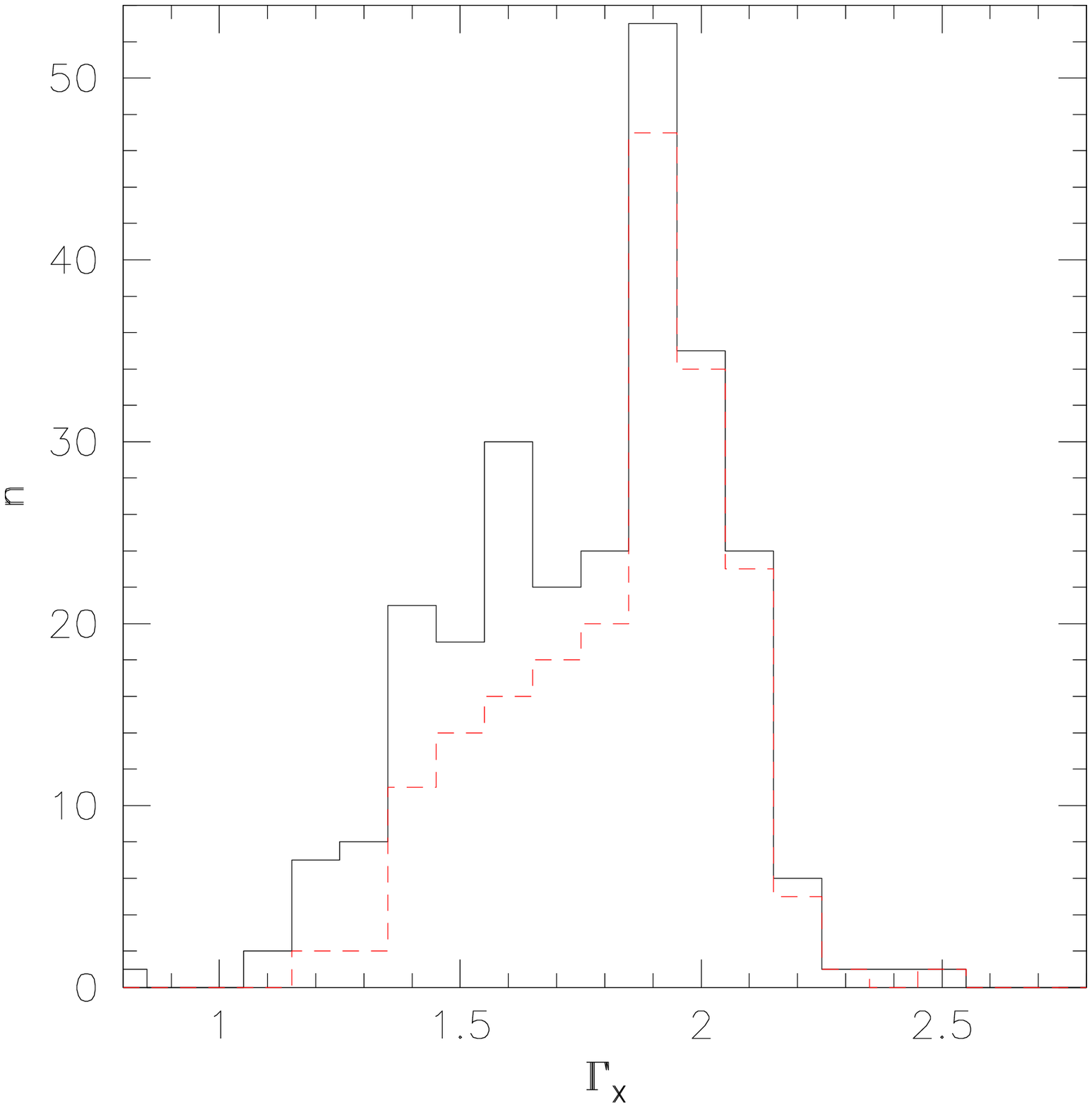}}}
\caption{{\it Left}: histogram of the distribution of the X-ray flux estimated in the 0.3--10 keV energy range for each observation of the sources in the sample. The fluxes are corrected for Galactic extinction. {\it Centre}: histogram of the distribution of the intrinsic 0.3--10 keV luminosity of the sources in the sample. {\it Right:} histogram of the distribution of the X-ray photon index estimated in the 0.3--10 keV energy range for the sources in the sample. The red dashed line refers to values with associated error less than 0.3, while solid black line refer to all values.}
\label{hist}
\end{center}
\end{figure*}
  
\subsection{Flux variability}
\label{Var_flux}

The X-ray fluxes (corrected for the Galactic extinction) of the $\gamma$-ray-emitting NLSy1 estimated in the 0.3--10 keV energy band  with {\em Swift}-XRT observations span the range between 1.1$\times$10$^{-12}$ and 5.6$\times$10$^{-11}$ erg cm$^{-2}$ s$^{-1}$ (see Fig.~\ref{hist}, left-panel). As a comparison, the X-ray fluxes estimated for a sample of radio-loud NLSy1 (excluding the $\gamma$-ray-emitting NLSy1) from {\em Swift}, {\em XMM--Newton}, {\em Chandra}, and ROSAT observations vary between 5.1$\times$10$^{-15}$ and 4.6$\times$10$^{-12}$ erg cm$^{-2}$ s$^{-1}$ \citep{foschini15}\footnote{The fluxes reported in \citet{foschini15} are not corrected for Galactic extinction. However, the $N_H$ values for these sources varies between 0.74 and 8.27$\times$10$^{20}$ cm$^{-2}$, therefore the correction for this effect is not significant.}. Therefore, the $\gamma$-ray-emitting NLSy1 usually show X-ray fluxes higher than the values observed for other radio-loud NLSy1. Fluxes lower than 5$\times$10$^{-12}$ erg cm$^{-2}$ s$^{-1}$ (i.e., the highest flux observed from the sample of radio-loud NLSy1 not detected in $\gamma$ rays) have been observed also in case of $\gamma$-ray-emitting NLSy1, but just for $\sim$10 per cent of the {\em Swift} observations.

\noindent Based on the XRT fluxes, we have also computed the 0.3--10 keV intrinsic luminosity for each observation of the $\gamma$-ray-emitting NLSy1 (see Fig.~\ref{hist}, central panel). The values vary between 6.3$\times$10$^{43}$ and 1.8$\times$10$^{46}$ erg s$^{-1}$, with an average luminosity of 1.3$\times$10$^{45}$ erg s$^{-1}$. The low-luminosity tail of the distribution is mainly due to the closest source, that is 1H 0323$+$342. As a comparison, the intrinsic X-ray luminosity of the other radio-loud NLSy1 studied in \citet{foschini15} with $z$ $<$ 0.585 (i.e., the redshift of the farthest $\gamma$-ray-emitting NLSy1 studied here) span the range between 2.0$\times$10$^{41}$ and 8.5$\times$10$^{44}$ erg s$^{-1}$, with an average luminosity of 2.6$\times$10$^{44}$ erg s$^{-1}$. This comparison indicates that $\gamma$-ray-emitting NLSy1 show higher X-ray luminosities with respect to the other radio-loud NLSy1 in the same range of redshift.  

In case of $\gamma$-ray-emitting NLSy1, together with the presence of unbeamed emission from the disc and corona, a beamed relativistic jet emission can contribute to the X-ray emission. The high $\gamma$-ray luminosity of these sources (see Table \ref{Swift_Obs}) is a clear indication of a small viewing angle of the relativistic jet with respect to the observers, therefore of a high beaming factor of the non-thermal emission from the jet. The range of X-ray fluxes and luminosities observed for $\gamma$-ray-emitting NLSy1 suggests a significant contribution of the jet emission also in the X-ray regime, with the Doppler boosting effects making the values higher with respect to the other radio-loud NLSy1, for long periods and not only for short activity periods.

In order to take into account the possible effect of the Doppler boosting on the X-ray emission of the $\gamma$-ray-emitting NLSy1, it is worth comparing their fluxes to the values observed for blazars, that is the other class of jetted AGN with small angles between our line of sight and the jet. In this context, the unabsorbed 0.3--10 keV flux of 22 bright $\gamma$-ray blazars (11 FSRQ and 11 BL Lac) monitored by {\em Swift}-XRT during 2004 December--2012 August varies between 3$\times$10$^{-13}$ and 6$\times$10$^{-9}$ erg cm$^{-2}$ s$^{-1}$ \citep{stroh13}. In particular, the fluxes of FSRQ range between 3$\times$10$^{-13}$ and 4$\times$10$^{-10}$ erg cm$^{-2}$ s$^{-1}$, while the fluxes of BL Lacs between 8$\times$10$^{-13}$ and 6$\times$10$^{-9}$ erg cm$^{-2}$ s$^{-1}$. The higher fluxes observed in BL Lacs may be related to the fact that in the 0.3--10 keV energy range covered by {\em Swift}-XRT we are observing the synchrotron peak, produced by the most energetic electrons, in BL Lac objects and the onset of the IC emission, produced by less energetic electrons, in FSRQ. The X-ray flux level of $\gamma$-ray-emitting NLSy1 seems to be similar to what is observed for FSRQ. This is in agreement with the fact that the synchrotron peak of these sources has been estimated in the 10$^{12}$--10$^{13}$\,Hz frequency range, except for 1H 0323+342 for which the synchrotron peak is estimated at 2$\times$10$^{15}$\,Hz, as reported in the fourth catalogue of AGN detected by {\em Fermi}-LAT \citep[4LAC;][]{4LAC}. 
 
\noindent The corresponding 0.3--10 keV intrinsic luminosity of the blazars studied by \citet{stroh13} with a redshift $z$ $<$ 0.585 ranges between 2.2$\times$10$^{43}$ and 3.5$\times$10$^{46}$ erg s$^{-1}$, similar to the range of luminosities observed in the $\gamma$-ray-emitting NLSy1. This is a further indication that both in blazars and $\gamma$-ray-emitting NLSy1 the X-ray emission is amplified by relativistic effects, as expected if the emission is produced by a jet observed at small angle of view with respect to the observers. 

Radio-quiet NLSy1 are usually highly variable in X-rays with high-amplitude outbursts \citep[e.g.,][]{grupe04, fabian12, parker17}, showing variability amplitude a factor of 10 and more on different time-scales. Different effects has been proposed to be the origin of the X-ray variability in those objects (e.g., intrinsic disc instabilities, coronal flaring, and hotspots) including also relativistic boosting or beaming of the emission \citep[e.g.,][]{leighly99}, as observed for blazars. For this reason, it is interesting to study the X-ray variability in these $\gamma$-ray-emitting NLSy1.  

\noindent Correcting for instrumental artifacts (i.e., hot pixels and bad columns on the CCD), pile-up, and after the background has been subtracted, a search for variability on short time-scales in X-rays by using segment binning results in four significant ($>$ 3-$\sigma$) increase of the count rate in consecutive XRT observation segments for 1H 0323$+$342 (on 2013 July 19), FBQS J1644$+$2619 (on 2015 April 9 and June 6), and PKS 2004--447 (on 2013 July 7), with a temporal distance between the two segments ($\tau$) of $\sim$6 ks for 1H 0323$+$342, $\sim$12 ks and $\sim$17 ks for FBQS J1644$+$2619, and $\sim$22 ks for PKS 2004--447. These short-term variability episodes are likely due to fast variability of the jet component. Based on causality argument, in this scenario, it is possible to constrain the intrinsic size of the emitting region to be $R$ $<$ $c$ $\delta$ $\tau$ /(1+$z$) = 1.8$\times$10$^{15}$\,cm for 1H 0323$+$342, 3.1$\times$10$^{15}$\,cm and 4.5$\times$10$^{15}$\,cm for FBQS J1644$+$2619, and 5.3$\times$10$^{15}$\,cm for PKS 2004--447 (assuming a typical Doppler factor $\delta$ = 10), suggesting that the X-ray emission is produced in compact regions within the jet. We cannot ruled out that, in the context of thermal Comptonization from the corona, these rapid events are driven by changes in accretion rate or in geometry of the inner accretion flow, as observed in other NLSy1 \citep[e.g.,][]{alston20}. However, in case of 1H 0323$+$342 the rapid X-ray variability has been observed during a $\gamma$-ray flaring activity \citep{paliya14}, confirming unambiguously as the jet be the origin of this episode. Moreover, the X-ray spectrum of PKS 2004--447 is well fitted in the 0.3--10 keV by a single power law with a hard spectrum representative of a relativistic jet component with no clear evidence of Seyfert-like features from the accretion flow, like the soft X-ray excess \citep[see Section \ref{summed} and][]{orienti15}.  

Among the sample studied here, optical intraday variability has been already reported for PMN J0948$+$0022 \citep{liu10, maune13, itoh13}, SBS~0846$+$513 \citep{maune14, paliya16}, and~1H 0323$+$342 \citep{itoh14, ojha19}. The detection of intraday variability in these three NLSy1 has provided evidence about the relativistically beamed synchrotron emission been the dominant mechanism, similar to what is observed for blazars \citep[e.g.,][]{heidt96, sagan04}. On the contrary, no optical intraday variability has been observed for PKS 1502$+$036 \citep{ojha19}. In optical and UV bands, a daily monitoring\footnote{A dedicated analysis of the individual segments collected for each {\em Swift}-UVOT observation of the six sources will be presented in a separate paper.} has been carried out by {\em Swift}-UVOT only for 1H 0323$+$342, for which a maximum increase of 0.45 mag in two consecutive days has been observed between 2010 November 26 and 27. Similar rapid changes in one day, of lower amplitudes, have been observed for the same source. On a weekly time-scale, significant changes has been observed for SBS 0846$+$513 ($\Delta$mag = 2.20 in 5 d, $b$ band), PMN J0948$+$0022 ($\Delta$mag = 1.95 in 5 d, $v$ band), and PKS 1502$+$036 ($\Delta$mag = 0.81 in 3 d, $w2$ band). These events are definitely related to an increase of the non-thermal emission from the relativistic jet.
 
In order to evaluate the long-term variability of these sources in optical and UV, the intrinsic maximum difference in magnitude (i.e., the difference between maximum and minimum observed magnitude, once subtracted the sum of the uncertainties related to the two measurements) has been calculated in all UVOT filters\footnote{In this calculation, the $v$ band for PKS 1502$+$036 and FBQS J1644$+$2619, and the $m2$ and $w2$ bands for PKS 2004--447 are excluded due to the low number of observations and large uncertainties.}. This value results in: $\Delta$mag$_{max}$ $>$ 0.4 (for 1H 0323$+$342 and FBQS J1644$+$2619), $\Delta$mag$_{max}$ $>$ 0.5 (for PKS 1502$+$036 and PKS 2004--447), $\Delta$mag$_{max}$ $>$ 1 (for PMN J0948$+$0022), and $\Delta$mag$_{max}$ $>$ 2 (for SBS 0846$+$513), in all optical and UV bands. These variations in optical and UV are significantly larger than the maximum variations observed in radio-quiet NLSy1 \citep{ai13}, where disc instability or variations in the accretion rate are responsible for the observed variability. This is a further indication that optical and UV variability in $\gamma$-ray-emitting NLSy1 is related to variations in the jet synchrotron emission, which thus represents the dominant contribution to the continuum flux in the optical--UV part of the spectrum.    
 
\begin{figure*}
\begin{center}
\rotatebox{0}{\resizebox{!}{120mm}{\includegraphics{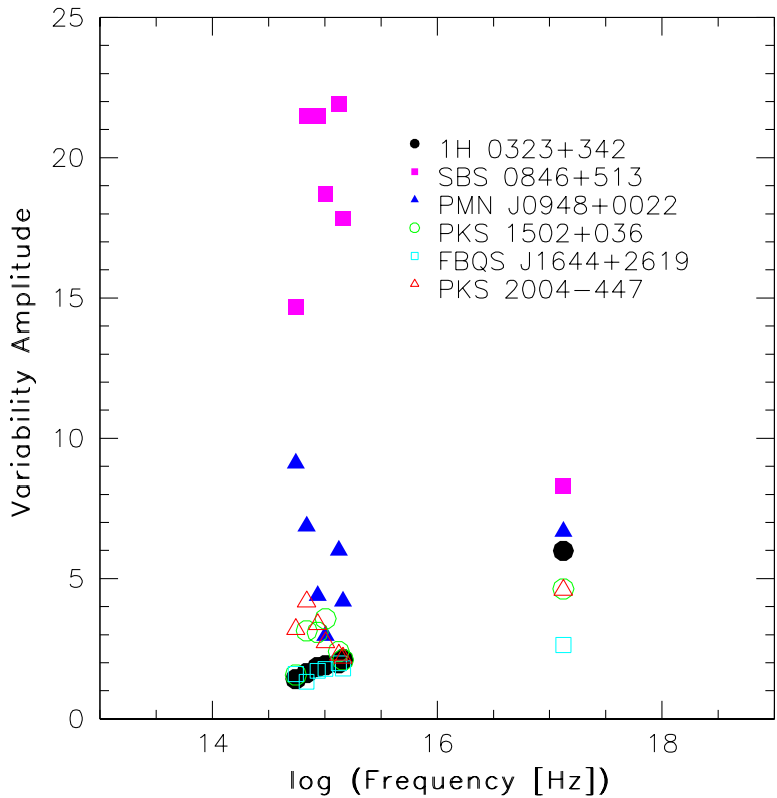}}}
\caption{Variability amplitude vs frequency in the optical ($v$, $b$, $u$), UV ($w1$, $m2$, $w2$), and X-ray bands for the six $\gamma$-ray-emitting NLSy1 well sampled.}
\label{amplitude}
\end{center}
\end{figure*}   

Moreover, in order to compare the variability at different frequencies, we quantify the observed variability of 1H 0323$+$342, SBS 0846$+$513, PMN J0948$+$0022, PKS 1502$+$036, FBQS J1644$+$2619, and PKS 2004--447 in optical, UV, and X-rays bands through the variability amplitude, $V_{\mathrm{amp}}$, calculated as the ratio of maximum to minimum flux corrected for Galactic extinction. The variability amplitude may depend on the sampling of the light curves at the different frequencies. However, at least for a comparison between different bands, the simultaneity of observations in optical, UV, and X-rays guaranteed by the {\em Swift} observations helps mitigate this problem. The variability amplitude in X-rays of the six sources varies between 2.7 and 8.5. These jetted NLSy1 are less variable than the other non-jetted NLSy1, suggesting that a different emission mechanism is driving the flux variability in these two subclasses of NLSy1. The relatively low amplitude observed in $\gamma$-ray-emitting NLSy1 may be partially due to the fact that, in the jet scenario, the X-ray emission being produced by scattering of the low-energy tail of the electron distribution is less variable, at least for the observations for which the photon index is lower than 2 (see Fig.~\ref{hist}, right-hand panel). This is something already observed in FSRQ \citep[e.g.,][]{dammando19b}. However, the FSRQ studied in \citet{stroh13} have shown a variability amplitude between 3 and 54. This can indicate that, although the jet emission should be the dominant contribution in the X-ray part of the spectrum, both jet and corona radiation are responsible for the X-ray emission in $\gamma$-ray-emitting NLSy1. 
 
In the optical bands, the variability amplitude of the six $\gamma$-ray-emitting NLSy1 ranges between 1.4/1.3/1.7 and 14.7/21.5/21.5 in the $v$, $b$, and $u$ bands, respectively, while the UV bands have shown a variability amplitude ranges between 1.8/2.0/1.8 and 18.7/17.9/21.8 in the $w1$, $m2$, and $w2$ bands, respectively. Radio-loud NLSy1 are usually found to be more variable than their radio-quiet counterparts in optical, with a correlation between the optical variability and radio power \citep[e.g.,][]{rakshit17b}. This can be due to the presence of non-thermal jet emission in most of radio-loud NLSy1. In this context, the large variability amplitude of $\gamma$-ray-emitting NLSy1 observed in optical and UV bands with {\em Swift}-UVOT confirms that the jet contribution is significantly larger in these sources with respect to other radio-loud NLSy1. The relatively smaller variability observed in the UVOT UV bands (in particular $w1$ and $m2$ bands) with respect to the optical bands can be a consequence of the fact that the accretion disc emission peaks in the UV bands in these sources \citep[e.g.,][]{dammando15, dammando16b}, and therefore the thermal radiation from the accretion disc partially dilutes the jet radiation reducing the increase of flux due to synchrotron emission.  
 
\begin{table*}
\caption{X-ray variability properties of the sample based on {\em Swift}-XRT observations.}
\begin{center}
\begin{tabular}{cccccc}
\hline 
\multicolumn{1}{c}{\textbf{Source name}} &
\multicolumn{1}{c}{\textbf{Minimum flux}} &
\multicolumn{1}{c}{\textbf{Maximum flux}} &
\multicolumn{1}{c}{\textbf{Median flux}} &
\multicolumn{1}{c}{\textbf{$F_{var}$}} &
\multicolumn{1}{c}{\textbf{$V_{\mathrm{amp}}$}} \\
\multicolumn{1}{c}{} &
\multicolumn{1}{c}{(10$^{-12}$ erg cm$^{-2}$ s$^{-1}$)} &
\multicolumn{1}{c}{(10$^{-12}$ erg cm$^{-2}$ s$^{-1}$)} &
\multicolumn{1}{c}{(10$^{-12}$ erg cm$^{-2}$ s$^{-1}$)} &
\multicolumn{1}{c}{} &
\multicolumn{1}{c}{} \\  
\hline
1H 0323$+$342     &  9.30 & 55.71 & 18.50 & 0.31 $\pm$ 0.02 & 5.99  \\
SBS 0846$+$513    &  0.24 &  1.99 &  6.16 & 0.37 $\pm$ 0.06 & 8.30  \\
PMN J0948$+$0022  &  2.23 & 14.89 &  4.63 & 0.44 $\pm$ 0.03 & 6.68  \\
PKS 1502$+$036    &  0.32 &  1.50 &  0.51 & 0.51 $\pm$ 0.08 & 4.63  \\
FBQS J1644$+$2619 &  1.12 &  2.94 &  1.56 & 0.27 $\pm$ 0.07 & 2.63  \\
PKS 2004--447     &  0.47 &  2.15 &  9.21 & 0.37 $\pm$ 0.05 & 4.60  \\
\hline
\end{tabular}
\end{center}
\label{XRT}
\end{table*}

In Fig.~\ref{amplitude}, the variability amplitude is plotted as a function of frequency for the six well-sampled $\gamma$-ray-emitting NLSy1. The highest values in optical, UV, and X-rays are observed for SBS 0846$+$513, while the lowest values are observed for FBQS J1644$+$2619. A low variability amplitude in optical and UV bands has been observed for 1H 0323$+$342 with relatively high corresponding amplitude in X-rays. In a jet-dominated scenario, this suggests that the thermal emission from the accretion disc can be significant in the optical and UV part of the spectrum, as suggested also by the high disc luminosity estimated by the SED modelling \citep[e.g., $L_{\rm disc}$ $\sim$ 10$^{45}$ erg s$^{-1}$;][]{paliya14}. 
A similar situation, although less evident, is observed also for PKS 1502$+$036, for which an accretion disc with $L_{\rm disc}$ $\sim$ 6$\times$10$^{44}$ erg s$^{-1}$ is identified in the UV part of the spectrum \citep{dammando16b}. On the contrary, in case of SBS 0846$+$513, the variability amplitude in optical and UV is higher than those observed in X-rays, in agreement with the lack of a strong accretion disc in this source \citep[$L_{\rm disc}$ $\sim$ 4$\times$10$^{43}$ erg s$^{-1}$;][]{dammando13b}. A comparable variability amplitude has been observed in optical--UV and X-rays for FBQS J1644$+$2619 and PKS 2004--447. A decreasing variability amplitude is observed from optical to UV in PMN J0948$+$0022, in agreement with a strong accretion disc peaked in UV \citep[$L_{\rm disc}$ = 5.7 $\times$ 10$^{45}$ erg s$^{-1}$;][]{dammando15}, and a lower variability in X-rays with respect to the optical ones. In a disc reprocessing scenario \citep[e.g.,][]{haardt91}, usually invoked for studying the variability in optical, UV, and X-rays in radio-quiet NLSy1 \citep[e.g.,][and the references therein]{lobban20}, the X-ray emission is expected to be more variable than the optical and UV emission. This rules out the possibility to interpret the multiband amplitude variability observed for SBS 0846$+$513, PMN J0948$+$0022, FBQS J1644$+$2619, and PKS 2004$-$447 in a disc reprocessing scenario, favouring the jet-dominated scenario.
   
In addition to the amplitude variability that provides the maximum of the flux variability observed at each energy band, we estimate the fractional variability parameter $F_{\mathrm{Var}}$, commonly used for investigating the degree of variability of a light curve taking into account also the uncertainties on the flux. Upper limits are not taken into account for the fractional variability calculations. We followed the prescription given by \cite{vaughan03}:   

\begin{equation}
  F_{\mathrm{var}} = \sqrt{\frac{S^2 -
      <\sigma_{\mathrm{err}}^2>}{<F_{\gamma}>^2}} \nonumber
\end{equation}
\noindent where  $<F_{\gamma}>$ denotes the average flux, $S$ the
standard deviation of the $N$ flux measurements and
\mbox{$<\sigma_{\mathrm{err}}^2>$} the mean-squared error. 
The uncertainty of $F_{\mathrm{var}}$ is estimated following \citet{poutanen08}:

\begin{equation}
\Delta F_{\mathrm{var}} = \sqrt{F^{2}_{\mathrm{var}} +
  err(\sigma^{2}_{\mathrm{NXS}})} -F_{\mathrm{var}} \nonumber
\end{equation}

\noindent where $err(\sigma^{2}_{\mathrm{NXS}})$ is given by equation 11 in \cite{vaughan03}: 
\begin{equation}
err(\sigma^{2}_{\mathrm{NXS}}) = \sqrt{\left(\sqrt{\frac{2}{N}}\frac{<\sigma^{2}_\mathrm{err}>}{<F_{\gamma}>^{2}}\right)^{2}  +
  \left(\sqrt{\frac{<\sigma^{2}_\mathrm{err}>}{N}}\frac{2F_{\mathrm{var}}}{<F_{\gamma}>}\right)^{2}} \nonumber
\end{equation}

The $F_{\mathrm{var}}$ values are reported in Tables~\ref{XRT} and \ref{UVOT}. In general, they confirm the results of the variability amplitude. Moreover, the $F_{\mathrm{var}}$ values demonstrate how the variability of the sources is not dominated by the uncertainties on the flux, except for the $v$ band in PKS 1502$+$036 and FBQS J1644$+$2619, and the $m2$ and $w2$ bands in PKS 2004--447. In those cases, corresponding to variability amplitude 2 or lower and large uncertainties, no values are reported in Table~\ref{UVOT}. It is worth mentioning that $F_{\mathrm{var}}$ is an estimator that is a function also of the monitoring length in the source rest frame \citep[e.g.,][]{vagnetti16} and this should be taken into account when comparing sources observed over different length time intervals. For 5 of the 6 sources studied here similar monitoring intervals have been considered, therefore this marginally affects the results. Following eq. (9) in \citet{vagnetti16}, we estimated the correction factor for the duration effect using a fixed rest-frame time interval of 2000 d and $b =$ 0.12. This correction factor (not applied to the values in tables) results in: 0.833 (1H 0323$+$342), 1.031 (SBS 0846$+$513), 1.034 (PMN J0948$+$0022), 1.005 (PKS 1502$+$036), 1.008 (FBQS J1644$+$2619), and 1.053 (PKS 2004--447).   

\begin{table*}
\caption{Optical and UV variability properties of the sample based on {\em Swift}-UVOT observations.}
\begin{center} 
\begin{tabular}{ccccccc}
\hline 
\multicolumn{1}{c}{\textbf{Source name}} &
\multicolumn{1}{c}{\textbf{Filter}} &
\multicolumn{1}{c}{\textbf{Minimum flux}} &
\multicolumn{1}{c}{\textbf{Maximum flux}} &
\multicolumn{1}{c}{\textbf{Median flux}} &
\multicolumn{1}{c}{\textbf{$F_{var}$}} &
\multicolumn{1}{c}{\textbf{$V_{\mathrm{amp}}$}} \\
\multicolumn{1}{c}{} &
\multicolumn{1}{c}{} &
\multicolumn{1}{c}{(10$^{-12}$ erg cm$^{-2}$ s$^{-1}$)} &
\multicolumn{1}{c}{(10$^{-12}$ erg cm$^{-2}$ s$^{-1}$)} &
\multicolumn{1}{c}{(10$^{-12}$ erg cm$^{-2}$ s$^{-1}$)} &
\multicolumn{1}{c}{} &
\multicolumn{1}{c}{} \\
\hline
1H 0323$+$342     & V  & 15.80 & 22.42 &  18.73 & 0.06 $\pm$ 0.01 & 1.42  \\
                  & B  & 14.58 & 23.70 &  19.35 & 0.10 $\pm$ 0.01 & 1.63  \\
                  & U  & 15.26 & 28.24 &  21.48 & 0.13 $\pm$ 0.01 & 1.85  \\
                  & W1 & 15.32 & 29.34 &  21.40 & 0.13 $\pm$ 0.01 & 1.91  \\
                  & M2 & 13.88 & 27.45 &  20.39 & 0.16 $\pm$ 0.01 & 1.98  \\
                  & W2 & 14.46 & 30.67 &  21.90 & 0.15 $\pm$ 0.01 & 2.12  \\
SBS 0846$+$513    & V  &  0.41 &  5.98 &  0.73  & 1.15 $\pm$ 0.05 & 14.68 \\
                  & B  &  0.27 &  5.84 &  0.45  & 1.48 $\pm$ 0.04 & 21.48 \\
                  & U  &  0.16 &  3.39 &  0.25  & 1.44 $\pm$ 0.04 & 21.50 \\
                  & W1 &  0.11 &  2.03 &  0.20  & 0.85 $\pm$ 0.07 & 18.71 \\
                  & M2 &  0.13 &  2.86 &  0.25  & 1.38 $\pm$ 0.04 & 21.91 \\
                  & W2 &  0.10 &  1.71 &  0.23  & 1.06 $\pm$ 0.04 & 17.84 \\
PMN J0948$+$0022  & V  &  1.07 &  9.74 &  1.92  & 0.74 $\pm$ 0.04 & 9.11  \\
                  & B  &  1.22 &  8.40 &  1.90  & 0.62 $\pm$ 0.03 & 6.86  \\
                  & U  &  1.28 &  5.62 &  1.79  & 0.37 $\pm$ 0.02 & 4.39  \\ 
                  & W1 &  1.60 &  4.75 &  2.07  & 0.27 $\pm$ 0.02 & 2.97  \\
                  & M2 &  1.39 &  8.34 &  2.23  & 0.49 $\pm$ 0.04 & 6.01  \\
                  & W2 &  1.70 &  7.12 &  2.21  & 0.42 $\pm$ 0.02 & 4.19  \\ 
PKS 1502$+$036    & V  &  0.72 &  1.13 &  0.81  & -               & 1.56  \\
                  & B  &  0.41 &  1.28 &  0.70  & 0.24 $\pm$ 0.08 & 3.14  \\
                  & U  &  0.34 &  1.04 &  0.55  & 0.25 $\pm$ 0.07 & 3.08  \\
                  & W1 &  0.30 &  1.08 &  0.62  & 0.18 $\pm$ 0.08 & 3.57  \\
                  & M2 &  0.46 &  1.10 &  0.73  & 0.23 $\pm$ 0.03 & 2.40  \\
                  & W2 &  0.47 &  0.99 &  0.66  & 0.18 $\pm$ 0.03 & 2.10  \\
FBQS J1644$+$2619 & V  &  1.75 &  2.80 &  2.49  & -               & 1.60  \\
                  & B  &  1.97 &  2.60 &  2.23  & 0.04 $\pm$ 0.03 & 1.32  \\
                  & U  &  1.92 &  3.28 &  2.30  & 0.18 $\pm$ 0.03 & 1.71  \\
                  & W1 &  1.73 &  3.09 &  2.42  & 0.17 $\pm$ 0.04 & 1.78  \\
                  & M2 &  1.65 &  3.28 &  2.48  & 0.24 $\pm$ 0.04 & 1.98  \\
                  & W2 &  1.89 &  3.40 &  2.68  & 0.20 $\pm$ 0.03 & 1.80  \\
PKS 2004--447     & V  &  0.31 &  0.99 &  0.74  & 0.20 $\pm$ 0.11 & 3.20  \\
                  & B  &  0.18 &  0.77 &  0.45  & 0.25 $\pm$ 0.11 & 4.18  \\
                  & U  &  0.20 &  0.67 &  0.38  & 0.25 $\pm$ 0.06 & 3.37  \\
                  & W1 &  0.14 &  0.38 &  0.21  & 0.12 $\pm$ 0.11 & 2.73  \\
                  & M2 &  0.12 &  0.27 &  0.16  & -               & 2.28  \\
                  & W2 &  0.10 &  0.22 &  0.16  & -               & 2.18  \\
\hline
\end{tabular}
\end{center}
\label{UVOT}
\end{table*}

\begin{figure*}
\begin{center}
{\includegraphics[width=0.75\textwidth]{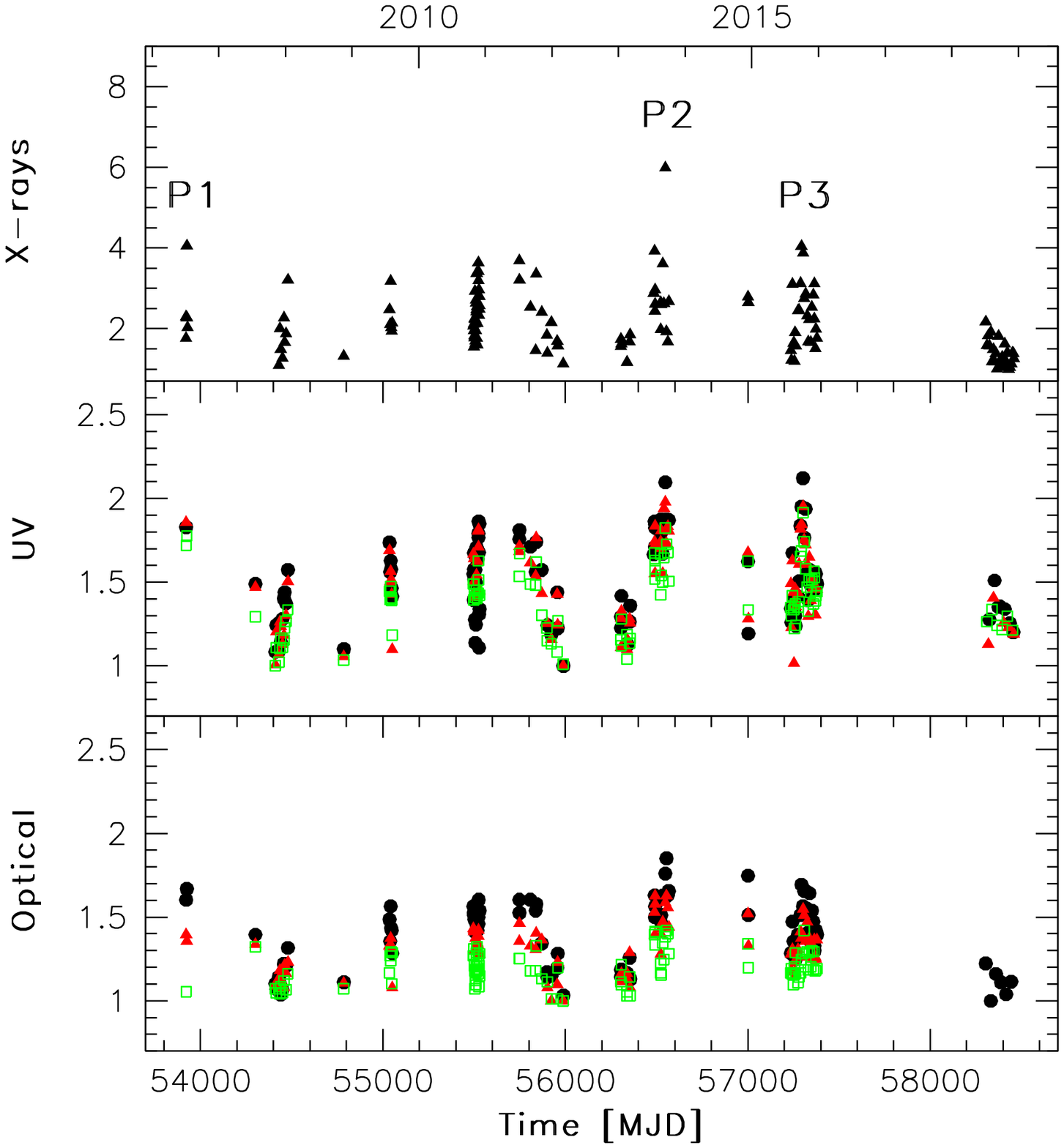}}
\caption{Multifrequency light curve of 1H 0323$+$342 normalized to the minimum value observed in X-rays (0.3--10 keV; top panel), UV bands ($w1$, open squares; $m2$, filled triangles; $w2$, filled circles; middle panel), and optical band ($v$, open squares; $b$, filled triangles; $u$, filled circles; bottom panel).}
\label{0323_all}
\end{center}
\end{figure*}

\begin{figure*}
\begin{center}
\rotatebox{0}{\resizebox{!}{85mm}{\includegraphics{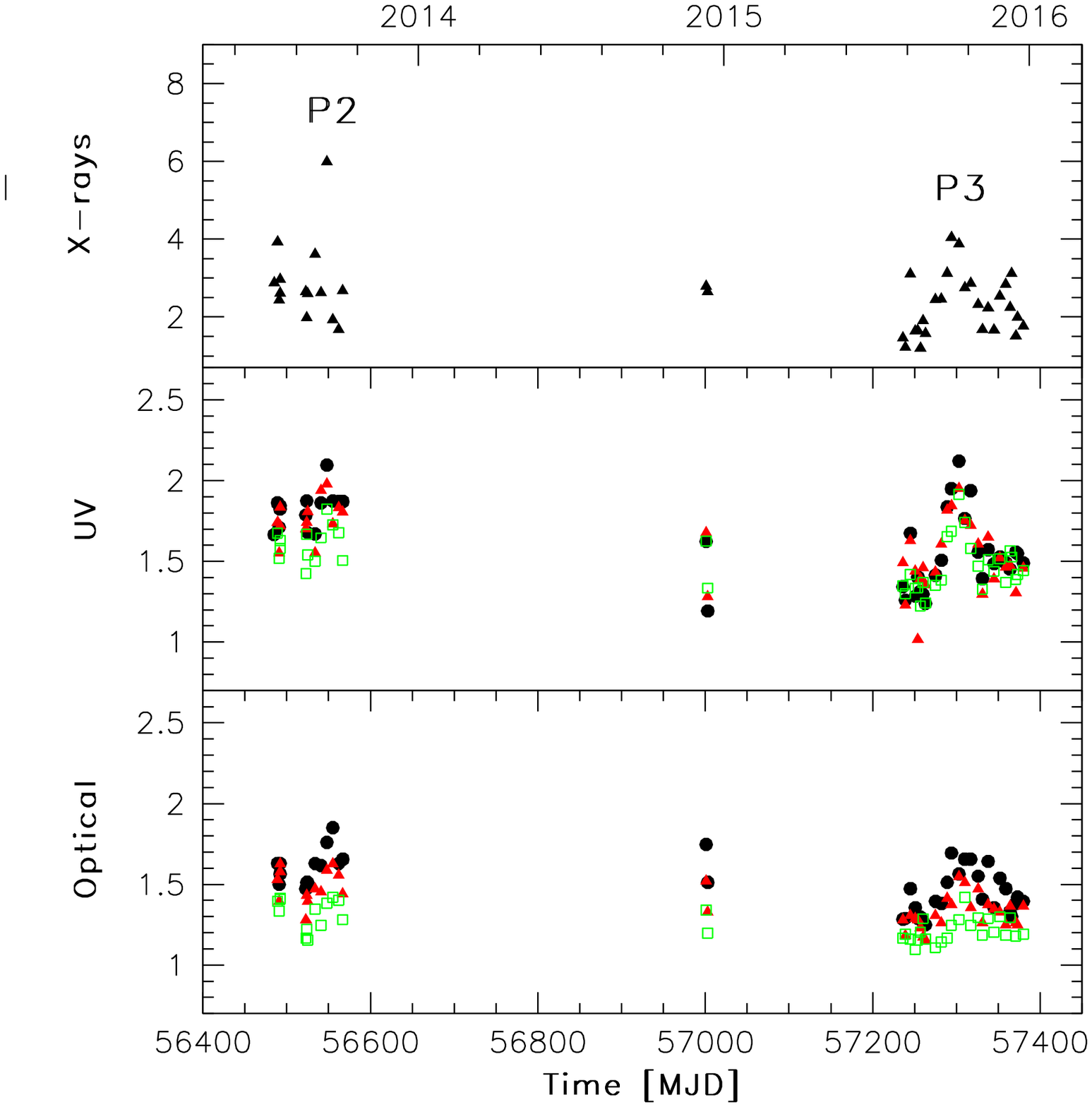}}}
\hspace{0.1cm}
\rotatebox{0}{\resizebox{!}{85mm}{\includegraphics{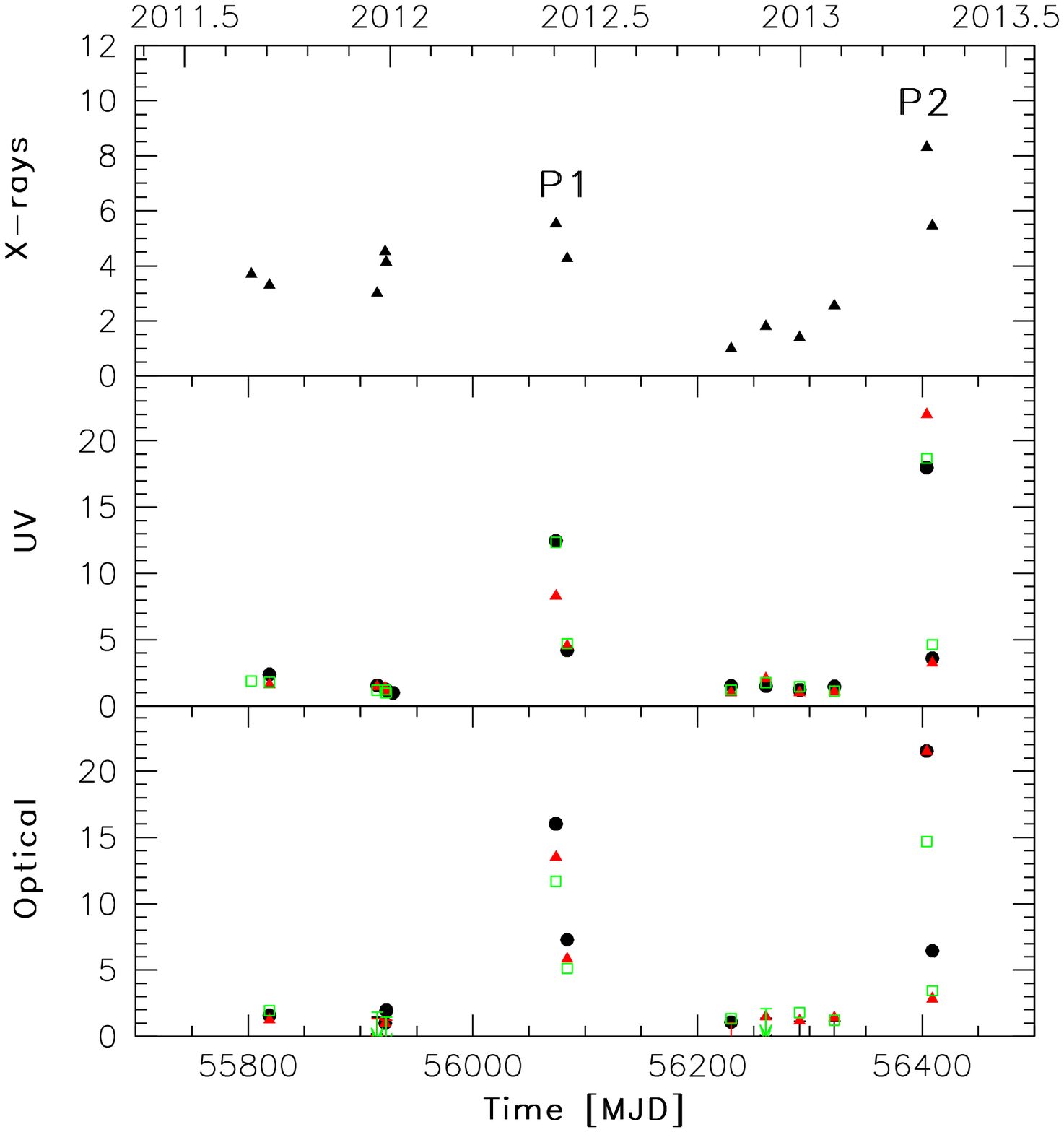}}}
\caption{Multifrequency light curve of 1H 0323$+$342 (left) and SBS 0846$+$513 (right) normalized to the minimum value observed in X-rays (0.3--10 keV; top panel), UV bands ($w1$, open squares; $m2$, filled triangles; $w2$, filled circles; middle panel), and optical band ($v$, open squares; $b$, filled triangles; $u$, filled circles; bottom panel) and focused on the high activity periods. The complete light curves of 1H 0323$+$342 and SBS 0846$+$513 are reported in Fig.~\ref{0323_all} and Appendix C, respectively.} 
\label{lc1}
\end{center}
\end{figure*}

\begin{figure*}
\begin{center}
\rotatebox{0}{\resizebox{!}{85mm}{\includegraphics{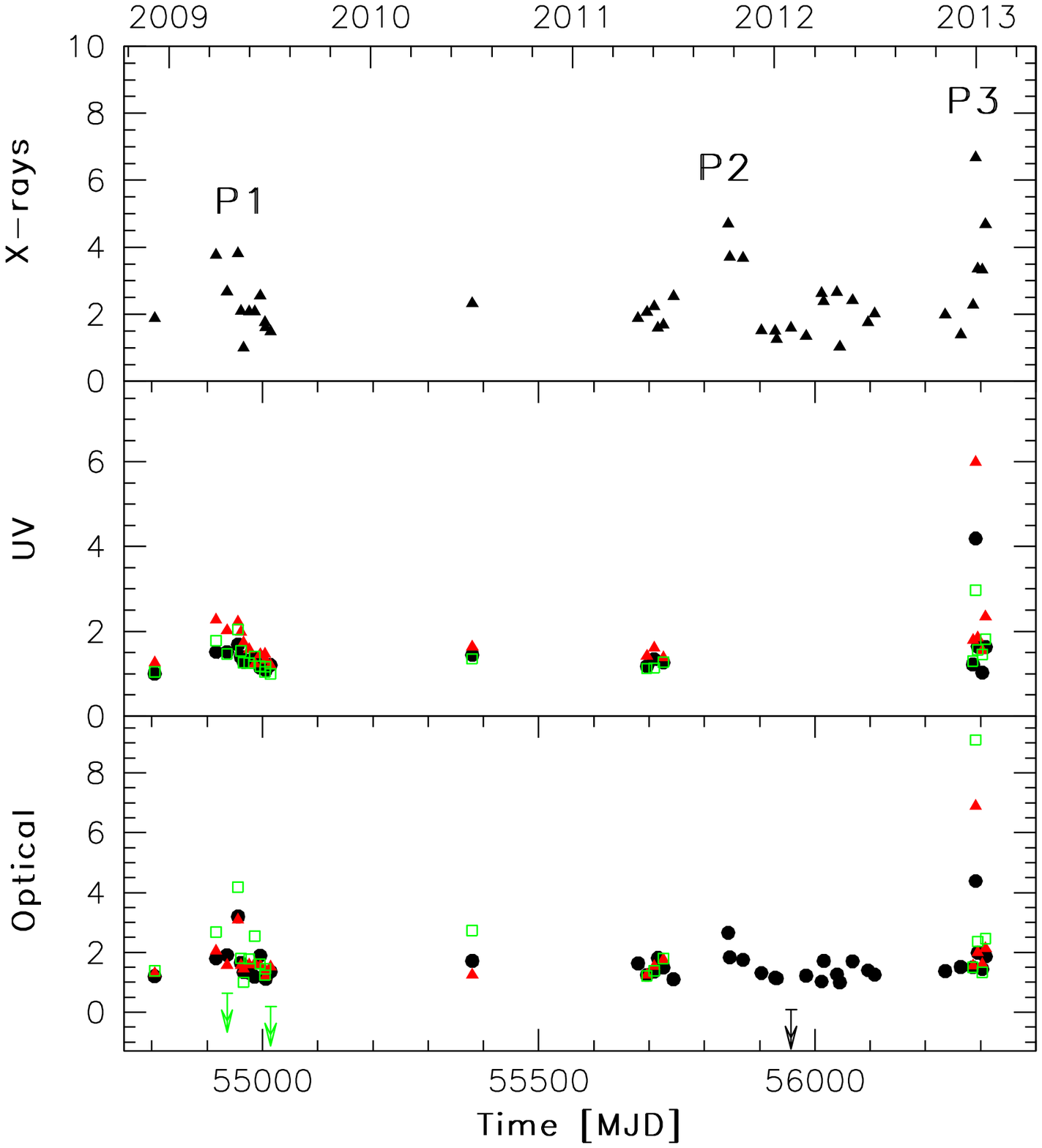}}}
\hspace{0.1cm}
\rotatebox{0}{\resizebox{!}{85mm}{\includegraphics{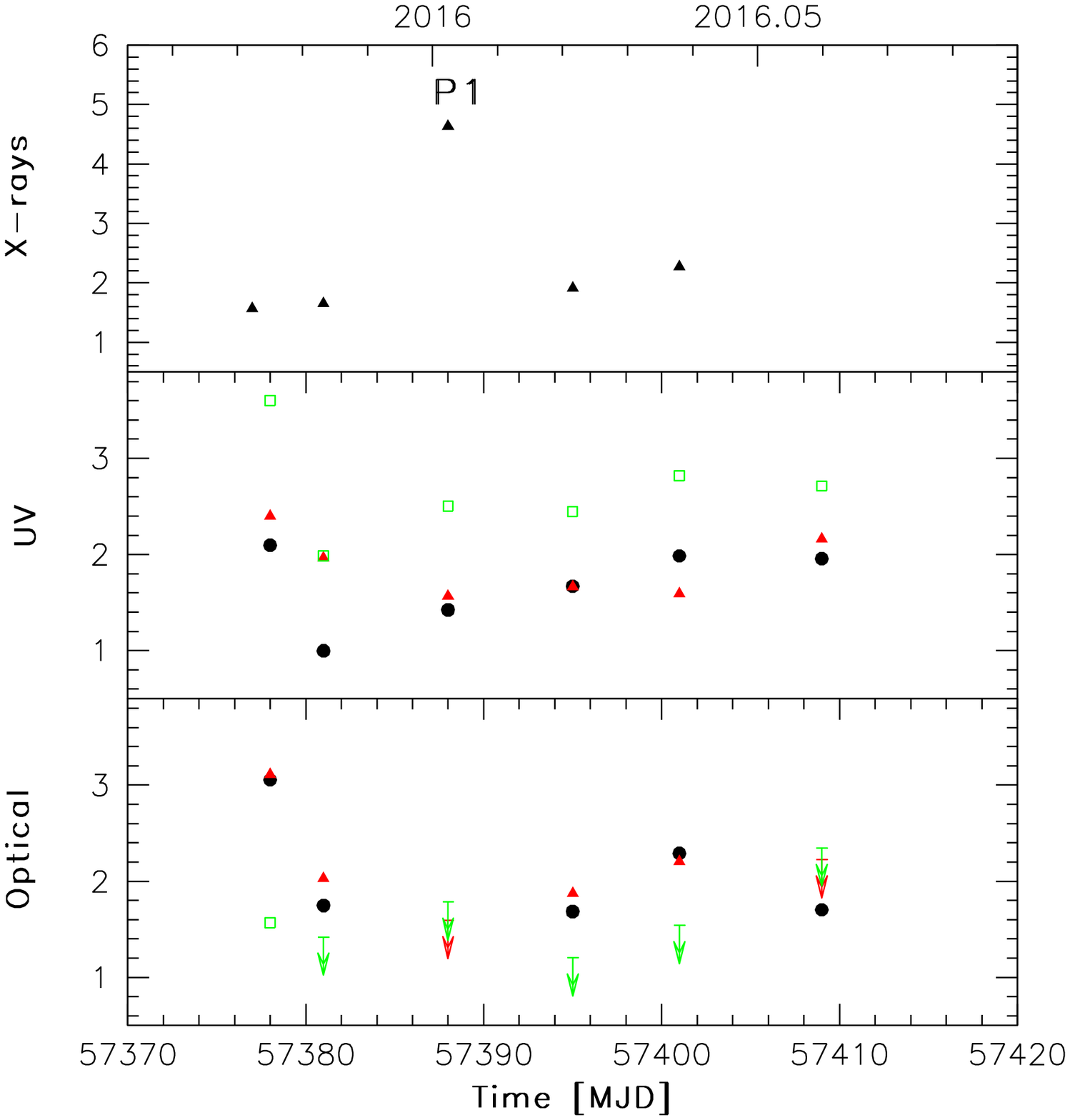}}}
\caption{Multifrequency light curve of PMN J0948$+$0022 (left) and PKS 1502$+$036 (right) normalized to the minimum value observed in X-rays (0.3--10 keV; top panel), UV bands ($w1$, open squares; $m2$, filled triangles; $w2$, filled circles; middle panel), and optical band ($v$, open squares; $b$, filled triangles; $u$, filled circles; bottom panel) and focused on the high activity periods. The complete light curves of PMN J0948$+$0022 and PKS 1502$+$036 are reported in Appendix C.}
\label{lc2} 
\end{center}
\end{figure*}

\begin{figure*}
\begin{center}
\rotatebox{0}{\resizebox{!}{85mm}{\includegraphics{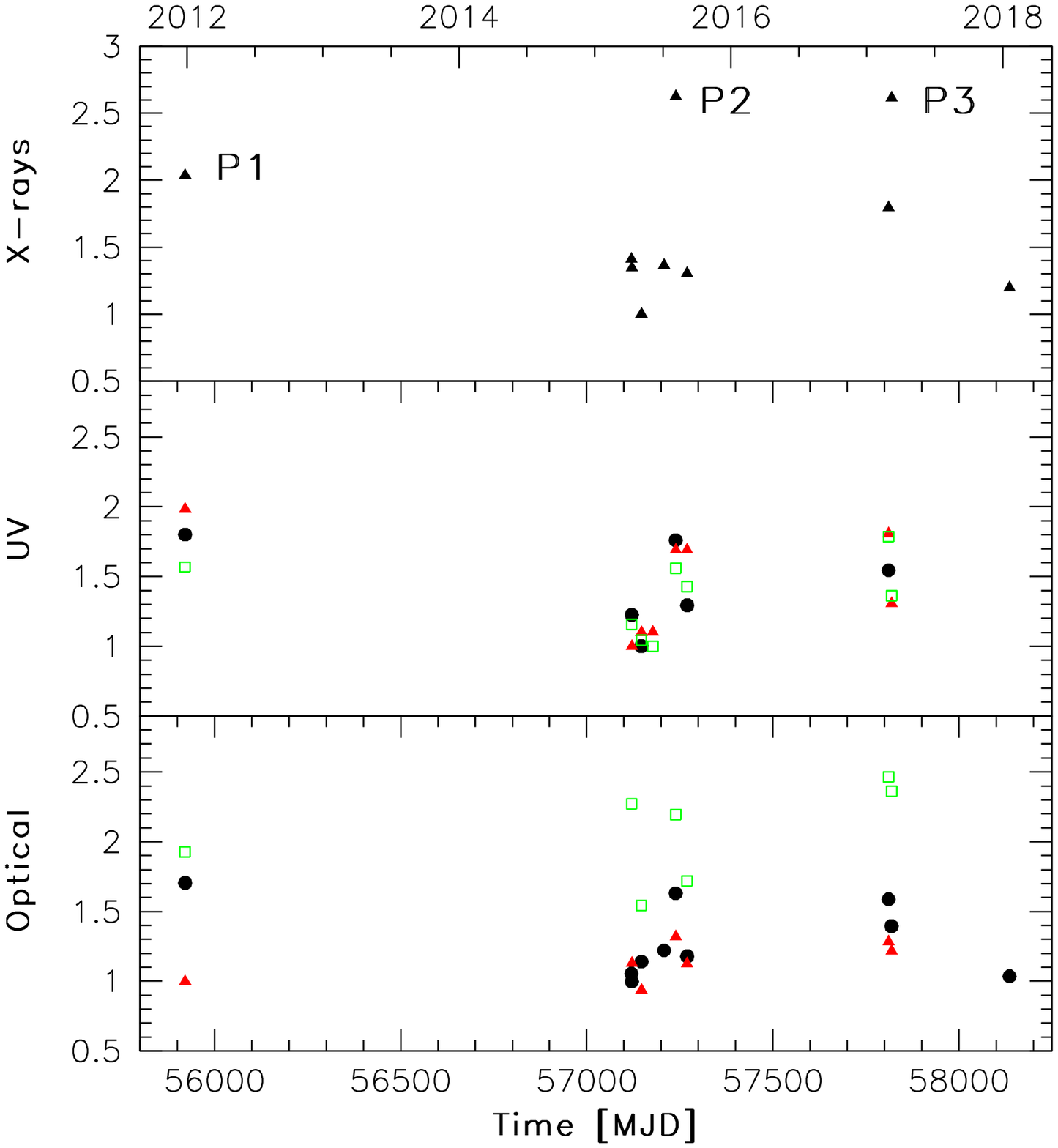}}}
\hspace{0.1cm}
\rotatebox{0}{\resizebox{!}{85mm}{\includegraphics{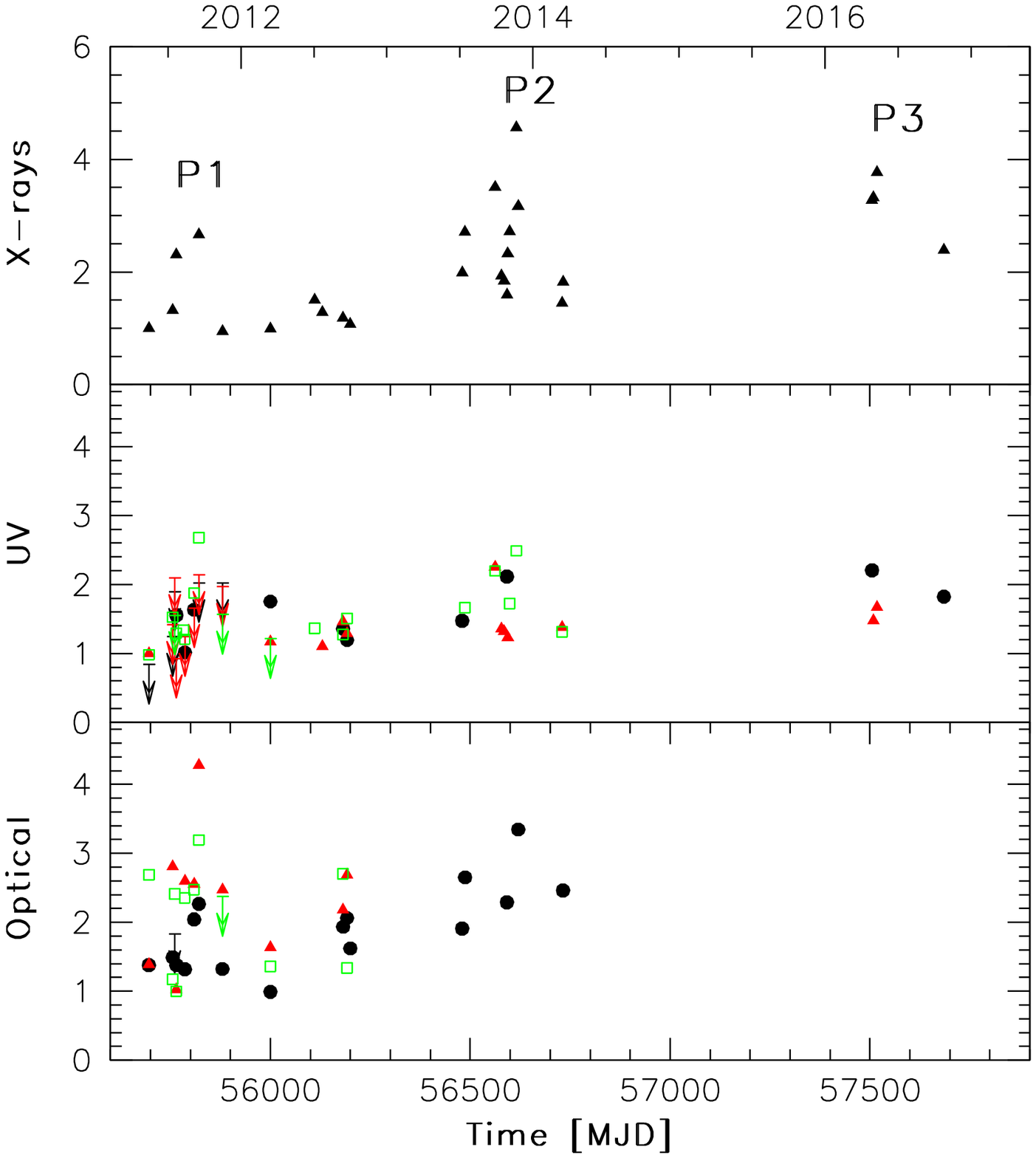}}}
\caption{Multifrequency light curve of FBQS J1644$+$2619 (left) and PKS 2004--447 (right) normalized to the minimum value observed in X-rays (0.3--10 keV; top panel), UV bands ($w1$, open squares; $m2$, filled triangles; $w2$, filled circles; middle panel), and optical band ($v$, open squares; $b$, filled triangles; $u$, filled circles; bottom panel).} 
\label{lc3}
\end{center}
\end{figure*}

As a further step, the relationship between optical ($u$ band), UV ($w2$ band), and X-ray radiation has been investigated using the Pearson correlation analysis. The Pearson correlation coefficient $r$ measures the linear relationship between two data sets, $r$ varies between -1 and +1 with 0 implying no correlation. The $p$-value indicates the probability of an uncorrelated system producing data sets that have a Pearson correlation $r$ as the one computed from these data sets. Upper limits are not considered in the calculation. The results are reported in Table~\ref{pearson}.

\noindent A strong positive correlation between X-ray, UV, and optical emission has been obtained for 1H 0323$+$342, SBS 0846$+$513, and PMN J0948$+$0022 at the 99 per cent confidence level. The correlation is stronger between optical and UV bands with respect to X-rays and optical or X-rays and UV. For FBQS J1644$+$2619, the correlation between the three bands is very strong, but only at the 95 per cent confidence level. In the case of PKS 1502$+$036 a strong positive correlation has been obtained only between optical and UV emission at the 99 per cent confidence level. For PKS 2004--447 a strong positive correlation has been obtained only between X-ray and optical emission, significant at the 95 per cent confidence level but not at the 99 per cent confidence level. However, the $p$-value is affected by the sample size, therefore this issue may influence the confidence level for FBQS J1644$+$2619, PKS 1502$+$036, and PKS 2004--447 that are less well sampled. In this context, the fact that only 95 per cent confidence level has been reached for FBQS J1644$+$2619 seems to be related to the low number of observations. This issue can be related also to the fact that for PKS 2004--447 and PKS 1502$+$036 a significant correlation has been estimated only for the case with the larger number of observations. A better sampling of the observations should be useful to test the significance of correlation between X-ray, UV, and optical emission for these three sources. 

A correlation between optical, UV, and X-rays is expected in case the jet radiation being the dominant mechanisms in the optical-to-X-ray part of the spectrum, with the synchrotron mechanism responsible for the optical and UV bands, and the IC mechanism for the X-ray band, as usually observed in FSRQ. A correlation between optical, UV, and X-ray emission is expected also for the disc reprocessing scenario, in which X-rays are produced primarily in the corona, illuminate the accretion disc and are reprocessed producing the optical and UV emission. This results in a delayed production of optical and UV photons with respect to the X-ray ones. The delay between X-ray and optical and UV emission is related to the disc structure and the BH mass, with larger masses producing longer delay. The latter effect should be more pronounced in $\gamma$-ray-emitting NLSy1 that have BH masses larger than radio-quiet NLSy1. On the other contrary in the jet-dominated scenario simultaneous flux variations in optical, UV, and X-rays are expected if the emission comes from the same region of the jet, as usually assumed for blazars \citep[but see e.g.][]{raiteri17,dammando19b}, and thus the same electrons produce both the synchrotron and IC fluxes. To check if the multiband variations are simultaneous or not in Figs.~\ref{0323_all},~\ref{lc1},~\ref{lc2}, and~\ref{lc3} we compare the light curves obtained in X-rays, UV, and optical bands for the six $\gamma$-ray-emitting NLSy1. All fluxes are normalized to the minimum value observed in the considered period in order to compare when and how much the flux increased in the different energy bands. X-ray peaks are labelled with P1, P2, and P3 in the plots. The complete light curves of SBS 0846$+$513, PMN J0948$+$0022, and PKS 1502$+$036 are shown in Appendix C. A simultaneous increase of the activity from optical to X-rays has been observed for 1H 0323$+$342, SBS 0846$+$513, PMN J0948$+$0022, FBQS J1644$+$2619, and PKS 2004--447 in agreement with the jet-dominated scenario. Notes on individual sources are reported below.       
 
\begin{table*}
\caption{Pearson correlation analysis results for the X-ray, UV ($w2$), and optical ($u$) bands including Pearson's correlation coefficient $r$, the $p$-value, and the number of observations considered for the analysis N$_{obs}$.}
\label{pearson} 
\begin{center}
\begin{tabular}{ccccccccccc}
\hline
\textbf{Source name}   & \multicolumn{3}{c}{\textbf{X vs W2}} & \multicolumn{3}{c}{\textbf{X vs U}} & \multicolumn{3}{c}{\textbf{W2 vs U}} \\
                       &          r    & p-value &  $N_{obs}$  &  r    & p-value &  $N_{obs}$  & r    & p-value &  $N_{obs}$  \\
\hline
1H 0323$+$342     &  0.688 & 4.99$\times$10$^{-17}$ & 112 & 0.656 & 7.23$\times$10$^{-13}$ &   94  & 0.842 & 8.25$\times$10$^{-25}$ & 88  \\
SBS 0846$+$513    &  0.781 & 2.53$\times$10$^{-6}$  & 26  & 0.833 & 9.63$\times$10$^{-6}$  &   19  & 0.973 & 7.26$\times$10$^{-13}$ & 20  \\
PMN J0948$+$0022  &  0.811 & 1.50$\times$10$^{-6}$  & 24  & 0.803 & 9.15$\times$10$^{-11}$ &   43  & 0.892 & 4.76$\times$10$^{-9}$  & 24  \\
PKS 1502$+$036    &  0.086 & 0.801                  & 11  & 0.325 & 0.360                  &   10  & 0.701 & 7.58$\times$10$^{-3}$  & 13  \\
FBQS J1644$+$2619 &  0.925 & 0.008                  &  6  & 0.759 & 0.011                  &   10  & 0.910 & 0.012                  & 6   \\
PKS 2004$-$447    &  0.481 & 0.274                  &  7  & 0.783 & 1.55$\times$10$^{-3}$  &   13  & 0.235 & 0.575                  & 8   \\
\hline
\end{tabular}
\end{center}
\end{table*}

{\it 1H 0323$+$342}: there is a very good agreement between the behaviour observed in optical, UV, and X-ray bands, with all three peaks (and more generally each increase of activity) in X-rays corresponding to a peak in the other bands. Less variability amplitude has been observed going from X-rays to optical bands. In the jet-dominated scenario, this can be due to the contribution of the thermal emission from the accretion disc that dilutes the synchrotron emission in UV and even more in optical bands. The highest peak, P2, has been observed at the time of an intense $\gamma$-ray flaring activity from the source \citep{paliya14}.

{\it SBS 0846$+$513}: a significant increase of activity has been simultaneously observed from optical to X-rays during both P1 and P2. These two peaks are observed during two $\gamma$-ray flaring activities, as reported in \citet{dammando13b} and \citet{paliya16}, respectively, indicating as the relativistic jet radiation is dominant over the entire electromagnetic spectrum in these two periods. In both cases, the increase observed in optical and UV bands is higher with respect to the X-ray band, ruling out the disc reprocessing scenario, and as expected if energetic electrons are responsible for the synchrotron emission that produces the optical--UV part of the spectrum, while the low-energy part of the electron distribution produces the X-ray part of the spectrum by means of IC mechanism.

{\it PMN J0948$+$0022}: there is a fair general agreement between the behaviour observed in X-ray, UV, and optical bands. Three peaks can be identified in the X-ray light curve. A similar increase has been observed in optical and X-ray bands during P1, with a less increase in UV, disfavouring the disc reprocessing scenario. In case of P2, only U data are available showing less increase in optical with respect to X-rays. A significant increase has been observed simultaneously from optical to X-rays during P3, at the time of the most powerful $\gamma$-ray flaring activity observed by the source so far \citep{dammando15}. This is an unquestionable indication that the dominant contribution to the optical-to-X-ray emission has been produced by the relativistic jet in that period. 

{\it PKS 1502$+$036}: the peak of activity P1 in X-rays is observed 10 d before the optical and UV peak. In case of optical and UV emission, the peak has been observed soon after the peak of the first $\gamma$-ray flaring activity detected from this source \citep{dammando16b}, indicating that the optical and UV emission is mainly produced by the relativistic jet. It is interesting to note that in 2012 there is an increase of the activity by a factor of 2 in UV (in particular in the $w1$ band) and in optical (in the $u$ band) with no corresponding activity in X-rays (see the complete light curve in Appendix C). Also in this case the $\gamma$-ray flux was relatively high \citep{dammando13a}, suggesting as the jet radiation is dominating the optical and UV spectrum in that period. The smaller variability observed during these two periods in X-rays with respect to the optical and UV bands is difficult to reconcile with the disc reprocessing scenario, in case of the jet-dominated scenario may be related to the less energy of the photons that produce the X-ray emission.

{\it FBQS J1644$+$2619}: no significant increase of activity from this source has been observed during the {\em Swift} monitoring. However, the three peaks in X-rays have a corresponding but smaller or comparable increase of activity in optical and UV bands, in line with a scenario in which the jet dominating the X-ray emission, while the disc dilutes the variability in optical and UV bands.

{\it PKS 2004--447}: the three peaks observed in X-rays correspond to an increase of activity in optical and UV bands, although the variability amplitude is not the same in the three cases. In particular, P1 has a comparable variability amplitude in X-rays and UV, with a larger amplitude in optical (in particular in the $b$ band), disfavouring a disc reprocessing scenario, while a higher variability amplitude has been observed in X-rays for P2 and P3.    
 
\subsection{Spectral variability}
\label{Var_spec}

Thanks to the {\em Swift} monitoring, it is possible to check the distribution of the X-ray photon index and its evolution in time for 1H 0323$+$342, SBS 0846$+$513, PMN J0948$+$0022, PKS 1502$+$036, FBQS J1644$+$2619, and PKS 2004--447. The average photon index considering all the XRT single observations of the six $\gamma$-ray-emitting NLSy1 in the 0.3--10 keV energy range is 1.77 $\pm$ 0.27, while taking into account only the observations for which relatively small uncertainties on the photon index ($< 0.3$) has been obtained the average photon index is 1.83 $\pm$ 0.23. As a comparison, the average photon index of a sample of radio-quiet NLSy1 and BLSy1 obtained with {\em Swift}-XRT data in the 0.3--10 keV energy range is 2.68 $\pm$ 0.51 and 2.11 $\pm$ 0.42, respectively \citep{grupe10}. This can be an indication that the X-ray spectrum of the radio-loud NLSy1 detected in $\gamma$ rays is mainly produced by IC radiation from a relativistic jet. In fact, the jet component usually produces an hard spectrum in FSRQ. On the other hand, the average photon index of the $\gamma$-ray-emitting NLSy1 is softer than the average values derived for FSRQ detected by {\em Fermi}-LAT \citep[$\Gamma$ $\sim$ 1.6;][]{williamson14}. This suggests that together with the jet component a contribution from the corona and the accretion disc may be still present in the X-ray spectra of $\gamma$-ray-emitting NLSy1, in agreement with the presence of a significant soft X-ray excess in most of them, although less pronounced that in radio-quiet NLSy1 (see Section \ref{summed}). The average X-ray photon index of $\gamma$-ray-emitting NLSy1 is significantly harder than the average values estimated for BL Lac objects \citep[$\Gamma$ $\sim$ 2.2;][]{williamson14}, confirming that $\gamma$-ray-emitting NLSy1 are more similar to FSRQ than BL Lacs, as suggested by previous works \citep[e.g.,][]{foschini15,dammando19}.  

\noindent Similarly to the average photon index of the entire sample, the median of the 0.3--10 keV X-ray spectral index obtained in the single XRT observations is lower than 2 for all six $\gamma$-ray-emitting NLSy1 studied in detail here, although for three sources (1H 0323$+$342, PKS 1502$+$036, and FBQS J1644$+$2619) in some observations the spectrum becomes softer than 2 reaching values of 2.2--2.4 (Table~\ref{XRT_statistics}), suggesting that in some periods the jet radiation can be less important in X-rays in these sources and the radiation from the corona can provide a contribution to the total X-ray spectrum. In Fig.~\ref{hist} (right-hand panel), it is shown the distribution of the X-ray photon indices of the sources if we consider all the {\em Swift}-XRT observation (in black) or only the observations with a photon index with an associated error less than 0.3 (in red). In both cases, the photon index distribution has a peak at around 1.8--2.0, lower than the values estimated with {\em Swift} data for radio-quiet NLSy1 \citep{grupe10}.

Using {\em XMM--Newton} data for a sample of radio-quiet NLSy1, \citet{bianchi09} and \citet{gliozzi20} found an average photon index of  2.00 $\pm$ 0.07 and  2.01 $\pm$ 0.05 in the 2--10 keV energy range, where the dominant mechanism in radio-quiet AGN should be the Comptonization of optical and UV photons from an accretion disc in a hot corona. This value is compatible within the uncertainties with the average X-ray photon index obtained here for the $\gamma$-ray-emitting NLSy1 in the 0.3--10 keV energy range. The limited statistics of the {\em Swift}-XRT spectra do not allow us to estimate the photon index in the 2--10 keV energy range with reasonable uncertainties for a large part of observations, and thus to have a direct comparison for all sources. As said, the 0.3--10 keV spectra of these NLSy1 should include a soft X-ray excess component below 2 keV, but the statistics of the single XRT observations do not allow us to test models more complex than a single power law. However, we noted that the ROSAT All-Sky Survey (RASS) 0.2--2 keV average photon index of radio-quiet NLSy1 reported in \citet{grupe10}\footnote{We have excluded RX J0134.2--4258, RX J0136.9--3510, and RX J1209.8$+$3217, for which the uncertainties on the photon index are $\geq$ 0.5, that is 6.94 $\pm$ 2.60, 4.90 $\pm$ 0.50, and 3.18 $\pm$ 1.18, respectively.} is 2.92 $\pm$ 0.39. It is reasonable to consider that the presence of a soft component in the 0.2--2 keV energy range leads to a softer photon index for the total spectrum in the 0.3--10 keV energy range. Therefore, the average photon index for the $\gamma$-ray emitting NLSy1 in the 2--10 keV energy range should be harder than 1.8, in agreement also with the 2--10 keV photon index obtained with {\em XMM--Newton} observations for PMN J0948$+$0022 \citep[1.48, see][]{dammando14} and the photon index obtained above the break for the summed XRT spectra in case of a broken power-law model (see Section~\ref{summed}). For the brightest X-ray source in our sample, i.e.~1H 0323$+$342, considering only the observations with uncertainties on the photon index lower than 0.3, the average 2--10 keV photon index obtained with {\em Swift} is 1.75 $\pm$ 0.17 with respect to 1.95 $\pm$ 0.14 obtained in the 0.3--10 keV energy range. This confirms an harder photon index limiting the range of energy to 2.0--10 keV, as hypothesized above.  
 
\begin{figure*}
\begin{center}
\rotatebox{0}{\resizebox{!}{70mm}{\includegraphics{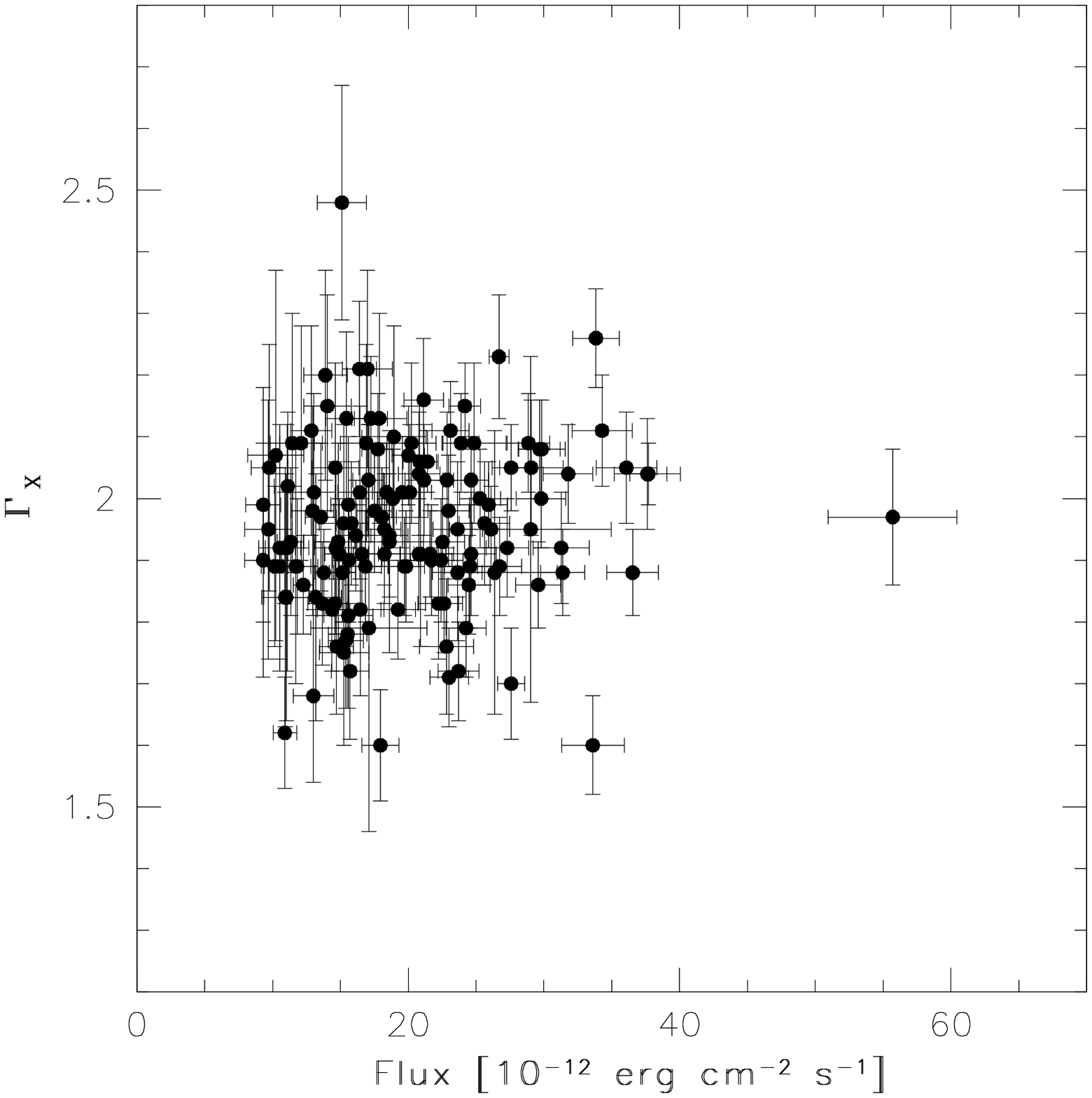}}}
\hspace{0.1cm}
\rotatebox{0}{\resizebox{!}{70mm}{\includegraphics{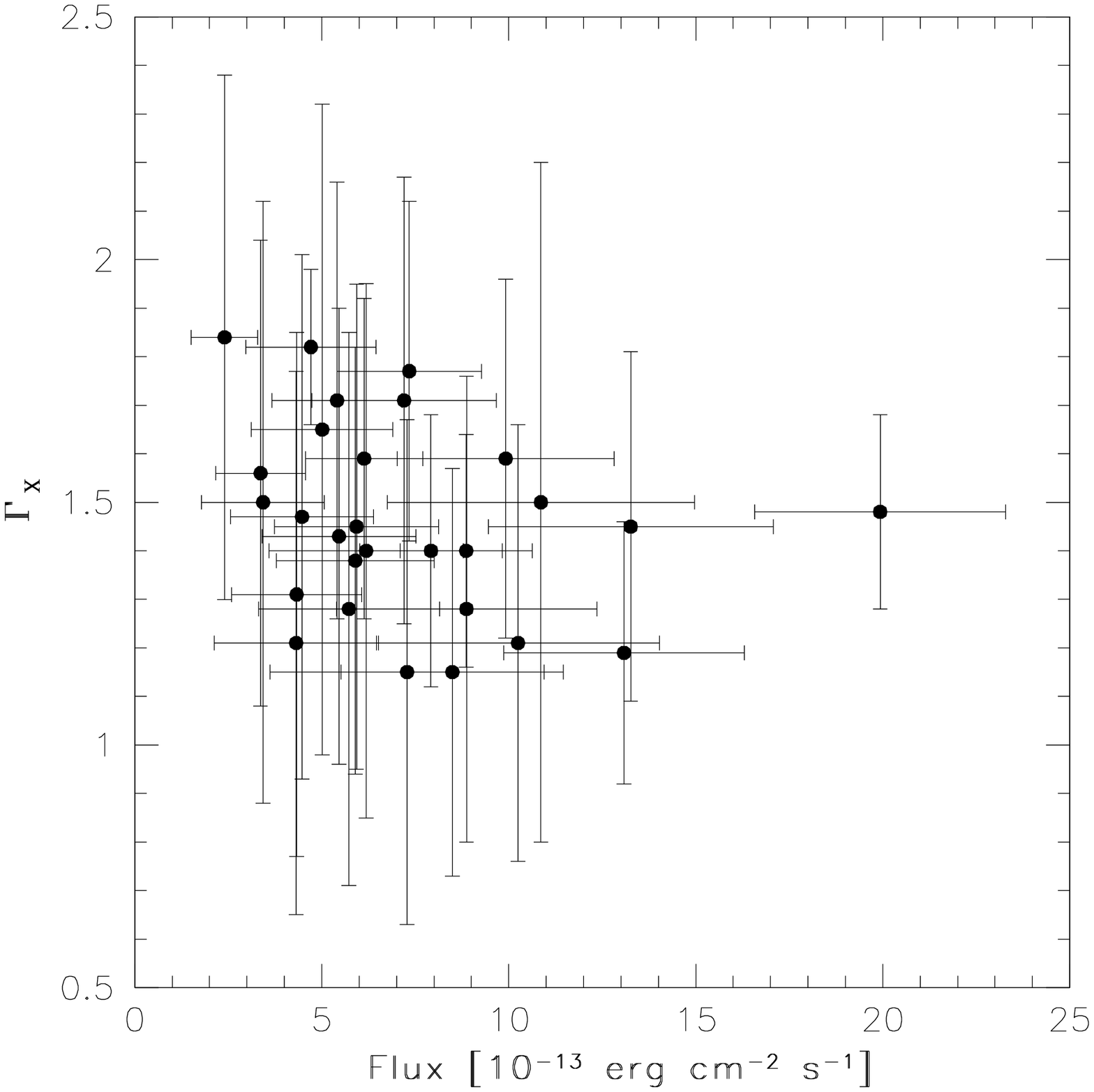}}}
\caption{{\em Swift}/XRT photon index as a function of the 0.3-10 keV unabsorbed flux of 1H 0323$+$342 ({\it left-hand panel}) and SBS 0846$+$513 ({\it right-hand panel}).}\label{XRThard1}
\end{center}
\end{figure*}

\begin{figure*}
\begin{center}
\rotatebox{0}{\resizebox{!}{70mm}{\includegraphics{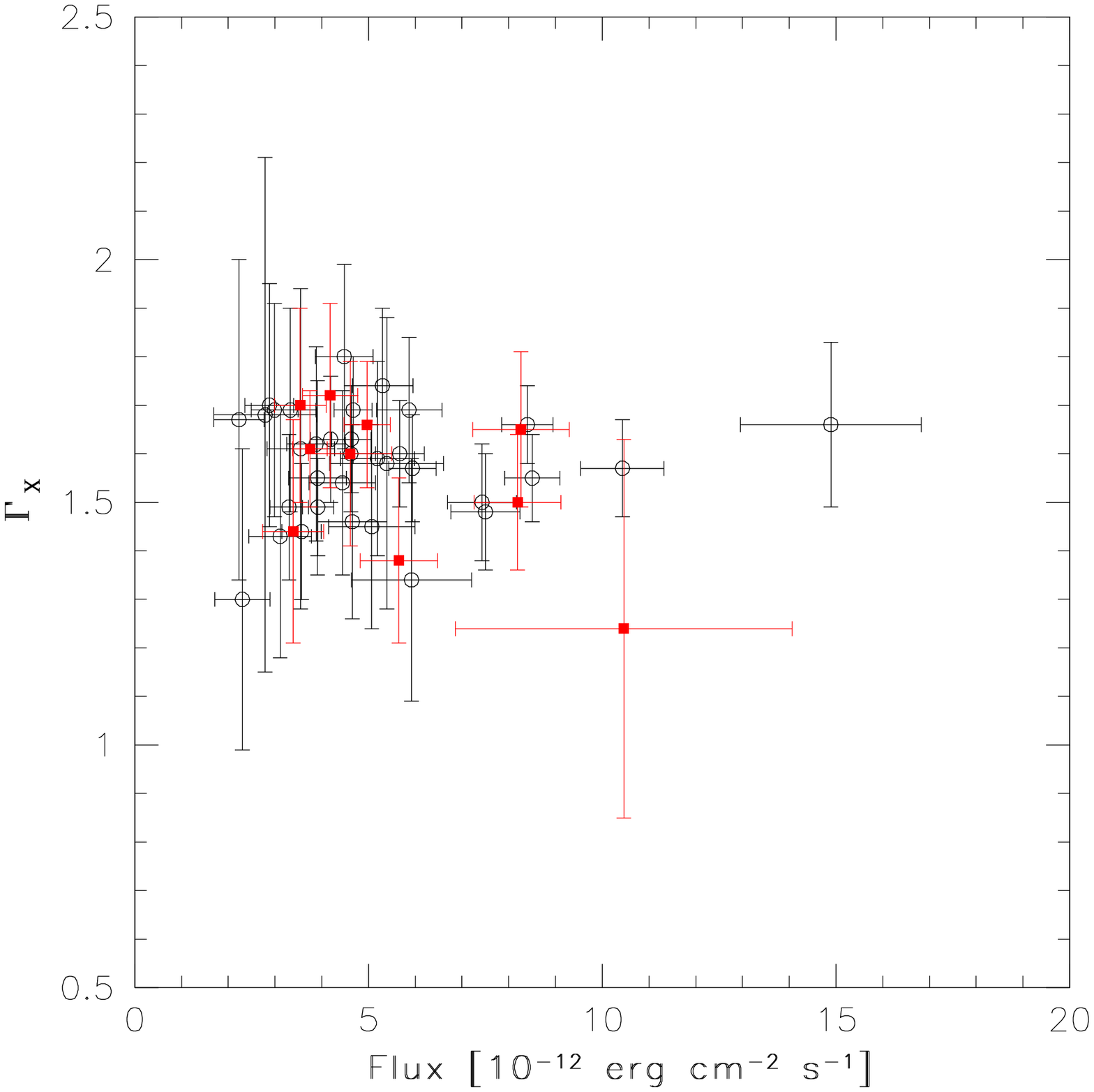}}}
\hspace{0.1cm}
\rotatebox{0}{\resizebox{!}{70mm}{\includegraphics{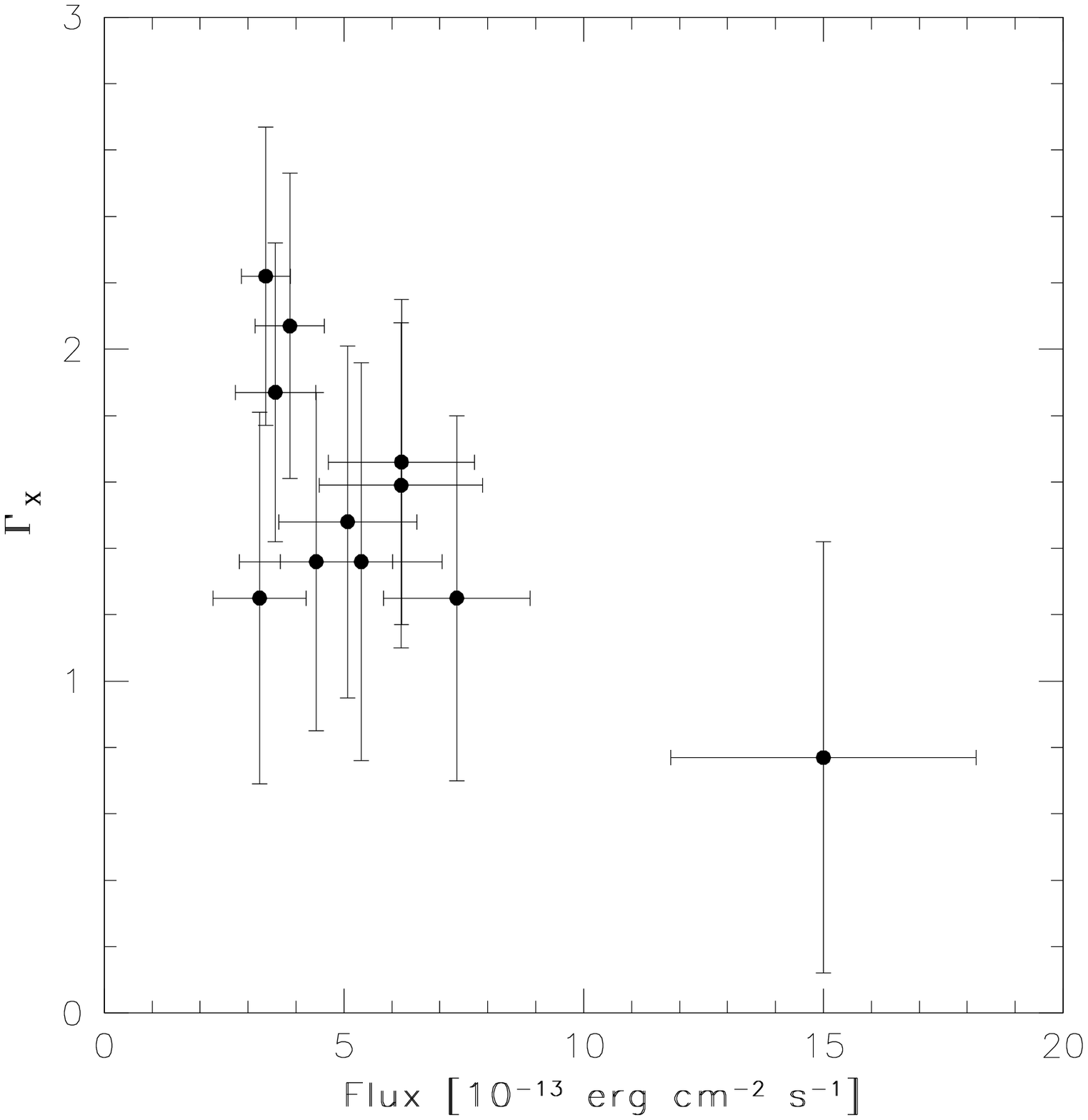}}}
\caption{{\em Swift}/XRT photon index as a function of the 0.3-10 keV unabsorbed flux of PMN J0948$+$0022 ({\it left-hand panel}) and PKS 1502$+$036 ({\it right-hand panel}). In the left panel, red squares refer to 2011 observations.}\label{XRThard2}
\end{center}
\end{figure*}

\begin{figure*}
\begin{center}
\rotatebox{0}{\resizebox{!}{70mm}{\includegraphics{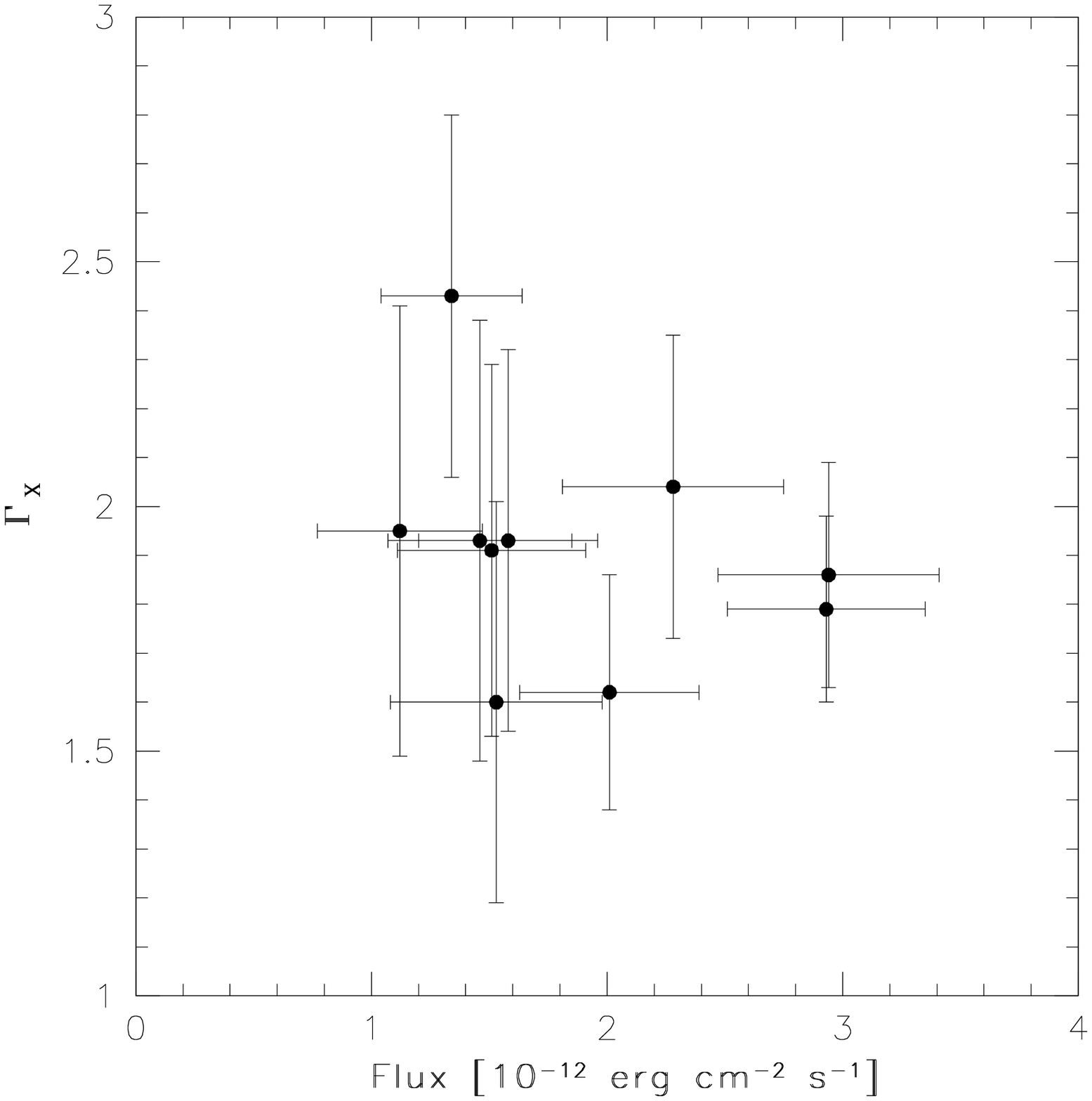}}}
\hspace{0.1cm}
\rotatebox{0}{\resizebox{!}{70mm}{\includegraphics{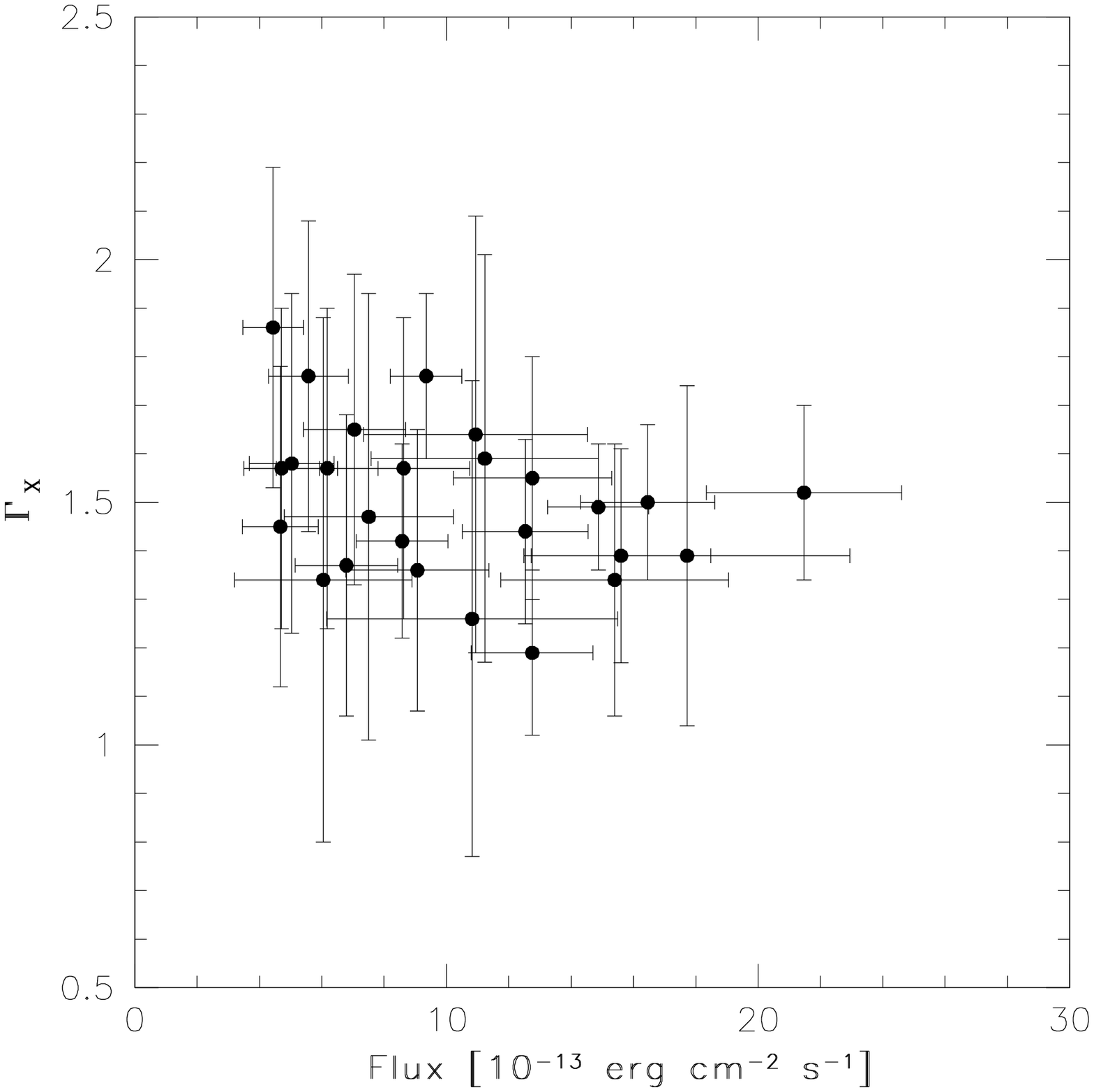}}}
\caption{{\em Swift}/XRT photon index as a function of the 0.3-10 keV unabsorbed flux of FBQS J1644$+$2619 ({\it left-hand panel}) and PKS 2004--447 ({\it right-hand panel}).}\label{XRThard3}
\end{center}
\end{figure*}

A hardening of the X-ray spectrum with the increase of the flux has been observed in some FSRQ \citep[e.g.,][]{dammando11, hayashida15, orienti20}, a spectral behaviour usually related to the competition between acceleration and cooling processes acting on relativistic electrons, with the injection of fresh energetic particles in the flow. On the other hand, AGN without jet display a softer-when-brighter behaviour \citep[e.g.,][]{serafinelli17}. Considering that there are multiple observations with sufficient exposure time for these NLSy1, a correlation between flux and spectral index has been searched. Figs.~\ref{XRThard1}, \ref{XRThard2}, and \ref{XRThard3} show the 0.3--10 keV X-ray flux and photon index values for each source, plotted against each other. No clear trend has been found in the six $\gamma$-ray-emitting NLSy1, except for PKS 1502$+$036, for which, based on a Pearson correlation coefficient, a strong anticorrelation between flux and spectral index (Pearson $r$-coefficient of $-$0.71) has been estimated at 99 per cent confidence level. A possible hint of hardening of the spectrum when the source is brighter (Pearson $r$-coefficient of $-$0.56 at 95 per cent confidence level) has been observed in PMN J0948$+$0022 by considering only the observations in 2011. This can be related to a change in the electron energy distribution of the emitting plasma in the relativistic jet as the main driver of the X-ray variability in that period. However, the large uncertainties do not allow us to obtain definitive conclusions.     
 
\begin{table}
\caption{X-ray spectral properties of the sample based on {\em Swift}-XRT observations.}
\begin{center}
\begin{tabular}{cccc}
\hline 
\multicolumn{1}{c}{\textbf{Source name}} &
\multicolumn{1}{c}{\textbf{$\Gamma_{\rm min}$}} &
\multicolumn{1}{c}{\textbf{$\Gamma_{\rm max}$}} &
\multicolumn{1}{c}{\textbf{$\Gamma_{\rm median}$}} \\
\hline
1H 0323$+$342     & 1.60 $\pm$ 0.08 & 2.48 $\pm$ 0.19 &  1.95 $\pm$ 0.14 \\ 
SBS 0846$+$513    & 1.15 $\pm$ 0.52 & 1.84 $\pm$ 0.54 &  1.45 $\pm$ 0.20 \\ 
PMN J0948$+$0022  & 1.24 $\pm$ 0.38 & 1.80 $\pm$ 0.19 &  1.60 $\pm$ 0.12 \\ 
PKS 1502$+$036    & 1.25 $\pm$ 0.56 & 2.22 $\pm$ 0.45 &  1.48 $\pm$ 0.41 \\ 
FBQS J1644$+$2619 & 1.60 $\pm$ 0.41 & 2.43 $\pm$ 0.37 &  1.92 $\pm$ 0.23 \\ 
PKS 2004--447     & 1.19 $\pm$ 0.17 & 1.86 $\pm$ 0.33 &  1.50 $\pm$ 0.16 \\ 
\hline
\end{tabular}
\end{center}
\label{XRT_statistics}
\end{table}

\section{Summed X-ray spectra}\label{summed}

\begin{table*}
\caption{X-ray fitting results of the summed spectra with a simple power-law model. Fluxes are corrected for the Galactic absorption. Asterisk marks cases in which a Cash statistics is used.}
\begin{center}
\begin{tabular}{cccccc}
\hline 
\multicolumn{1}{c}{\textbf{Source name}} &
\multicolumn{1}{c}{\textbf{Net exposure time}} &
\multicolumn{1}{c}{\textbf{Counts}} &
\multicolumn{1}{c}{\textbf{$\Gamma$}} &
\multicolumn{1}{c}{\textbf{$\chi^{2}$/d.o.f.}} &
\multicolumn{1}{c}\textbf{{Flux$_{\rm\,0.3-10\,keV}$}}  \\
\multicolumn{1}{c}{} &
\multicolumn{1}{c}{(s)} &
\multicolumn{1}{c}{} &
\multicolumn{1}{c}{} &
\multicolumn{1}{c}{} &
\multicolumn{1}{c}{(10$^{-12}$ erg cm$^{-2}$ s$^{-1}$)} \\
\hline
1H 0323$+$342     &  324586   & 100646  &  1.96 $\pm$ 0.01       &  1241.25/643  & 19.81 $\pm$ 0.01   \\
SBS 0846$+$513    &  100810   &   1200  &  1.43 $\pm$ 0.06       &  71.53/53     & 0.69 $\pm$ 0.05    \\
PMN J0948$+$0022  &  120913   &  11836  &  1.59 $\pm$ 0.02       &  365.32/319   & 4.92 $\pm$ 0.59    \\
IERS B1303$+$515  &    4158   &     12  &  1.29 $^{+0.87}_{-0.81}$  &  11.37/10*    & 0.18 $\pm$ 0.12     \\
B3 1441$+$476     &    6503   &     30  &  1.51 $\pm$ 0.55       &  35.87/27*    & 0.21 $\pm$ 0.09     \\
PKS 1502$+$036    &   43395   &    348  &  1.53 $\pm$ 0.16       &  15.36/15     & 0.45 $\pm$ 0.05     \\ 
FBQS J1644$+$2619 &   18830   &    742  &  1.89 $\pm$ 0.11       &  38.07/33     & 1.77 $\pm$ 0.12     \\ 
PKS 2004--447     &  151938   &   2705  &  1.50 $\pm$ 0.05       & 111.66/111    & 0.93 $\pm$ 0.04     \\ 
\hline
\end{tabular}
\end{center}
\label{XRTsummed_pow}
\end{table*}

\begin{figure}
\begin{center}
\rotatebox{270}{\resizebox{!}{70mm}{\includegraphics{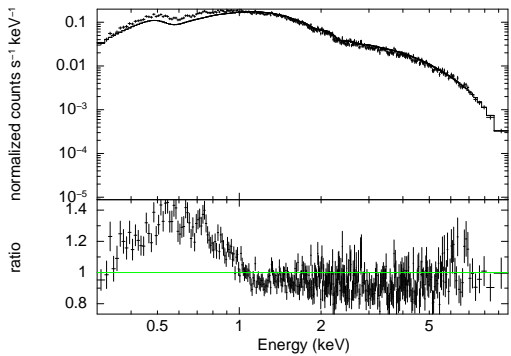}}}
\caption{XRT data of 1H 0323$+$342 fitted in the 0.3--10 keV energy range by a simple power law with a photon index $\Gamma$ = 1.81, as estimated above 2 keV. In the lower panel, the ratio of data to model is shown.}
\label{0323_excess}
\end{center}
\end{figure} 

\begin{table*}
\caption{X-ray fitting results of the summed spectra with broken power-law models, being $\Gamma_{1}$ and $\Gamma_{2}$ the photon index below and above the energy break $E_{break}$. Luminosities are corrected for the Galactic absorption.} 
\begin{center}
\begin{tabular}{ccccccc}
\hline 
\multicolumn{1}{c}{\textbf{Source name}} &
\multicolumn{1}{c}{\textbf{$\Gamma_{1}$}} &
\multicolumn{1}{c}{\textbf{$E_{break}$}} &
\multicolumn{1}{c}{\textbf{$\Gamma_{2}$}} &
\multicolumn{1}{c}{\textbf{$\chi^{2}$/d.o.f.}} &
\multicolumn{1}{c}{\textbf{F-test p-value}} &
\multicolumn{1}{c}{\textbf{Lum$_{\rm\,2-10\,keV}$}} \\
\multicolumn{1}{c}{} &
\multicolumn{1}{c}{} &
\multicolumn{1}{c}{(keV)} &
\multicolumn{1}{c}{} &
\multicolumn{1}{c}{} &
\multicolumn{1}{c}{} &
\multicolumn{1}{c}{}(erg s$^{-1}$) \\
\hline
1H 0323$+$342     & 2.12 $\pm$ 0.02          & 1.49 $\pm$ 0.11                  & 1.83 $\pm$ 0.02    &  947.67/641             & 2.73$\times$10$^{-38}$ & 8.88 $\times$10$^{43}$ \\
SBS 0846$+$513    & 1.07 $^{+0.15}_{-0.16}$  & 1.54 $^{+0.23}_{-0.27}$          & 1.77 $\pm$ 0.15    &  50.81/50               & 1.63$\times$10$^{-5}$  & 4.69 $\times$10$^{44}$ \\
PMN J0948$+$0022  & 1.69 $\pm$ 0.06          & 1.23 $^{+0.27}_{-0.23}$          & 1.55 $\pm$ 0.03    &  301.86/317             & 7.64$\times$10$^{-12}$ & 3.69 $\times$10$^{45}$ \\
PKS 1502$+$036    & 2.27 $^{+0.75}_{-0.57}$  & 0.73 $^{+0.21}_{-0.28}$          & 1.43 $\pm$0.17     & 13.09/13                & 0.35                   & 1.57 $\times$10$^{44}$ \\ 
FBQS J1644$+$2619 & 2.12 $\pm$ 0.18          & 1.41 $^{+0.37}_{-0.34}$          & 1.67 $\pm$ 0.18    & 31.56/31                & 0.05                   & 5.62 $\times$10$^{43}$ \\ 
PKS 2004--447     & 1.35 $^{+0.10}_{-0.11}$  & 1.58 $^{+0.42}_{-0.31}$          & 1.64 $\pm$ 0.10    & 103.93/110              & -                      & 9.33 $\times$10$^{43}$ \\ 
\hline
\end{tabular}
\end{center}
\label{XRTsummed_bkn}
\end{table*}

\begin{table*}
\caption{X-ray fitting results of the summed spectra with a power-law plus a black body model. $kT_{bb}$ is the temperature at inner disc radius.}
\begin{center}
\begin{tabular}{cccc}
\hline 
\multicolumn{1}{c}{\textbf{Source name}} &
\multicolumn{1}{c}{\bf{$\Gamma$}} &
\multicolumn{1}{c}{\bf{$kT_{bb}$}} &
\multicolumn{1}{c}{\bf{$\chi^{2}$/d.o.f.}} \\
\multicolumn{1}{c}{} &
\multicolumn{1}{c}{} &
\multicolumn{1}{c}{(keV)} &
\multicolumn{1}{c}{} \\   
\hline
1H 0323$+$342     &  1.78 $\pm$ 0.02          & 0.154 $\pm$ 0.004           & 760.32/641  \\     
SBS 0846$+$513    &  1.43 $\pm$ 0.07          & 0.224 $^{+0.431}_{-0.224}$  &  46.06/51   \\     
PMN J0948$+$0022  &  1.53 $\pm$ 0.04          & 0.214 $^{+0.043}_{-0.045}$  & 298.54/317  \\   
PKS 1502$+$036    &  1.42 $^{+0.21}_{-0.26}$  & 0.106 $^{+0.178}_{-0.100}$  & 13.23/13    \\   
FBQS J1644$+$2619 &  1.69 $\pm$ 0.20          & 0.140 $^{+0.048}_{-0.073}$  & 32.18/31    \\  
PKS 2004--447     &  1.50 $\pm$ 0.05          & 0.185 $^{+0.242}_{-0.164}$  & 103.48/109  \\  
\hline                                       
\end{tabular}
\end{center}
\label{XRTsummed_bbody}
\end{table*}

The summed X-ray spectra are then analysed for their higher signal-to-noise ratio compared to the single XRT observations. The spectra are first fitted with a single power law over the 0.3--10 keV energy range. The $\chi^{2}$ statistics has been used for all sources, except for IERS B1303$+$515 and B3 1441$+$476, for which due to the low statistics we used the Cash statistics. All spectra show a photon index lower than 2, in agreement with the hypothesis that the emission from relativistic jet dominates the X-ray spectrum in these sources. However, a good fit (0.9 $<$ $\chi^{2}_{\rm\,red}$ $<$ 1.1) has been obtained only for PKS 1502$+$036 and PKS 2004--447 (see Table~\ref{XRTsummed_pow}), suggesting the presence of an additional component to be considered in the 0.3--10 keV energy range. When the power-law model fitted in the 2.0--10 energy range is extrapolated to 0.3--10 keV, the presence of the soft X-ray excess is visible in the residuals of 1H 0323$+$342, SBS 0846$+$513, PMN J0948$+$0022, and FBQS J1644$+$2619 (the best example is 1H 0323$+$342, Fig.~\ref{0323_excess}). In case of PKS 1502$+$036, the lack of evidence of the soft X-ray excess may be related to the too low statistics. In fact, based on a relatively short {\em XMM--Newton} observation, a possible hint of the presence of a soft X-ray excess in this source as been reported in \citet{dammando16b}. On the other hand, long {\em XMM--Newton} observations of PKS 2004--447 in 2012 have shown that its X-ray spectrum is well described by a single power-law without significant soft excess \citep{orienti15,kreikenbohm16}, while only a hint of a possible soft excess has been seen in a {\em XMM--Newton} observation in 2004 \citep{gallo06}. No improvement of the fit has been obtained for any source when an intrinsic neutral absorber at the redshift of the source (\texttt{ztbabs} model) has been added to the model. This can be related to the fact that intrinsic absorption along the line of sight has been swept by the relativistic jet. 

Considering that for most of the sources a simple power law is not a good fit to the data, a broken power-law model is applied to the data. In case of IERS B1303$+$515 and B3 1441$+$476 the statistic is not enough to apply more complex model with respect to a simple power law. The significance of the broken power law model has been tested by applying the $F$-test. Fit results for the broken power law are summarized in Table~\ref{XRTsummed_bkn}. According to the $F$-test, the broken power law is significantly preferred over a single power law for 1H 0323$+$342, SBS 0846$+$513, and PMN J0948$+$0022. In case of FBQS J1644$+$2619 the broken power law improves the fit but it is not significantly preferred at 99 per cent confidence level over a single power law, in agreement with the results obtained with a long {\em XMM--Newton} observation \citep{larsson18}. No improvement or only a marginal improvement of the fit has been obtained for PKS 2004--447 and PKS 1502$+$036, respectively.

Leaving the column density of the Galactic absorption free to vary, a significant improvement of the fit ($\Delta$\,$\chi^{2}$ = 153 and $F$-test probability $>$ 99.99\%) has been obtained for 1H 0323$+$342, with N$_{\rm\,HI}$ = 2.19 $^{+0.15}_{-0.14}$ $\times$10$^{21}$ cm$^{-2}$, $\Gamma_{1}$ = 2.90 $\pm$ 0.12, $E_{break}$ = 1.28 $\pm$ 0.03, and $\Gamma_{2}$ = 1.91 $\pm$ 0.02. The N$_{\rm\,HI}$ value is higher than the value reported by \citet{kalberla05} but compatible with the total (H$_{\rm\,I}$ plus H$_{2}$) Galactic column of 2.17 $\times$ 10$^{21}$ cm$^{-2}$ reported by \citet{willingale13}. A similar value has been found by \citet{kynoch18} analysing {\em XMM--Newton} data of the source. Further high-resolution observations are needed to investigate in detail this issue. 

As already cited in Section \ref{Var_spec}, a likely explanation for the X-ray spectra of these $\gamma$-ray-emitting NLSy1 is that the underlying Seyfert emission, originating from the corona and accretion disc system, has a noticeable contribution at low energies, in particular the~so-called soft X-ray excess. Such a Seyfert component is a typical feature in the X-ray spectra of radio-quiet NLSy1, but it is quite unusual in jet-dominated AGN, even if not unique \citep[e.g., PKS 1510--089, 3C 273, and 4C $+$04.42;][]{grandi04,kataoka08,derosa08}. The~soft excess is often exceptionally strong in radio-quiet NLSy1, making it plausible that it would be detectable in the $\gamma$-ray-emitting radio-loud NLSy1 even though the jet emission is strong. In this context, the broken power law is a phenomenological representation of a two-component spectrum, with the coronal emission dominating below the energy break and the relativistic jet above the energy break. For all sources, the photon index above the energy break is quite hard ($\Gamma$ = 1.4--1.8), confirming the jet radiation as the dominant mechanism in the spectrum of $\gamma$-ray-emitting NLSy1 above $\sim$2 keV, in agreement with the results reported in \citet{larsson18}. In this context, the very hard photon index below the break energy obtained for SBS 0846$+$513 ($\Gamma$ $\sim$ 1.1) is not easily attributable to a Seyfert component. Deeper X-ray observations will be needed to investigate the spectrum of this source. 

Alternatively to the broken power-law model, to reproduce the soft X-ray excess a black body component (\texttt{zbbody} in \textsc{XSPEC}) in addition to the simple power law has been added. Similar results have been obtained for all sources by considering a spectrum from an accretion disc consisting of multiple black body components (\texttt{diskbb} in \textsc{XSPEC}). Adding a black body component to account for the soft excess improves the fit with respect to the simple power-law model (Table~\ref{XRTsummed_bbody}). As in the case of the broken power-law model, the photon index of the power law is quite hard ($\Gamma$ = 1.4--1.8), in agreement with the presence of a relativistic jet above 2 keV. In case of 1H 0323$+$342 the improvement of the fit is larger than what obtained with a broken power-law. A further improvement of the fit has been obtained by leaving the column density of the Galactic absorption free to vary ($\chi^{2}$/d.o.f. = 695.55/640). In that case the fit results in an N$_{\rm\,HI}$ = 1.81 $^{+0.14}_{-0.13}$ $\times$10$^{21}$ cm$^{-2}$, $\Gamma$ = 1.84 $\pm$ 0.02, with a black body temperature $kT_{bb}$ = 0.130 $\pm$ 0.004 keV. The temperatures of the accretion disc obtained for the six sources range values between 0.106 and 0.224 keV, although the values are quite well constrained only for 1H 0323$+$342, PMN J0948$+$0022, and FBQS J1644$+$2619. The temperatures are in agreement with the values detected for different AGN samples with different BH masses and accretion rates \citep[i.e., 0.1--0.2 keV, e.g.,][]{gierlinski04, crummy06}, suggesting that we are observing the same Seyfert-like component in both the radio-quiet and $\gamma$-ray-emitting NLSy1. However, the nature of the soft excess is still debated also for radio-quiet NLSy1, therefore it is not easy to understand the connection between the jet component and the soft excess in $\gamma$-ray-emitting NLSy1. Compelling models can be distinguished with joint observations with {\em XMM--Newton} and {\em NuSTAR} and future observations with the {\em Athena} and eXTP satellites.  
    
Considering that the X-ray spectrum of some $\gamma$-ray-emitting NLSy1 shows a Seyfert component like the soft X-ray excess it is reasonable searching for other features typical of radio-quiet Seyfert galaxies like the iron K$\alpha$ line. To reproduce the possible presence of an iron K$\alpha$ line in the X-ray spectrum of the sources a Gaussian emission-line profile ($\texttt{zgauss}$ in \textsc{XSPEC}) has been added to the broken power-law model. The energy of the line has been fixed to 6.4 keV, and the width of the line to 10 eV (i.e., lower than the energy resolution of {\em Swift}-XRT). In case of 1H 0323$+$342 the fit improves from a $\chi^{2}$/d.o.f. of 947.67/641 to 937.67/640, and the equivalent width (EW) of the iron line is 57 eV \footnote{Applying an F-test a p-value of 9.2$\times$10$^{-3}$ is obtained, suggesting an improvement. However see \citet{protassov02} for the use of the F-test in case of adding an emission line.}, in agreement with the detection of a weak iron line reported by \citet{kynoch18} analysing {\em XMM--Newton} EPIC-pn data. However, the {\em Swift}-XRT data alone are not good enough to safely claim the presence of this emission line. 

\noindent No significant improvement of the fit has been obtained in case of PMN J0948$+$0022 and PKS 2004--447 with an upper limit on the EW of 18 and 15 eV, respectively. The fit is not well constrained by adding the iron line in case of SBS 0846$+$513, PKS 1502$+$036, and FBQS J1644$+$2619, with a negligible normalization of the Gaussian emission line. According to the X-ray Baldwin effect, an anticorrelation between the EW of the iron K$\alpha$ line and the 2--10 keV luminosity has been observed in radio-quiet and radio-loud AGN \citep{grandi06,bianchi09}. A similar anti-correlation has been observed moving from the 1H 0323$+$342, at lower luminosity, to PMN J0948$+$0022 at higher luminosity. Following the relationship between the EW of the narrow iron line and the 2--10 keV luminosity based on radio-quiet type 1 AGN reported in \citet{bianchi07}, that is log(EW$_{\rm\,iron}$) = (1.73 $\pm$ 0.03) + (-0.17 $\pm$ 0.03) log(L$_{\rm\,X,44}$), where EW$_{\rm\,iron}$ is the EW of the neutral iron line in eV and L$_{\rm\,X,44}$ is the 2--10 keV X-ray luminosity in units of 10$^{44}$ erg s$^{-1}$, an EW of 55, 29, and 54 eV is expected for 1H 0323$+$342, PMN J0948$+$0022 and PKS 2004--447, respectively. The lower value obtained for PMN J0948$+$0022 with respect to what expected for a radio-quiet AGN can be related to a higher dilution of the thermal accretion flow from the non-thermal emission  of the relativistic jet. In this context, the even lower upper limit on EW obtained for PKS 2004--447 is in agreement with the fact that in this source even a clear soft excess is missing in the X-ray spectrum, suggesting that the X-ray emission is almost completely dominated by the jet radiation.
  
\section{Summary}\label{summary}

In this work, we have presented the results of a homogeneous analysis of the optical, UV, and X-ray data of eight of the nine $\gamma$-ray-emitting NLSy1 included in the 4FGL catalogue based on all {\em Swift} observations carried out up to 2019 April, for a total of 286 observations. The flux variability and spectral changes of the sources has been investigated, in particular for six of these eight sources (for IERS 1305+515 and B3 1441+476 there is not enough statistics for a detailed investigation). 

The 0.3--10 keV intrinsic luminosities of the $\gamma$-ray-emitting NLSy1 by {\em Swift}-XRT vary between 6.3$\times$10$^{43}$ and 1.8$\times$10$^{46}$ erg s$^{-1}$, with an average luminosity of 1.3$\times$10$^{45}$ erg s$^{-1}$, a range of values higher than what observed in other radio-loud NLSy1 and comparable to the luminosities of blazars in the same range of redshift. This is a strong indication that jet emission amplified by Doppler boosting effects due to the small angle of view with respect to the observers makes the luminosities higher in these sources with respect to the other radio-loud NLSy1. Similar conclusions have been obtained considering fluxes instead of luminosities. However, the relatively lower variability amplitude observed in X-rays (F$_{var}$ = 2.7--8.5) with respect to blazars is an indication that, even if the jet emission produces the dominant contribution, also the corona radiation is responsible for the X-ray emission in $\gamma$-ray-emitting NLSy1.

\noindent Rapid variability in X-rays has been found for 1H 0323$+$342, FBQS J1644$+$2619, and PKS 2004--447 with time-scales varying between $\sim$6 ks and $\sim$22 ks, suggesting that the emission is produced in compact regions within the jet, although we cannot ruled that these events are related to changes in accretion rate or in the disc--corona structure in the disc reprocessing scenario, in particular for FBQS J1644$+$2619. In fact, for 1H 0323$+$342 the X-ray flaring episode happened simultaneously to a $\gamma$-ray flare, making the jet as the most likely origin of this episode. In case of PKS 2004--447 no clear evidence of Seyfert-like features has been identified in its X-ray spectrum disfavouring the disc--corona emission as dominant mechanism.

The average 0.3--10 keV photon index, considering the six $\gamma$-ray-emitting NLSy1, is 1.77 $\pm$ 0.27 (1.83 $\pm$ 0.23 taking into account only the values with uncertainties lower than 0.3), harder than the photon index of a sample of radio-quiet NLSy1 and BLSy1 studied by \citet{grupe10}, suggesting that the X-ray spectrum of these sources is mainly produced by IC radiation from a relativistic jet. However, the average photon index of $\gamma$-ray-emitting NLSy1 is softer than the values obtained for FSRQ, indicating that a contribution from the corona and the accretion disc may be still present in the X-ray spectra of jetted NLSy1, in agreement with the presence of a significant soft X-ray excess in four of the six sources (i.e., 1H 0323$+$342, SBS 0846$+$513, PMN J0948$+$0022, and FBQS J1644$+$2619). The average 0.3--10 keV photon index of the sample of $\gamma$-ray-emitting NLSy1 is compatible with an average 2--10 keV photon index of $\sim$2 obtained by \citet{gliozzi20} and \citet{bianchi09} for radio-quiet NLSy1 using {\em XMM--Newton} data. The limited statistics of the {\em Swift}-XRT observations do not allow us to have robust estimation of the photon index in the 2--10 keV energy range for all sources for a direct comparison. However, the presence of a soft component in the 0.2--2 keV range should lead to a softer photon index for the total spectrum of NLSy1 in the 0.3--10 keV energy range, therefore the average photon index for the $\gamma$-ray emitting NLSy1 in the 2$-$10 keV should be harder than 1.8. For 1H 0323$+$342, that is the brightest X-ray source in our sample, the average 2--10 keV photon index obtained with Swift is 1.75 $\pm$ 0.17 with respect to 1.95 $\pm$ 0.14 obtained in the 0.3$-$10 keV energy range, confirming an harder spectrum reducing the energy  range from 0.3--10 to 2--10 keV.   

\noindent Similar to the average photon index of the sample studied of $\gamma$-ray-emitting NLSy1, the median of the 0.3$-$10 keV X-ray spectral index obtained in the single XRT observations is lower than 2 for all six NLSy1. For three sources (1H 0323$+$342, PKS 1502$+$036, and FBQS J1644$+$2619) the spectrum becomes softer in some observations, reaching values of 2.2$-$2.4. This indicates that in some periods for these sources the jet radiation can be less important in X-rays and the disc--corona component can be more easily detectable in the spectrum. No clear trend between flux and spectral index has been found in the six $\gamma$-ray-emitting NLSy1, except for PKS 1502$+$036, and a possible hint of harder-when-brighter behaviour observed in 2011 for PMN J0948$+$0022.

In optical and UV bands, significant changes on time-scale of one day (for 1H 0323$+$342) and a few days (for SBS 0846$+$513, PMN J0948$+$0022, and PKS 1502$+$036) have been observed during the {\em Swift} monitoring, clearly related to an increase of the jet synchrotron emission. On a long-term scale, a large variability amplitude, significantly higher with respect to the other radio-loud NLSy1, has been observed for all six $\gamma$-ray-emitting NLSy1. The variability amplitude in the optical and UV bands increases with increasing frequency in quasars \citep[e.g.,][]{vandenberk04}, differently from what is observed in these NLSy1. These results confirm the dominance of the jet radiation in this part of the spectrum of $\gamma$-ray-emitting NLSy1. Variable synchrotron is more contaminated by thermal emission from the accretion disc at higher frequencies emission, in agreement with the relatively smaller variability observed in the UV bands, at least in the cases in which a strong accretion disc is present (i.e., 1H 0323$+$342, PMN J0948$+$0022, and PKS 1502$+$036). 
 
A correlation between optical, UV, and X-ray bands is expected in case the jet radiation being the dominant mechanisms in the optical-to-X-ray part of the spectrum, as usually observed in blazars. We found a strong positive correlation between X-ray, UV, and optical emission for 1H 0323$+$342, SBS 0846$+$513, PMN J0948$+$0022 (at the 99 per cent confidence level), and for FBQS J1644$+$2619 (at the 95 per cent confidence level). The lack of significant correlation for PKS 2004$-$447 and PKS 1502$+$036, as well as the lower significance obtained for FBQS J1644$+$2619, can be related to the limited sampling of these three sources. The observed correlation can be in agreement also with the disc reprocessing scenario usually proposed for interpreting the multiband variability in radio-quiet AGN. However, in that scenario a larger amplitude variability is expected in X-rays with respect to optical and UV bands. This is different from what is observed in SBS 0846$+$513, PMN J0948$+$0022, FBQS J1644$+$2619, and PKS 2004--447, where a lower or comparable X-ray variability has been observed, ruling out the disc reprocessing scenario in favour of the jet-dominated scenario. Moreover, simultaneous flux variations has been observed in optical, UV and X-rays (and in some cases also in $\gamma$ rays) for 1H 0323$+$342, SBS 0846$+$513, PMN J0948$+$0022, and, although with variations of minor entity, for FBQS J1644$+$2619 and PKS 2004--447, as expected in the jet-dominated scenario. A delayed optical and UV emission with respect of the X-ray emission is expected in the disc reprocessing scenario, disfavoring this scenario to interpret the multiband variability of these sources. In case of PKS 1502$+$036, the almost simultaneous increase of activity observed in optical, UV, and $\gamma$ rays is a clear indication that the synchrotron emission is the dominant mechanism during that activity period. It is intriguing the delay of $\sim$10 days of the X-ray peak with respect to the optical and UV ones, not easy to be explained in both the jet-dominated and disc reprocessing scenarios. 

Fitting the summed XRT spectra with a simple power-law model in the 0.3--10 keV energy range a good fit (0.9 $< \chi^2_{\rm\,red} <$ 1.1) has been obtained only for PKS 2004--447 and to a lesser extent for PKS 1502$+$036, suggesting the presence of an additional component in the spectrum of the other sources. In case of PKS 2004--447 the 0.3--10 keV spectrum is well fitted by a single power law with a hard photon index of 1.5, similar to the typical jet-dominated spectra of FSRQ. An improvement of the fit has been obtained using a broken power law for 1H 0323$+$342, SBS 0846$+$513, and PMN J0948$+$0022, and less significantly for FBQS J1644$+$2619. A likely explanation for the X-ray spectra of these $\gamma$-ray-emitting NLSy1 is that the underlying Seyfert emission, originating from the corona and accretion disc, has a noticeable contribution at low energies, in particular the so-called soft X-ray excess. For those four sources and PKS 1502$+$036, the photon index above the break ($\Gamma$ = 1.4--1.8) is significantly harder than in radio-quiet NLSy1 and instead similar to FSRQ, showing that IC emission from the jet is dominating the spectrum. A black body component, in addition to a hard power law reproducing the jet emission, provides a fit comparable to the broken power-law model. The distribution of black body temperatures of the accretion disc (0.106--0.224 keV) is in the typical range 0.1--0.2 keV observed also for other AGN with soft excess, suggesting that the same Seyfert-like component is present in both the radio-quiet and $\gamma$-ray-emitting NLSy1. 

A slight improvement of the fit adding an iron $K\alpha$ line in the X-ray spectrum to the broken power-law model has been obtained only for 1H 0323$+$342, in agreement with the weak emission line  reported in \citet{kynoch18}, with an EW of 57 eV. An upper limit of 18 and 15 eV on the EW of the iron line has been obtained for PMN J0948$+$0022 and PKS 2004$-$447, respectively. An anticorrelation between the EW of the iron line and the 2-10 keV luminosity has been observed for 1H 0323$+$324 and PMN J0948$+$0022, in agreement with the X-ray Baldwin effect observed for radio-quiet AGN type 1, although the value obtained for PMN J0948$+$0022 is lower than what expected for the anticorrelation obtained for radio-quiet AGN. This can be related to a higher dilution of the thermal accretion flow from the non-thermal emission of the relativistic jet. In the same way, the lower upper limit on the EW of the line obtained for PKS 2004$-$447 with respect to the expected value is in agreement with the lack of any signs of Seyfert components, including the soft excess, in the X-ray spectrum of this source.

Long joint {\em XMM--Newton} and {\em NuSTAR} observations will be important to better characterize the spectra of this small class of beamed-jetted AGN. A further step forward to place stronger constraints on the models, study the connection between the jet and accretion flow, and understand the origin of feedback and its relation with the AGN activity, in particular with the relativistic jets, will be obtained with {\em Athena} observations of these sources \citep{barcons17}.    
 
\section*{Acknowledgements}

This work is dedicated to the memory of Neil Gehrels. FD thank the {\em Swift} team for making these observations possible, the duty scientists, and science planners. This research has made use of the XRT Data Analysis Software (XRTDAS). This work made use of data supplied by the UK Swift Science Data Centre at the University of Leicester. This research has made use of data obtained through the High Energy Astrophysics Science Archive Research Center Online Service, provided by the NASA/Goddard Space Flight Center. This research has made use of the NASA/IPAC Extragalactic Database (NED), which is operated by the Jet Propulsion Laboratory, California Institute of Technology, under contract with the National Aeronautics and Space Administration. FD acknowledges financial contribution from the agreement ASI-INAF n. 2017-14-H.0. FD thank the anonymous referee for constructive comments that improved the clarity of the paper, and Andrew Lobban, Maurizio Paolillo, and Ciro Pinto for useful suggestions during the revision of the manuscript.

\appendix

\onecolumn
 
\section{Swift-XRT results}\label{XRT_Appendix}

\setcounter{table}{0}
\begin{table*}
\caption{Log and fitting results of {\em Swift}-XRT observations of 1H 0323$+$342 using a PL model with $N_{\rm H}$ fixed to Galactic absorption,  that is 1.27$\times$10$^{21}$ cm$^{-2}$. Fluxes are corrected for the Galactic absorption.} 
\label{XRT_0323}
\begin{center}
\begin{tabular}{ccccc}
\hline
\multicolumn{1}{c}{\textbf{Date}} &
\multicolumn{1}{c}{\textbf{MJD}} &
\multicolumn{1}{c}{\textbf{Net exposure time}} &
\multicolumn{1}{c}{\textbf{Photon index}}      &
\multicolumn{1}{c}{\textbf{Flux$_{\rm\,0.3-10\,keV}$}}  \\
\multicolumn{1}{c}{(UT)} &
\multicolumn{1}{c}{} &
\multicolumn{1}{c}{(s)} &
\multicolumn{1}{c}{($\Gamma_{\rm\,X}$)}  &
\multicolumn{1}{c}{(10$^{-11}$ erg cm$^{-2}$ s$^{-1}$)}  \\
\hline
2006-07-05  &  53921  &  1516  &    2.01 $\pm$  0.14   &    1.64 $\pm$  0.16  \\
2006-07-06  &  53922  &  8274  &    2.06 $\pm$  0.05   &    2.14 $\pm$  0.08  \\
2006-07-09  &  53925  &  2390  &    2.03 $\pm$  0.09   &    2.12 $\pm$  0.14  \\  
2006-07-09  &  53925  &  5379  &    2.04 $\pm$  0.05   &    3.78 $\pm$  0.14  \\
2006-07-14  &  53930  &   827  &    2.10 $\pm$  0.18   &    1.89 $\pm$  0.24  \\
2007-11-25  &  54429  &  2215  &    1.89 $\pm$  0.13   &    1.02 $\pm$  0.10  \\
\hline
\end{tabular}
\end{center}
\end{table*}

\addtocounter{table}{-1}
\begin{table*}
\begin{center}
\begin{tabular}{ccccc}
\hline
\multicolumn{1}{c}{\textbf{Date}} &
\multicolumn{1}{c}{\textbf{MJD}} &
\multicolumn{1}{c}{\textbf{Net exposure time}} &
\multicolumn{1}{c}{\textbf{Photon index}}      &
\multicolumn{1}{c}{\textbf{Flux$_{\rm\,0.3-10\,keV}$}}  \\
\multicolumn{1}{c}{(UT)} &
\multicolumn{1}{c}{} &
\multicolumn{1}{c}{(s)} &
\multicolumn{1}{c}{($\Gamma_{\rm\,X}$)}  &
\multicolumn{1}{c}{(10$^{-11}$ erg cm$^{-2}$ s$^{-1}$)}  \\
\hline
2012-02-02  &  55959  &  1304  &    1.92 $\pm$  0.14   &    1.47 $\pm$  0.16  \\ 
2012-03-03  &  55989  &  1558  &    1.89 $\pm$  0.15   &    1.05 $\pm$  0.12  \\ 
2013-01-13  &  56305  &  3654  &    1.83 $\pm$  0.08   &    1.46 $\pm$  0.09 \\ 
2013-01-14  &  56306  &  2767  &    1.94 $\pm$  0.09   &    1.61 $\pm$  0.11  \\ 
2013-01-15  &  56307  &  3711  &    1.88 $\pm$  0.08   &    1.51 $\pm$  0.09  \\ 
2013-02-15  &  56338  &  4103  &    1.62 $\pm$  0.09   &    1.09 $\pm$  0.09  \\
\hline
\end{tabular}
\end{center}
\end{table*} 

\addtocounter{table}{-2}
\begin{table*}
\begin{center}
\begin{tabular}{ccccc}
\hline
\multicolumn{1}{c}{\textbf{Date}} &
\multicolumn{1}{c}{\textbf{MJD}} &
\multicolumn{1}{c}{\textbf{Net exposure time}} &
\multicolumn{1}{c}{\textbf{Photon index}}      &
\multicolumn{1}{c}{\textbf{Flux$_{\rm\,0.3-10\,keV}$}}  \\
\multicolumn{1}{c}{(UT)} &
\multicolumn{1}{c}{} &
\multicolumn{1}{c}{(s)} &
\multicolumn{1}{c}{($\Gamma_{\rm\,X}$)}  &
\multicolumn{1}{c}{(10$^{-11}$ erg cm$^{-2}$ s$^{-1}$)}  \\
\hline
2018-08-16  &  58346  &  1049  &    2.20 $\pm$  0.17   &    1.39 $\pm$  0.16  \\
2018-08-23  &  58353  &   867  &    1.92 $\pm$  0.20   &    1.11 $\pm$  0.16  \\
2018-08-30  &  58360  &  1004  &    2.11 $\pm$  0.17   &    1.29 $\pm$  0.15  \\
2018-09-06  &  58367  &  1031  &    1.99 $\pm$  0.19   &    0.93 $\pm$  0.13  \\
2018-09-13  &  58374  &   909  &    2.09 $\pm$  0.16   &    1.69 $\pm$  0.19  \\
2018-09-20  &  58381  &  1049  &    2.07 $\pm$  0.30   &    1.02 $\pm$  0.21  \\
2018-09-27  &  58388  &  1074  &    2.09 $\pm$  0.21   &    1.15 $\pm$  0.16  \\
2018-10-04  &  58395  &   969  &    2.09 $\pm$  0.19   &    1.21 $\pm$  0.16  \\
2018-10-11  &  58402  &  1009  &    2.05 $\pm$  0.20   &    0.98 $\pm$  0.13  \\
2018-10-18  &  58409  &  1004  &    2.48 $\pm$  0.19   &    1.51 $\pm$  0.18  \\
2018-10-25  &  58416  &  1079  &    1.84 $\pm$  0.21   &    1.09 $\pm$  0.17  \\
2018-11-01  &  58423  &  1006  &    1.84 $\pm$  0.20   &    1.32 $\pm$  0.19  \\
2018-11-08  &  58430  &  1046  &    1.90 $\pm$  0.19   &    0.93 $\pm$  0.13  \\
2018-11-15  &  58437  &   872  &    1.95 $\pm$  0.21   &    0.97 $\pm$  0.17  \\
2018-11-22  &  58444  &   887  &    1.92 $\pm$  0.20   &    1.05 $\pm$  0.15  \\
2018-11-29  &  58451  &   999  &    2.01 $\pm$  0.16   &    1.31 $\pm$  0.16  \\
2018-12-06  &  58458  &   964  &    1.89 $\pm$  0.19   &    1.17 $\pm$  0.16  \\ 
2018-12-13  &  58455  &  1029  &    1.98 $\pm$  0.17   &    1.29 $\pm$  0.16  \\
\hline
\end{tabular}
\end{center}
\end{table*} 

\newpage

\setcounter{table}{1}
\begin{table*}
\caption{Log and fitting results of {\em Swift}-XRT observations of SBS 0846$+$513 using a PL model with $N_{\rm H}$ fixed to Galactic absorption,  that is 2.91$\times$10$^{20}$ cm$^{-2}$. Fluxes are corrected for the Galactic absorption.} 
\label{XRT_0846}
\begin{center}
\begin{tabular}{ccccc}
\hline
\multicolumn{1}{c}{\textbf{Date}} &
\multicolumn{1}{c}{\textbf{MJD}} &
\multicolumn{1}{c}{\textbf{Net exposure time}} &
\multicolumn{1}{c}{\textbf{Photon index}}      &
\multicolumn{1}{c}{\textbf{Flux$_{\rm\,0.3-10\,keV}$}}  \\  
\multicolumn{1}{c}{(UT)} &
\multicolumn{1}{c}{} &
\multicolumn{1}{c}{(s)} &
\multicolumn{1}{c}{($\Gamma_{\rm\,X}$)}  &
\multicolumn{1}{c}{(10$^{-13}$ erg cm$^{-2}$ s$^{-1}$)}  \\
\hline
2011-08-30 & 55803 &  6738 & 1.40 $\pm$ 0.24 &   8.86 $\pm$  1.77 \\ 
2011-09-15 & 55819 &  5439 & 1.40 $\pm$ 0.28 &   7.92 $\pm$  1.90 \\  
2011-12-20 & 55915 &  1968 & 1.71 $\pm$ 0.46 &   7.20 $\pm$  2.47 \\  
2011-12-27 & 55922 &  1526 & 1.50 $\pm$ 0.70 &  10.86 $\pm$  4.11 \\  
2011-12-28 & 55923 &  2540 & 1.59 $\pm$ 0.37 &   9.92 $\pm$  2.90 \\  
2012-05-27 & 56074 &  1965 & 1.45 $\pm$ 0.36 &  13.26 $\pm$  3.81 \\
\hline
\end{tabular}
\end{center}
\end{table*}  
 
\setcounter{table}{2}
\begin{table*}
\caption{Log and fitting results of {\em Swift}-XRT observations of PMN J0948$+$0022 using a PL model with $N_{\rm H}$ fixed to Galactic absorption, that is 5.08$\times$10$^{20}$ cm$^{-2}$. Fluxes are corrected for the Galactic absorption.} 
\label{XRT_0948}
\begin{center}
\begin{tabular}{ccccc}
\hline
\multicolumn{1}{c}{\textbf{Date}} &
\multicolumn{1}{c}{\textbf{MJD}} &
\multicolumn{1}{c}{\textbf{Net exposure time}} &
\multicolumn{1}{c}{\textbf{Photon index}}      &
\multicolumn{1}{c}{\textbf{Flux$_{\rm\,0.3-10\,keV}$}}  \\  
\multicolumn{1}{c}{(UT)} &
\multicolumn{1}{c}{} &
\multicolumn{1}{c}{(s)} &
\multicolumn{1}{c}{($\Gamma_{\rm\,X}$)}  &
\multicolumn{1}{c}{(10$^{-12}$ erg cm$^{-2}$ s$^{-1}$)}  \\
\hline
2008-12-05 &  54805  &   4195  & 1.63 $\pm$ 0.13 &  4.19 $\pm$ 0.43 \\
2009-03-26 &  54916  &   4850  & 1.66 $\pm$ 0.08 &  8.40 $\pm$ 0.55 \\ 
2009-04-15 &  54936  &   4393  & 1.57 $\pm$ 0.11 &  5.94 $\pm$ 0.51 \\
2009-05-05 &  54956  &   4817  & 1.55 $\pm$ 0.09 &  8.50 $\pm$ 0.59 \\
2009-05-10 &  54961  &   4925  & 1.69 $\pm$ 0.11 &  4.67 $\pm$ 0.41 \\
2009-05-15 &  54966  &   1416  & 1.67 $\pm$ 0.33 &  2.23 $\pm$ 0.54 \\
\hline
\end{tabular}
\end{center}
\end{table*} 

\setcounter{table}{3}
\begin{table*}
\caption{Log and fitting results of {\em Swift}-XRT observations of PKS 1502$+$036 using a PL model with $N_{\rm H}$ fixed to Galactic absorption, that is 3.93$\times$10$^{20}$ cm$^{-2}$. Fluxes are corrected for the Galactic absorption.} 
\label{XRT_1502}
\begin{center}
\begin{tabular}{ccccc}
\hline
\multicolumn{1}{c}{\textbf{Date}} &
\multicolumn{1}{c}{\textbf{MJD}} &
\multicolumn{1}{c}{\textbf{Net exposure time}} &
\multicolumn{1}{c}{\textbf{Photon index}}      &
\multicolumn{1}{c}{\textbf{Flux$_{\rm\,0.3-10\,keV}$}}  \\  
\multicolumn{1}{c}{(UT)} &
\multicolumn{1}{c}{} &
\multicolumn{1}{c}{(s)} &
\multicolumn{1}{c}{($\Gamma_{\rm\,X}$)}  &
\multicolumn{1}{c}{(10$^{-13}$ erg cm$^{-2}$ s$^{-1}$)}  \\
\hline
2009-07-25  &  55037  &  4672  &  1.25 $\pm$ 0.56  & 3.24 $\pm$ 0.97 \\
2012-04-25  &  56042  &  4822  &  1.36 $\pm$ 0.51  & 4.42 $\pm$ 1.60 \\
2012-05-25  &  56072  &  4657  &  1.87 $\pm$ 0.45  & 3.57 $\pm$ 0.84 \\
2012-06-25  &  56103  &  5154  &  2.22 $\pm$ 0.45  & 3.37 $\pm$ 0.51 \\
2012-08-07  &  56146  &  3946  &  2.07 $\pm$ 0.46  & 3.87 $\pm$ 0.72 \\
2015-12-21  &  57377  &  2782  &  1.48 $\pm$ 0.53  & 5.08 $\pm$ 1.44 \\
2015-12-25  &  57381  &  2215  &  1.36 $\pm$ 0.60  & 5.36 $\pm$ 1.69 \\
2016-01-01  &  57388  &  1733  &  0.77 $\pm$ 0.65  & 15.0 $\pm$ 3.18 \\
2016-01-08  &  57395  &  2225  &  1.66 $\pm$ 0.49  & 6.20 $\pm$ 1.52 \\
2016-01-14  &  57401  &  2700  &  1.25 $\pm$ 0.55  & 7.35 $\pm$ 1.53 \\
2016-06-24  &  57563  &  2283  &  1.59 $\pm$ 0.49  & 6.19 $\pm$ 1.70 \\
\hline
\end{tabular}
\end{center}
\end{table*} 

\setcounter{table}{4}
\begin{table*}
\caption{Log and fitting results of {\em Swift}-XRT observations of FBQS J1644$+$2619 using a PL model with $N_{\rm H}$ fixed to Galactic absorption, that is 5.14$\times$10$^{20}$ cm$^{-2}$. Fluxes are corrected for the Galactic absorption.} 
\label{XRT_1644}
\begin{center}
\begin{tabular}{ccccc}
\hline
\multicolumn{1}{c}{\textbf{Date}} &
\multicolumn{1}{c}{\textbf{MJD}} &
\multicolumn{1}{c}{\textbf{Net exposure time}} &
\multicolumn{1}{c}{\textbf{Photon index}}      &
\multicolumn{1}{c}{\textbf{Flux$_{\rm\,0.3-10\,keV}$}}  \\  
\multicolumn{1}{c}{(UT)} &
\multicolumn{1}{c}{} &
\multicolumn{1}{c}{(s)} &
\multicolumn{1}{c}{($\Gamma_{\rm\,X}$)}  &
\multicolumn{1}{c}{(10$^{-12}$ erg cm$^{-2}$ s$^{-1}$)}  \\
\hline
2011-12-26 & 55921 &  1361 & 2.04 $\pm$ 0.31 & 2.28 $\pm$ 0.47 \\
2015-04-09 & 57121 &  1656 & 1.93 $\pm$ 0.39 & 1.58 $\pm$ 0.38 \\
2015-05-05 & 57147 &  1556 & 1.91 $\pm$ 0.38 & 1.51 $\pm$ 0.40 \\
2015-06-05 & 57178 &  1326 & 1.95 $\pm$ 0.46 & 1.12 $\pm$ 0.35 \\
2015-07-05 & 57208 &  1523 & 1.60 $\pm$ 0.41 & 1.53 $\pm$ 0.45 \\
2015-08-05 & 57239 &  2000 & 1.86 $\pm$ 0.23 & 2.94 $\pm$ 0.47 \\
2015-09-05 & 57270 &  1503 & 1.93 $\pm$ 0.45 & 1.46 $\pm$ 0.39 \\
2017-02-27 & 57811 &  2909 & 1.62 $\pm$ 0.24 & 2.01 $\pm$ 0.38 \\
2017-03-07 & 57819 &  2705 & 1.79 $\pm$ 0.19 & 2.93 $\pm$ 0.42 \\
2018-01-18 & 58136 &  2013 & 2.43 $\pm$ 0.37 & 1.34 $\pm$ 0.30 \\
\hline
\end{tabular}
\end{center}
\end{table*}

\setcounter{table}{5} 
\begin{table*}
\caption{Log and fitting results of {\em Swift}-XRT observations of PKS 2004--447 using a PL model with $N_{\rm H}$ fixed to Galactic absorption, that is 3.17$\times$10$^{20}$ cm$^{-2}$. Fluxes are corrected for the Galactic absorption.} 
\label{XRT_2004}
\begin{center}
\begin{tabular}{ccccc}
\hline
\multicolumn{1}{c}{\textbf{Date}} &
\multicolumn{1}{c}{\textbf{MJD}} &
\multicolumn{1}{c}{\textbf{Net exposure time}} &
\multicolumn{1}{c}{\textbf{Photon index}}      &
\multicolumn{1}{c}{\textbf{Flux$_{\rm\,0.3-10\,keV}$}}  \\
\multicolumn{1}{c}{(UT)} &
\multicolumn{1}{c}{} &
\multicolumn{1}{c}{(s)} &
\multicolumn{1}{c}{($\Gamma_{\rm\,X}$)}  &
\multicolumn{1}{c}{(10$^{-13}$ erg cm$^{-2}$ s$^{-1}$)}  \\
\hline
2011-05-15 & 55696 &  6788  &  1.57 $\pm$ 0.33  &   4.71 $\pm$ 1.21 \\
2011-07-14 & 55756 &  4935  &  1.57 $\pm$ 0.33  &   6.18 $\pm$ 1.63 \\
2011-07-23 & 55765 &  1523  &  1.26 $\pm$ 0.49  &  10.82 $\pm$ 4.67 \\
2011-09-17 & 55821 &  7362  &  1.44 $\pm$ 0.19  &  12.53 $\pm$ 2.02 \\
2011-11-15 & 55880 &  7115  &  1.86 $\pm$ 0.33  &   4.44 $\pm$ 0.97 \\
2012-03-14 & 56000 &  7327  &  1.45 $\pm$ 0.33  &   4.67 $\pm$ 1.22 \\
\hline
\end{tabular}
\end{center}
\end{table*} 

\clearpage

\newpage 

\section{{\em Swift}-UVOT results}\label{UVOT_Appendix}

\setcounter{table}{0}
\begin{table*}
\caption{Observed magnitude of 1H 0323$+$324 obtained by {\em Swift}-UVOT.}
\label{UVOT_0323}
\begin{center}
\begin{tabular}{cccccccc}
\hline
\multicolumn{1}{c}{\textbf{Date (UT)}} &
\multicolumn{1}{c}{\textbf{MJD}}       &
\multicolumn{1}{c}{\textbf{$v$}}       &
\multicolumn{1}{c}{\textbf{$b$}}       &
\multicolumn{1}{c}{\textbf{$u$}}       &
\multicolumn{1}{c}{\textbf{$w1$}}      &
\multicolumn{1}{c}{\textbf{$m2$}}      &
\multicolumn{1}{c}{\textbf{$w2$}}      \\
\hline
2006-07-06  &  53922  &  15.79 $\pm$ 0.09  &  16.11 $\pm$ 0.07  &  15.17 $\pm$ 0.06 & 15.32 $\pm$ 0.07  & 15.63 $\pm$ 0.09  &   15.57 $\pm$ 0.07  \\
2006-07-09  &  53925  &          $-$       &  16.14 $\pm$ 0.07  &  15.13 $\pm$ 0.06 & 15.29 $\pm$ 0.07  &       $-$         &           $-$       \\
2007-07-20  &  54301  &  15.54 $\pm$ 0.07  &  16.16 $\pm$ 0.07  &  15.33 $\pm$ 0.07 & 15.63 $\pm$ 0.08  & 15.88 $\pm$ 0.09  &   15.79 $\pm$ 0.07  \\
2007-11-04  &  54410  &  15.77 $\pm$ 0.05  &  16.37 $\pm$ 0.06  &  15.58 $\pm$ 0.06 & 15.91 $\pm$ 0.07  & 16.29 $\pm$ 0.09  &   16.14 $\pm$ 0.07  \\
2007-11-11  &  54417  &  15.80 $\pm$ 0.06  &  16.39 $\pm$ 0.06  &  15.61 $\pm$ 0.06 & 15.80 $\pm$ 0.07  & 16.10 $\pm$ 0.09  &   15.99 $\pm$ 0.06  \\
2007-11-25  &  54429  &  15.77 $\pm$ 0.06  &  16.34 $\pm$ 0.06  &  15.54 $\pm$ 0.06 & 15.89 $\pm$ 0.07  & 16.23 $\pm$ 0.08  &   16.13 $\pm$ 0.06  \\
2007-12-01  &  54435  &  15.80 $\pm$ 0.08  &  16.30 $\pm$ 0.07  &  15.58 $\pm$ 0.07 & 15.80 $\pm$ 0.08  & 16.05 $\pm$ 0.10  &   15.98 $\pm$ 0.07  \\
\hline
\end{tabular}
\end{center}
\end{table*} 

\addtocounter{table}{-1}
\begin{table*}
 \begin{center}
\begin{tabular}{cccccccc}
\hline
\multicolumn{1}{c}{\textbf{Date (UT)}} &
\multicolumn{1}{c}{\textbf{MJD}}       &
\multicolumn{1}{c}{\textbf{$v$}}       &
\multicolumn{1}{c}{\textbf{$b$}}       &
\multicolumn{1}{c}{\textbf{$u$}}       &
\multicolumn{1}{c}{\textbf{$w1$}}      &
\multicolumn{1}{c}{\textbf{$m2$}}      &
\multicolumn{1}{c}{\textbf{$w2$}}      \\
\hline 
2012-02-02  &  55959  &  15.65 $\pm$ 0.07  &  16.25 $\pm$ 0.06  &  15.42 $\pm$ 0.06 & 15.65 $\pm$ 0.08  & 16.06 $\pm$ 0.07  &   16.00 $\pm$ 0.07  \\
2012-03-03  &  55989  &  15.85 $\pm$ 0.06  &  16.48 $\pm$ 0.06  &  15.65 $\pm$ 0.06 & 15.90 $\pm$ 0.07  & 16.30 $\pm$ 0.08  &   16.22 $\pm$ 0.07  \\
2013-01-13  &  56305  &  15.73 $\pm$ 0.07  &  16.35 $\pm$ 0.07  &  15.54 $\pm$ 0.05 & 15.75 $\pm$ 0.08  & 16.19 $\pm$ 0.10  &   16.00 $\pm$ 0.07  \\
2013-01-14  &  56306  &  15.64 $\pm$ 0.07  &  16.35 $\pm$ 0.07  &  15.50 $\pm$ 0.06 & 15.79 $\pm$ 0.08  & 16.03 $\pm$ 0.09  &   15.95 $\pm$ 0.07  \\
2013-01-15  &  56307  &  15.75 $\pm$ 0.08  &  16.31 $\pm$ 0.08  &  15.53 $\pm$ 0.05 & 15.64 $\pm$ 0.08  & 15.99 $\pm$ 0.09  &   15.84 $\pm$ 0.07  \\
2013-02-15  &  56338  &  15.79 $\pm$ 0.06  &  16.32 $\pm$ 0.06  &  15.52 $\pm$ 0.06 & 15.72 $\pm$ 0.07  & 16.14 $\pm$ 0.08  &   16.10 $\pm$ 0.07  \\
2013-02-15  &  56338  &  15.81 $\pm$ 0.06  &  16.34 $\pm$ 0.06  &  15.54 $\pm$ 0.06 & 15.87 $\pm$ 0.07  & 16.20 $\pm$ 0.08  &   16.09 $\pm$ 0.07  \\
\hline
\end{tabular}
\end{center}
\end{table*} 

\addtocounter{table}{-2}
\begin{table*}
 \begin{center}
\begin{tabular}{cccccccc}
\hline
\multicolumn{1}{c}{\textbf{Date (UT)}} &
\multicolumn{1}{c}{\textbf{MJD}}       &
\multicolumn{1}{c}{\textbf{$v$}}       &
\multicolumn{1}{c}{\textbf{$b$}}       &
\multicolumn{1}{c}{\textbf{$u$}}       &
\multicolumn{1}{c}{\textbf{$w1$}}      &
\multicolumn{1}{c}{\textbf{$m2$}}      &
\multicolumn{1}{c}{\textbf{$w2$}}      \\  
\hline                                                                                                          
2018-08-16  &  58346  &         $-$         &         $-$         &         $-$       &        $-$        & 15.93 $\pm$ 0.07  &          $-$        \\
2018-08-23  &  58353  &         $-$         &         $-$         &         $-$       &        $-$        &        $-$        &   15.78 $\pm$ 0.06  \\
2018-08-30  &  58360  &         $-$         &         $-$         &  15.52 $\pm$ 0.05 &        $-$        &        $-$        &          $-$        \\
2018-09-06  &  58367  &         $-$         &         $-$         &         $-$       & 15.68 $\pm$ 0.06  &        $-$        &          $-$        \\
2018-09-13  &  58374  &         $-$         &         $-$         &         $-$       &        $-$        &        $-$        &   15.94 $\pm$ 0.07  \\    
2018-09-20  &  58381  &         $-$         &         $-$         &         $-$       &        $-$        &        $-$        &   15.89 $\pm$ 0.06  \\
2018-09-27  &  58388  &         $-$         &         $-$         &  15.57 $\pm$ 0.05 &        $-$        &        $-$        &          $-$        \\
\hline
\end{tabular}
\end{center}
\end{table*}

\setcounter{table}{1}
\begin{table*}
\caption{Observed magnitude of SBS 0846$+$513 obtained by {\em Swift}-UVOT.}
\label{UVOT_0846}
\begin{center}
\begin{tabular}{cccccccc}
\hline
\multicolumn{1}{c}{\textbf{Date (UT)}} &
\multicolumn{1}{c}{\textbf{MJD}}       &
\multicolumn{1}{c}{\textbf{$v$}}       &
\multicolumn{1}{c}{\textbf{$b$}}       &
\multicolumn{1}{c}{\textbf{$u$}}       &
\multicolumn{1}{c}{\textbf{$w1$}}      &
\multicolumn{1}{c}{\textbf{$m2$}}      &
\multicolumn{1}{c}{\textbf{$w2$}}      \\  
\hline
2011-08-30 & 55803 &        $-$        &        $-$        &        $-$        &  19.42 $\pm$ 0.17 &        $-$        &          $-$          \\
2011-09-15 & 55819 &  18.58 $\pm$ 0.38 &  19.88 $\pm$ 0.42 &  19.35 $\pm$ 0.35 &  19.46 $\pm$ 0.29 &  19.43 $\pm$ 0.28 &    19.37 $\pm$ 0.19   \\
2011-12-20 & 55915 &  $>$ 18.64        &  $>$ 19.81        &  $>$ 19.44        &  19.90 $\pm$ 0.22 &  19.53 $\pm$ 0.39 &    19.84 $\pm$ 0.29   \\ 
2011-12-27 & 55922 &        $-$        &  20.13 $\pm$ 0.42 &  19.85 $\pm$ 0.43 &  19.88 $\pm$ 0.35 &  19.58 $\pm$ 0.23 &    20.06 $\pm$ 0.31   \\ 
2011-12-28 & 55923 &  $>$ 18.91        &  $>$ 19.92        &  19.12 $\pm$ 0.31 &  20.11 $\pm$ 0.29 &  19.84 $\pm$ 0.40 &    20.11 $\pm$ 0.34   \\
2012-01-03 & 55929 &        $-$        &        $-$        &        $-$        &        $-$        &        $-$        &    20.32 $\pm$ 0.39   \\
2012-05-27 & 56074 &  16.64 $\pm$ 0.11 &  17.30 $\pm$ 0.09 &  16.84 $\pm$ 0.10 &  17.37 $\pm$ 0.13 &  17.65 $\pm$ 0.26 &    17.58 $\pm$ 0.11   \\
\hline
\end{tabular}
\end{center}
\end{table*}

\setcounter{table}{2}
\begin{table*}
\caption{Observed magnitude of PMN J0948$+$0022 obtained by {\em Swift}-UVOT.}
\label{UVOT_0948}
\begin{center}
\begin{tabular}{cccccccc}
\hline
\multicolumn{1}{c}{\textbf{Date (UT)}} &
\multicolumn{1}{c}{\textbf{MJD}}       &
\multicolumn{1}{c}{\textbf{$v$}}       &
\multicolumn{1}{c}{\textbf{$b$}}       &
\multicolumn{1}{c}{\textbf{$u$}}       &
\multicolumn{1}{c}{\textbf{$w1$}}      &
\multicolumn{1}{c}{\textbf{$m2$}}      &
\multicolumn{1}{c}{\textbf{$w2$}}      \\  
\hline
2008-12-05 &  54805  &  18.06 $\pm$ 0.25 &  18.41 $\pm$ 0.17 &  17.61 $\pm$ 0.13 &  17.48 $\pm$ 0.13 &   17.53 $\pm$ 0.15 &  17.58 $\pm$ 0.11  \\
2009-03-26 &  54916  &  17.34 $\pm$ 0.18 &  17.91 $\pm$ 0.14 &  17.17 $\pm$ 0.13 &  16.90 $\pm$ 0.12 &   16.89 $\pm$ 0.15 &  17.12 $\pm$ 0.10  \\
2009-04-15 &  54936  &  $>$ 16.50        &  18.20 $\pm$ 0.15 &  17.10 $\pm$ 0.11 &  17.11 $\pm$ 0.12 &   17.02 $\pm$ 0.12 &  17.13 $\pm$ 0.10  \\
2009-05-05 &  54956  &  16.85 $\pm$ 0.12 &  17.46 $\pm$ 0.11 &  16.54 $\pm$ 0.09 &  16.75 $\pm$ 0.10 &   16.91 $\pm$ 0.12 &  17.01 $\pm$ 0.09  \\
2009-05-10 &  54961  &  17.77 $\pm$ 0.22 &  18.20 $\pm$ 0.16 &  17.25 $\pm$ 0.12 &  17.06 $\pm$ 0.12 &   17.04 $\pm$ 0.13 &  17.21 $\pm$ 0.10  \\
2009-05-15 &  54966  &  18.41 $\pm$ 0.42 &  18.29 $\pm$ 0.21 &  17.49 $\pm$ 0.17 &  17.27 $\pm$ 0.15 &   17.18 $\pm$ 0.15 &  17.31 $\pm$ 0.11  \\
2009-05-25 &  54976  &  17.78 $\pm$ 0.19 &  18.18 $\pm$ 0.14 &  17.50 $\pm$ 0.13 &  17.28 $\pm$ 0.11 &   17.29 $\pm$ 0.12 &  17.28 $\pm$ 0.09  \\
\hline
\end{tabular}
\end{center}
\end{table*}  

\setcounter{table}{3}
\begin{table*}
\caption{Observed magnitude of PKS 1502$+$036 obtained by {\em Swift}-UVOT.}
\label{UVOT_1502}
\begin{center}
\begin{tabular}{cccccccc}
\hline
\multicolumn{1}{c}{\textbf{Date (UT)}} &
\multicolumn{1}{c}{\textbf{MJD}}       &
\multicolumn{1}{c}{\textbf{$v$}}       &
\multicolumn{1}{c}{\textbf{$b$}}       &
\multicolumn{1}{c}{\textbf{$u$}}       &
\multicolumn{1}{c}{\textbf{$w1$}}      &
\multicolumn{1}{c}{\textbf{$m2$}}      &
\multicolumn{1}{c}{\textbf{$w2$}}      \\  
\hline
2009-07-25  & 55037 &  $>$ 18.69        &  19.19  $\pm$ 0.27  &   18.47 $\pm$ 0.21  &  18.47 $\pm$ 0.20   &    18.20 $\pm$ 0.19   &    18.40 $\pm$ 0.14  \\
2012-04-25  & 56042 &  18.62 $\pm$ 0.35 &  19.76 $\pm$ 0.40   &   19.11 $\pm$ 0.33  &  18.89 $\pm$ 0.24   &    18.74 $\pm$ 0.13   &    18.49 $\pm$ 0.15  \\
2012-05-25  & 56072 &  18.67 $\pm$ 0.35 &  19.59 $\pm$ 0.32   &   18.60 $\pm$ 0.08  &  18.22 $\pm$ 0.16   &    18.45 $\pm$ 0.19   &    18.18 $\pm$ 0.12  \\
2012-06-25  & 56103 &  18.74 $\pm$ 0.39 &  19.23 $\pm$ 0.33   &   18.72 $\pm$ 0.31  &  18.40 $\pm$ 0.10   &    18.16 $\pm$ 0.18   &    18.51 $\pm$ 0.16  \\
2012-08-07  & 56146 &  18.61 $\pm$ 0.35 &  $>$ 19.94          &   19.20 $\pm$ 0.33  &  18.54 $\pm$ 0.20   &    18.71 $\pm$ 0.21   &    18.37 $\pm$ 0.13  \\ 
2012-08-08  & 56147 &  $>$ 18.81        &  19.15 $\pm$ 0.33   &   18.82 $\pm$ 0.36  &  19.12 $\pm$ 0.34   &    18.40 $\pm$ 0.20   &    18.45 $\pm$ 0.15  \\
2015-12-22  & 57378 &  18.26 $\pm$ 0.22 &  18.52 $\pm$ 0.18   &   17.89 $\pm$ 0.15  &  17.74 $\pm$ 0.13   &    17.79 $\pm$ 0.10   &    17.93 $\pm$ 0.11  \\
\hline
\end{tabular}
\end{center}
\end{table*}

\setcounter{table}{4}
\begin{table*}
\caption{Observed magnitude of FBQS J1644$+$2619 obtained by {\em Swift}-UVOT.}
\label{UVOT_1644}
\begin{center}
\begin{tabular}{cccccccc}
\hline
\multicolumn{1}{c}{\textbf{Date (UT)}} &
\multicolumn{1}{c}{\textbf{MJD}}       &
\multicolumn{1}{c}{\textbf{$v$}}       &
\multicolumn{1}{c}{\textbf{$b$}}       &
\multicolumn{1}{c}{\textbf{$u$}}       &
\multicolumn{1}{c}{\textbf{$w1$}}      &
\multicolumn{1}{c}{\textbf{$m2$}}      &
\multicolumn{1}{c}{\textbf{$w2$}}      \\
\hline
2011-12-26 & 55921  & 17.64 $\pm$ 0.22 & 18.18 $\pm$ 0.16 & 16.80 $\pm$ 0.10 & 16.97 $\pm$ 0.11 & 16.88 $\pm$ 0.12 & 16.85 $\pm$ 0.08  \\
2015-04-08 & 57120  &       $-$        &       $-$        & 17.32 $\pm$ 0.07 &       $-$        &       $-$        &       $-$         \\
2015-04-09 & 57121  & 17.46 $\pm$ 0.22 & 18.05 $\pm$ 0.17 & 17.38 $\pm$ 0.16 & 17.30 $\pm$ 0.17 & 17.62 $\pm$ 0.19 & 17.27 $\pm$ 0.09  \\ 
2015-05-05 & 57147  & 17.88 $\pm$ 0.28 & 18.26 $\pm$ 0.19 & 17.24 $\pm$ 0.15 & 17.42 $\pm$ 0.11 & 17.52 $\pm$ 0.18 & 17.49 $\pm$ 0.14  \\
2015-06-05 & 57178  &       $-$        &       $-$        &      $-$         & 17.46 $\pm$ 0.11 & 17.52 $\pm$ 0.12 &       $-$         \\
2015-07-05 & 57208  &       $-$        &       $-$        & 17.16 $\pm$ 0.07 &       $-$        &       $-$        &       $-$         \\
2015-08-05 & 57239  & 17.50 $\pm$ 0.19 & 17.88 $\pm$ 0.16 & 16.85 $\pm$ 0.12 & 16.98 $\pm$ 0.11 & 17.05 $\pm$ 0.14 & 16.88 $\pm$ 0.11  \\ 
2015-09-05 & 57270  & 17.77 $\pm$ 0.18 & 18.06 $\pm$ 0.16 & 17.20 $\pm$ 0.13 & 17.08 $\pm$ 0.12 & 17.05 $\pm$ 0.12 & 17.21 $\pm$ 0.10  \\ 
2017-02-27 & 57811  & 17.38 $\pm$ 0.14 & 17.91 $\pm$ 0.10 & 16.88 $\pm$ 0.08 & 16.83 $\pm$ 0.10 & 16.98 $\pm$ 0.10 & 17.02 $\pm$ 0.09  \\
2017-03-07 & 57819  & 17.42 $\pm$ 0.13 & 17.97 $\pm$ 0.11 & 17.02 $\pm$ 0.09 & 17.12 $\pm$ 0.11 & 17.33 $\pm$ 0.12 & 17.19 $\pm$ 0.09  \\ 
2018-01-18 & 58136  &       $-$        &       $-$        & 17.34 $\pm$ 0.05 &       $-$        &       $-$        &       $-$         \\
\hline
\end{tabular}
\end{center}
\end{table*}

\setcounter{table}{5}
\begin{table*}
\caption{Observed magnitude of PKS 2004--447 obtained by {\em Swift}-UVOT.}
\label{UVOT_2004}
\begin{center}
\begin{tabular}{cccccccc}
\hline
\multicolumn{1}{c}{\textbf{Date (UT)}} &
\multicolumn{1}{c}{\textbf{MJD}}       &
\multicolumn{1}{c}{\textbf{$v$}}       &
\multicolumn{1}{c}{\textbf{$b$}}       &
\multicolumn{1}{c}{\textbf{$u$}}       &
\multicolumn{1}{c}{\textbf{$w1$}}      &
\multicolumn{1}{c}{\textbf{$m2$}}      &
\multicolumn{1}{c}{\textbf{$w2$}}      \\  
\hline
2011-05-15 &   55696  &   18.56 $\pm$ 0.19 &  20.25 $\pm$ 0.33  &  19.27 $\pm$ 0.21  &  19.90 $\pm$ 0.38 &  20.09 $\pm$ 0.47 &  $>$ 20.51        \\
2011-07-14 &   55756  &   19.45 $\pm$ 0.50 &  19.48 $\pm$ 0.25  &  19.19 $\pm$ 0.25  &  19.42 $\pm$ 0.35 &  $>$ 19.71        &  $>$ 20.09        \\
2011-07-19 &   55761  &   18.67 $\pm$ 0.31 &        $-$         &  $>$ 18.96         &  $>$ 19.40        &  $>$ 19.29        &  $>$ 19.63        \\
2011-07-23 &   55765  &   19.63 $\pm$ 0.45 &  20.58 $\pm$ 0.47  &  19.27 $\pm$ 0.22  &  19.60 $\pm$ 0.32 &  $>$ 20.18        &  19.84 $\pm$ 0.33 \\
2011-08-10 &   55783  &         $-$        &        $-$         &        $-$         &  19.57 $\pm$ 0.24 &  $-$              &        $-$        \\
2011-08-13 &   55786  &  18.69 $\pm$ 0.33  &  19.56 $\pm$ 0.31  &  19.32 $\pm$ 0.34  &  19.68 $\pm$ 0.37 &  $>$ 19.86        &  20.31 $\pm$ 0.35 \\
2011-09-05 &   55809  &  18.64 $\pm$ 0.47  &  19.58 $\pm$ 0.51  &  18.85 $\pm$ 0.40  &  19.20 $\pm$ 0.42 &  $>$ 19.55        &  19.79 $\pm$ 0.37 \\
\hline
\end{tabular}
\end{center}
\end{table*}

\clearpage

\section{Multifrequency light curves}\label{MWL_lc}

\newpage

\begin{figure*}
\begin{center}
{\includegraphics[width=0.75\textwidth]{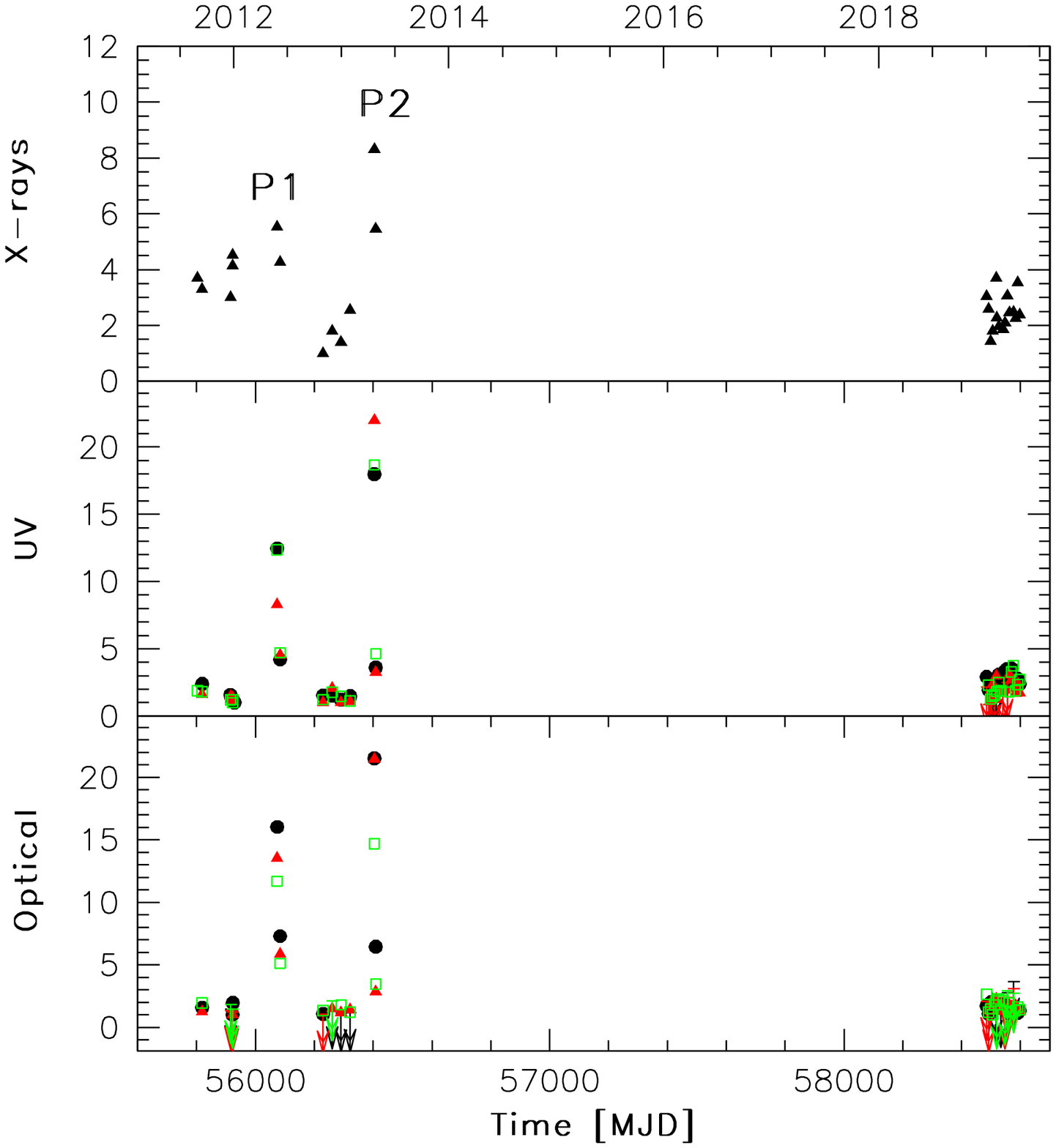}}
\caption{Multifrequency light curve of SBS 0846$+$513 normalized to the minimum value observed in X-rays (0.3--10 keV; top panel), UV bands ($w1$, open squares; $m2$, filled triangles; $w2$, filled circles; middle panel), and optical band ($v$, open squares; $b$, filled triangles; $u$, filled circles; bottom panel).}
\label{0846_online}
\end{center}
\end{figure*}

\begin{figure*}
\begin{center}
{\includegraphics[width=0.75\textwidth]{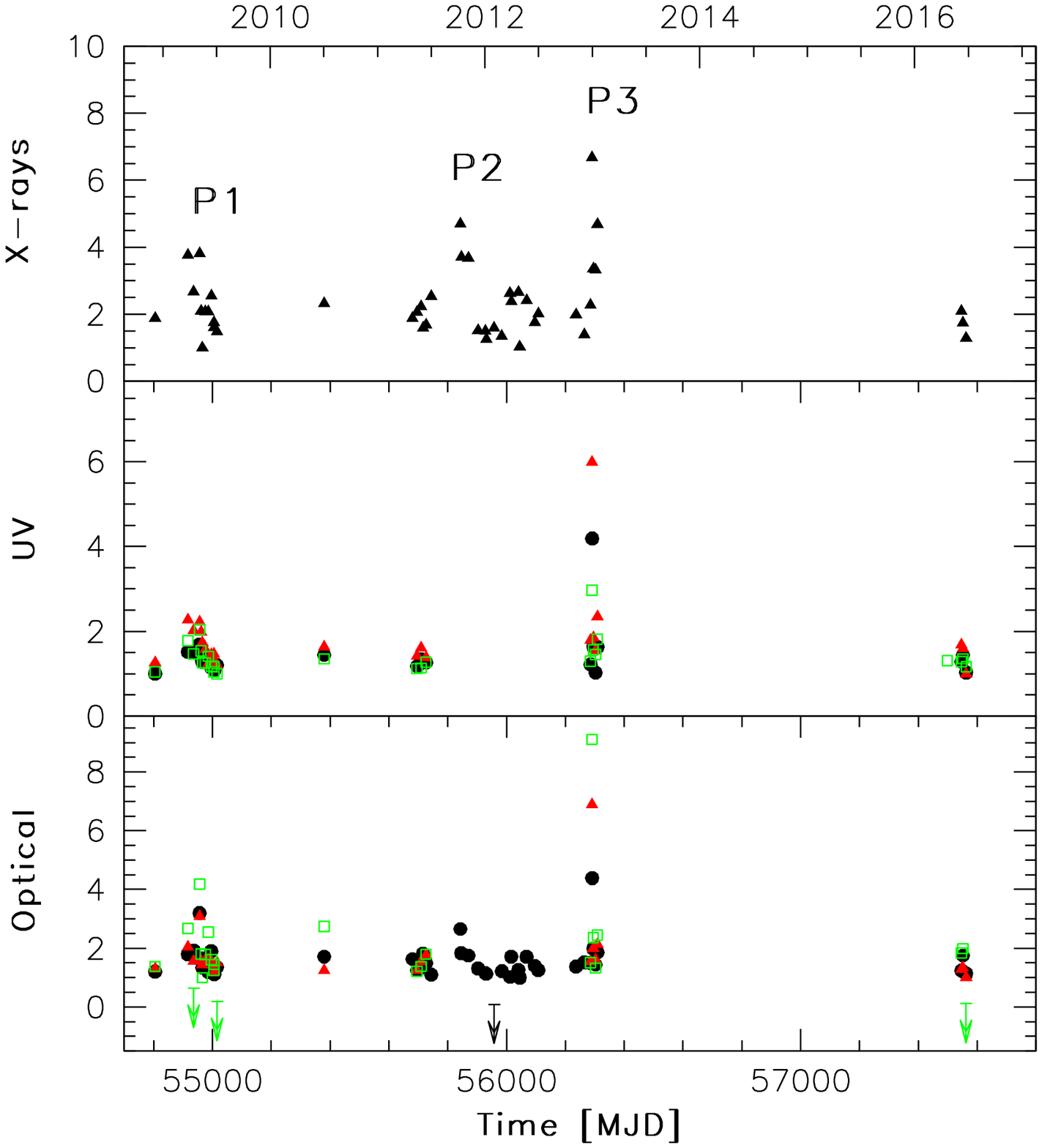}}
\caption{Multifrequency light curve of PMN J0948$+$0022 normalized to the minimum value observed in X-rays (0.3--10 keV; top panel), UV bands ($w1$, open squares; $m2$, filled triangles; $w2$, filled circles; middle panel), and optical band ($v$, open squares; $b$, filled triangles; $u$, filled circles; bottom panel).}
\label{0948_online}
\end{center}
\end{figure*}

\begin{figure*}
\begin{center}
{\includegraphics[width=0.75\textwidth]{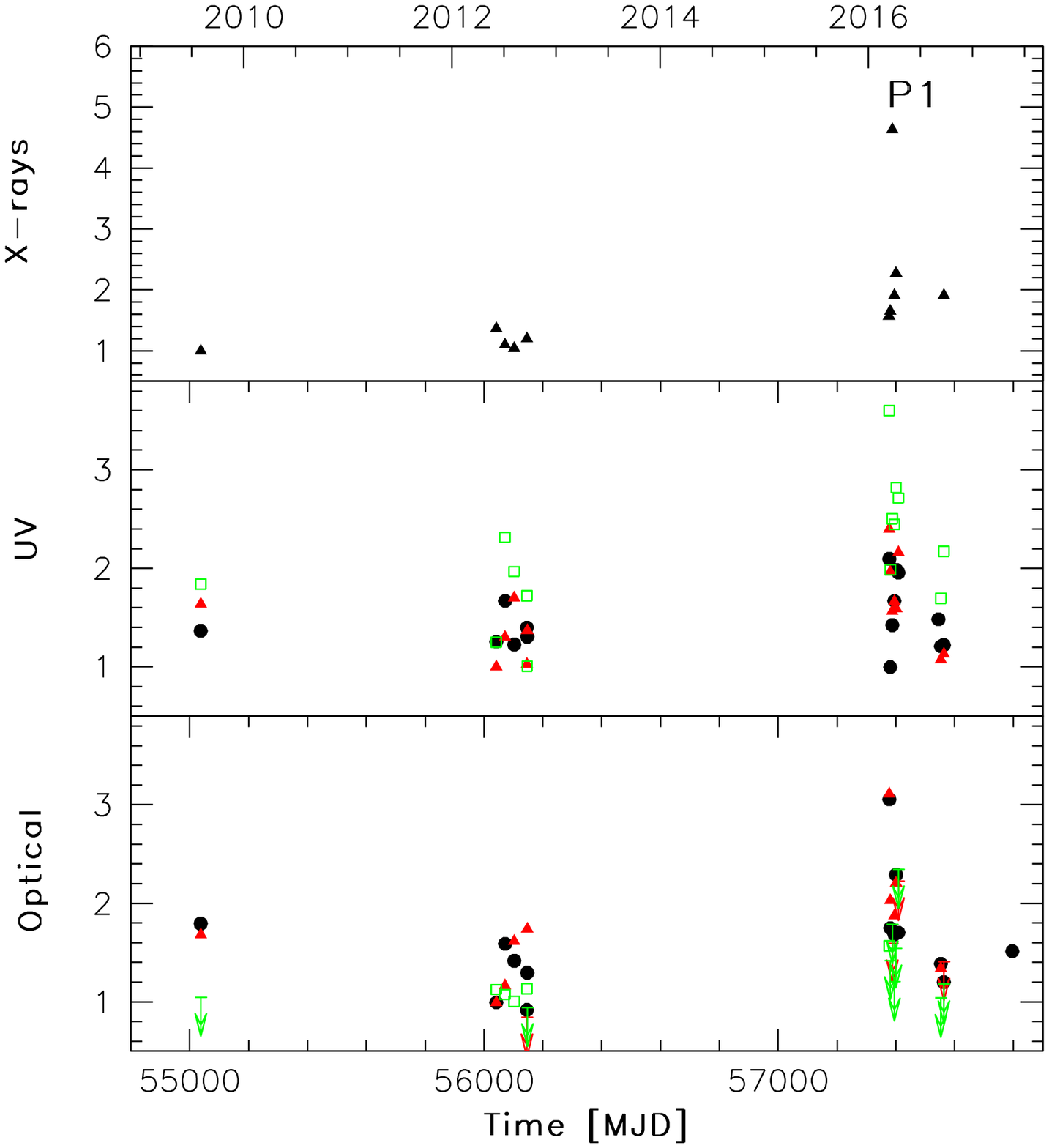}}
\caption{Multifrequency light curve of PKS 1502$+$036 normalized to the minimum value observed in X-rays (0.3--10 keV; top panel), UV bands ($w1$, open squares; $m2$, filled triangles; $w2$, filled circles; middle panel), and optical band ($v$, open squares; $b$, filled triangles; $u$, filled circles; bottom panel).}
\label{1502_online}
\end{center}
\end{figure*}

\label{lastpage}

\end{document}